\documentclass[12pt,preprint]{aastex}
\usepackage{graphicx}
\slugcomment{Accepted by ApJS}
\shorttitle{Spitzer Atlas}
\shortauthors{Ardila et al.}
\begin{document}
\title{The {\it Spitzer} Atlas of Stellar Spectra (SASS)}
\author{David R. Ardila}
\affil{NASA Herschel Science Center, California Institute of Technology, Mail Code 100-22, Pasadena, CA 91125}
\email{ardila@ipac.caltech.edu}
\author{Schuyler D. Van Dyk, Wojciech Makowiecki, John Stauffer}
\affil{Spitzer Science Center, California Institute of Technology}
\author{Inseok Song}
\affil{University of Georgia at Athens, Department of Physics and Astronomy}
\author{Jeonghee Rho, Sergio Fajardo-Acosta, D.W. Hoard, Stefanie Wachter}
\affil{Spitzer Science Center, California Institute of Technology}

\begin{abstract}
We present the {\it Spitzer} Atlas of Stellar Spectra (SASS), which includes 159 stellar spectra (5 to 32 $\mu$m; R$\sim$100) taken with the Infrared Spectrograph on the {\it Spitzer Space Telescope}. This Atlas gathers representative spectra of a broad section of the Hertzsprung-Russell diagram, intended to serve as a general stellar spectral reference in the mid-infrared. It includes stars from all luminosity classes, as well as Wolf-Rayet (WR) objects. Furthermore, it includes some objects of intrinsic interest, like blue stragglers and certain pulsating variables. All the spectra have been uniformly reduced, and all are available online. 

For dwarfs and giants, the spectra of early-type objects are relatively featureless, dominated by Hydrogen lines around A spectral types. Besides these, the most noticeable photospheric features correspond to water vapor and silicon monoxide in late-type objects and methane and ammonia features at the latest spectral types.  Most supergiant spectra in the Atlas present evidence of circumstellar gas. The sample includes five M supergiant spectra, which show strong dust excesses and in some cases PAH features. Sequences of WR stars present the well-known pattern of lines of HeI and HeII, as well as forbidden lines of ionized metals. The characteristic flat-top shape of the [Ne III] line is evident even at these low spectral resolutions. Several Luminous Blue Variables and other transition stars are present in the Atlas and show very diverse spectra, dominated by circumstellar gas and dust features. We show that the [8]-[24] {\it Spitzer} colors (IRAC and MIPS) are poor predictors of spectral type for most luminosity classes. 

\end{abstract}


\section{Introduction}

The cryogenic phase of the {\it Spitzer Space Telescope} came to a close on 15 May 2009. The Infrared Spectrograph (IRS, see \citealt{hou04}), one of the three scientific instruments in the {\it Spitzer} payload, allowed for spectroscopic observations from 5-40 $\mu$m, at resolutions R$ =\lambda/\Delta \lambda \sim$ 60 - 600. The end of ``cold {\it Spitzer}'' marked the end of the astronomical community's access to this spectral region from space until the launch of the James Webb Space Telescope ({\it JWST}). 

Here we present the {\it Spitzer} Atlas of Stellar Spectra (SASS), composed of objects observed with the IRS in low-resolution mode (R $\sim$ 60 - 130), throughout its almost six year lifetime. As detailed below, most of the 159 stars in this Atlas have been chosen because they are typical representatives of their spectral class. Beyond this general prescription, a few have been chosen for their intrinsic interest. All the spectra in the Atlas have been uniformly reduced and all are available to the astronomical community from the NASA/IPAC Infrared Science Archive (IRSA) and Vizier. They will also be temporarily available from the first author's webpage \footnote{http://web.ipac.caltech.edu/staff/ardila/Atlas/}. The Appendix lists the full content of the SASS. 

The operational goal of this project is to compile representative spectra of a broad section of the Hertzsprung-Russell  (HR) diagram, in order to understand the MK spectral sequence (which is a blue-optical classification) at mid-infrared wavelengths. The data are intended to serve as a general stellar spectral reference in the mid-infrared and to aid in the interpretation of observations from other facilities such as the Stratospheric Observatory For Infrared Astronomy (SOFIA) and the Herschel Space Observatory. In addition, given its population of massive stars, the Atlas may serve to refine galactic synthesis models. 

This Atlas inherits a rich history of efforts in collecting and classifying the mid-infrared spectra of stars. Most relevant for our work, \citet{oln86} produced an Atlas of 5425 spectra using the Low-Resolution Spectrometer (LRS; 7.7 - 22.7 $\mu$m and R$\sim$20 - 60) on board the {\it Infrared Astronomical Satellite} ({\it IRAS}) and \citet{kwo97} compiled LRS spectra of $\sim$11,200 sources. Those {\it IRAS} data are complemented by spectra obtained with the Short Wavelength Spectrometer (SWS; 2.4 - 45.4 $\mu$m, R$\sim$400) on the {\it Infrared Space Observatory} ({\it ISO}). \citet{kra02} and \citet{slo03} compiled $\sim$1000 {\it ISO} stellar spectra in order to derive a system of infrared (IR) spectral classification. Additional compilations based on {\it IRAS} and {\it ISO} mid-infrared data can be found in the literature (e.g. \citealt{vol89,hod04,eng04}).

The samples on those catalogs were limited by the sensitivity of the facilities used and did not provide complete coverage of the spectral sequence.  In partcular, they were dominated by giants or by evolved stars with strong circumstellar emission. On the other hand, the emphasis of the SASS is on naked photospheres although, as we will show, it includes stars with circumstellar material as well. 

The IRAS and ISO observations have served to develop infrared spectral classifications. The original {\it IRAS}-LRS classification scheme \citep{iras1985} sorted spectra according to a continuum spectral index or to absorption or emission features. \citet{vol89} devised a parallel classification for {\it IRAS}-LRS spectra, which  \citet{kwo97} used to classify the sources in their catalog. The {\it IRAS}-LRS data have also been classified using Autoclass, a Bayesian-based automatic technique \citep{goeb1989}. The stars from the SASS have been classified within the ISO-based KSPW system, as described by  \citet{kra02} and \citet{slo03} (see Section \ref{contents}).

Beyond its utility as a general reference, the Atlas' characteristics of completeness, uniformity, and availability, make it useful for exploring the role of effective temperature and surface gravity on the development of infrared spectroscopic features. We discuss here in some detail the behavior of the SiO fundamental feature and the H$_2$O bending mode in late type giants, as well as in the circumstellar environment of M supergiants. 

In this paper, we will focus only on a general description of the sample. The paper is organized as follows: In section \ref{spitzer} we summarize the characteristics of the IRS on {\it Spitzer}. We then present a description of the sample (section \ref{selec}) , the data reduction (section \ref{reduc}), and the reliability and limitations of the reduction (section \ref{relab}). While the contents of the Atlas in the Appendix are listed in conventional luminosity class order (WR, Class I, Class II, etc) a better context can be established by describing the sample in the opposite order. Therefore, in the text we analyze the SASS contents (section \ref{contents}) from Main Sequence (section \ref{ms}) to WR (section \ref{WR}). Section \ref{summary} presents a summary.

\section{The Infrared Spectrograph for the {\it Spitzer Space Telescope} \label{spitzer}}

Here we provide a short summary of the IRS as relevant for the SASS.  \citet{hou04} provide a detailed description of the instrument. More details about the observing modes and the data processing pipeline are available from the IRS Instrument Handbook \footnote{http://ssc.spitzer.caltech.edu/irs/irsinstrumenthandbook/}.

The IRS had two low-resolution gratings. The Short-Low grating (SL) covered the spectral region from 5 to 14 $\mu$m, while the Long-Low (LL) grating nominally covered the region from 14 to 40 $\mu$m. The resolution ranged from R $\sim$60 to $\sim$130, increasing linearly with wavelength in each module. For each grating, the complete spectral orders 1 and 2 were imaged. In addition, a short spectral segment belonging to order 1 (the so-called 'bonus' order) provided spectral redundancy with the long wavelength region of the second order. A different detector was used to image the spectra from the SL and LL gratings. 

The spectral slits were 57" $\times$ 3.7"  (SL) and 168" $\times$ 10.7"  (LL) and Nyquist-sampled at the longest wavelengths in each module. All the data included in the SASS were obtained in the IRS Staring mode. In this pointing mode, the telescope `nods' along the slit, and each observation results in two spectral traces separated by 1/3 of the slit length (20'' for SL, 55'' for LL). 

The data presented here were reduced with the default parameters of the Spitzer Science Center (SSC) IRS point-source pipeline S18.7.0. The pipeline averages multiple images at the same nod and subtracts the average of the images taken at one nod from the average at the other. While this method of sky subtraction works well for isolated point sources, it will fail for large extended sources. The nod-subtracted spectra are extracted by the default pipeline using a tapered aperture with a wavelength-dependent width. Finally, the pipeline divides the extracted spectra by a low-order polynomial (the `fluxcon') to obtain calibrated spectra. Therefore, the IRS Staring Mode observations result in two sky-subtracted, calibrated spectra per order for a total of 12 spectra per target. 

In the S18.7.0 pipeline, the low-resolution modules of the IRS are calibrated using a single standard star, the K1 giant HR~7341. During calibration observations, all the individual extracted spectra of this star are averaged and the ratio between the observations and the MARCS model of the star \citep{dec04} is fit with a low-order polynomial. The result of the fit becomes the fluxcon. 

\section{Sample Selection \label{selec}}

The goal of this project is to sample a broad region of the HR diagram, and produce spectra from 5 to 35 $\mu$m of typical members from a given class. From all the IRS Staring observations that became public before 31 May 2009, we primarily selected those stellar targets that had been observed with all of the low-resolution IRS modules. 

During our exploration of the {\it Spitzer} archive it became clear that some spectral classes had not been observed with the low-resolution modules. In addition, certain unique objects such as RR~Lyrae stars, or certain classes like Cepheid variables and blue stragglers had not been observed  with the low-resolution modules of the IRS. We requested, and subsequently received, a total of 29.6 hours of Director's Discretionary Time in order to obtain the IRS observations of missing or incomplete classes of stars (DDT 485, PI Ardila).

Because the Atlas' spectra should serve as templates to better understand the spectral classes to which they belong, we did not include known young stars with circumstellar material, stars known to harbor debris disks, or objects classified in SIMBAD as RS CVn stars, Be stars, or eclipsing binaries. These objects do not have a characteristic spectrum or a prototype that could describe the class as a whole. For example, there is no prototypical debris disk spectrum that describes all of them, nor a prototypical spectrum of an eclipsing binary that would describe all of its phases. In addition, rejecting those objects allows us, for the most part, to classify stars without reference to their evolutionary state, as a given place in the HR diagram will generally be occupied by only one class of objects. 

\begin{figure}
\epsscale{1}
\plotone{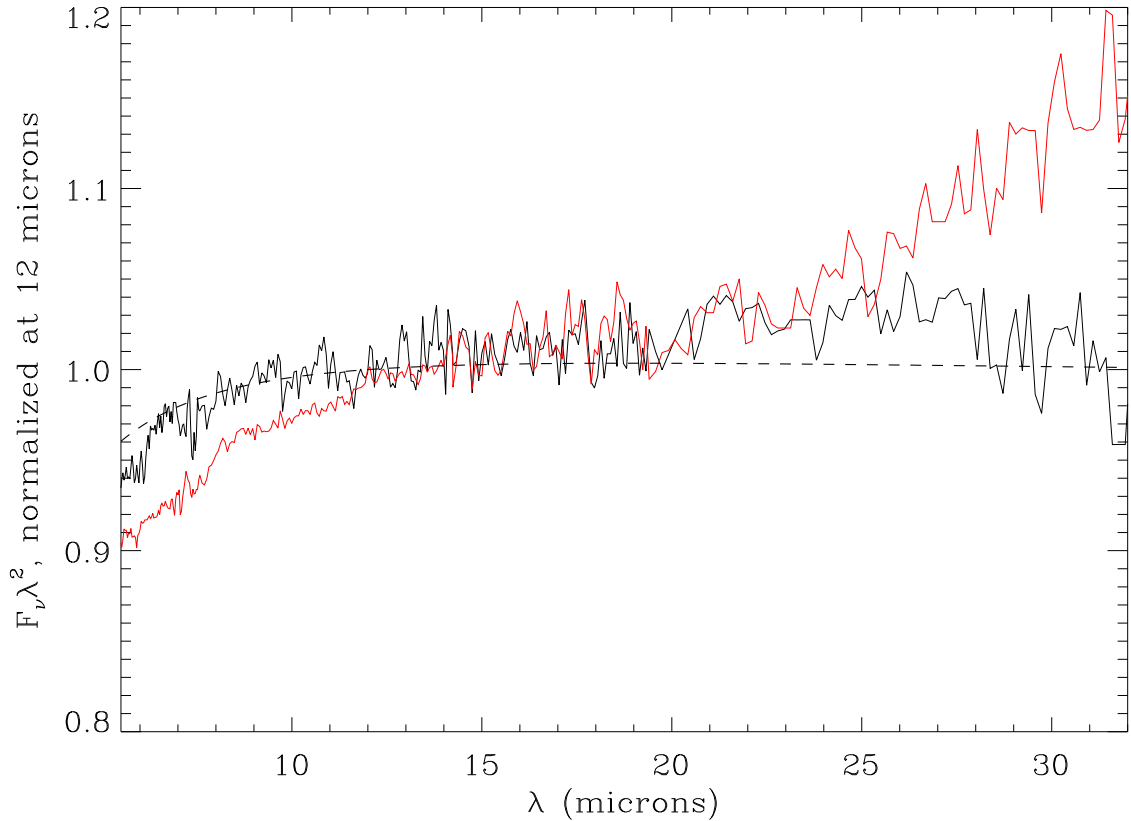} 
\caption{When plotted in the so-called "Rayleigh-Jeans units" ($\propto$ F$_\nu\lambda^2$), most stellar spectra should be relatively flat and excesses become easily noticeable. The black traces show the spectrum of the G1V star HD~168009 (solid) and of the Engelke function (dashed) that best matches its spectrum. The red trace shows the spectrum of the F0V star HD~27290. The latter has a clear spectral excess. We include it in the Atlas because it is the prototype for the $\gamma$ Dor variables.\label{engelke}}
\end{figure}

In the case of bright giants (luminosity class II), giants (class III), subgiants (class IV), and dwarfs (class V), we generally rejected targets presenting IR excesses. Beyond $\sim10 \mu$m, the photospheric IR continuum is relatively flat for spectral types earlier than mid-M. For each candidate star, we compared its spectrum with (depending on the effective temperature) a blackbody function, available stellar templates (\citealt{cohen2003, dec04}), or an Engelke function (\citealt{eng92}; Figure \ref{engelke}). Our determination of excess is qualitative, and low-level excess may still be present in the spectra. 

In the case of very high mass and/or evolved stars there are few objects presenting a pure photospheric spectrum. Pulsations and winds, both common occurrences in high mass stars, will often result in an IR excess. Extinction by interstellar dust along the line of sight will also result in a non-photospheric spectrum.  In the case of supergiants, most spectra included in the Atlas are clearly non-photospheric at the wavelengths considered here. In some of them, the presence of emission lines indicates the existence of gas in the vicinity of the star  (section \ref{supergiants_gen_desc}).  

Beyond prototypes for each class, the Atlas also includes a few stars specifically selected for their intrinsic interest, either because they represent optical prototypes of some variable types, or because they have not been well studied in the IR. These were included regardless of their IR excess and even if the Atlas already contained another star with the same spectral type. Some examples are: HD~205021 (B1IIIev, the prototype for the $\beta$ Cepheid variables), HD~182989 (F5III, the RR Lyrae prototype), V836 Oph (M4III, Mira Ceti type variable), HD~27396 (B3V, another example of $\beta$ Cep type), HD~27290  (F0V, the $\gamma$ Doradus prototype), and HD~88923, HD~106516, HD~35863 (some of the few blue stragglers amenable to IRS observations, see \citealt{and95}). In general, we have avoided classes that have been fully described with {\it IRAS}, {\it ISO}, or {\it Spitzer}, such as Asymptotic Giant Branch stars (AGBs, e.g. \citealt{sloan2008}).

Although the IRS range extends to 40 $\mu$m, the S/N of the targets included in the SASS decreases rapidly beyond $\sim$32 $\mu$m (see Section \ref{reduc}). The spectra in the Atlas are presented to 35$\mu$m, although we only describe them to 32 $\mu$m. When multiple observations of targets with a given spectral type were present in the archive, we selected the highest signal-to-noise (S/N) representative for inclusion in the Atlas.

Not all SASS members have coverage in all IRS wavelengths. While we aimed for most targets to have full spectral coverage, this proved impossible at the faintest spectral classes (late M, L and T dwarfs). These objects are unique enough that we have included them here, independently of whether or not the complete spectral range beyond 14 $\mu$m is available. For the same reason, we have included some WR stars for which the long wavelength modules are unusable (due to extended nebulosity which precludes the sky subtraction) or not present in the archive.

For each target, we searched the literature to find its spectral type. In general, the spectral types were taken from (in order of priority): NStED\footnote{http://nsted.ipac.caltech.edu/}, NStars\footnote{http://nstars.nau.edu/nau\_nstars/about.htm}, the Tycho-2 Spectral Type Catalog \citep{wri03}, and SIMBAD. We assigned luminosity classes in the same priority. NStED bases its spectral types primarily on the Michigan Spectral Type Catalog, as does NStars. The spectral types on the Tycho-2 catalog are determined by cross-referencing the catalog with primary and secondary spectral type references, including the Catalogue of Stellar Spectra Classified in the Morgan-Keenan System \citep{jas1964}, the Fifth Fundamental Catalogue (\citealt{fri1988, fri1991}), etc. SIMBAD uses all these and complements them by current literature observations. For certain types of objects, specialized catalogs in the literature served as the source of the spectral types. For example, the spectral types from WR stars were taken from the 7th Catalog of Galactic Wolf-Rayet stars \citep{vanderhucht01}. Other references are indicated in the Appendix.

Figure \ref{graph} presents a general overview of the Atlas contents.  This should be taken only as a cartoon representation of the HR diagram. In particular, the luminosities have generally been taken from the expected luminosities based on the stellar spectral type, not from actual brightness measurements.   The full content of the Atlas is presented in the Appendix.

\section{Data Reduction \label{reduc}}

For the data presented here, we used the regions of the spectrum indicated in Table \ref{trim}. While we kept data up to 35 $\mu$m in LL1, fluxes beyond 32 $\mu$m fall on a lower sensitivity and significantly damaged region of the detector. The spectra at the longest wavelengths present very high levels of unstable, rapidly varying (``rogue'') pixels. In order to reduce the number of rogues, the bias level of the LL detector was reduced on 29 October 2007\footnote{http://ssc.spitzer.caltech.edu/irs/features/}. While reducing the number of rogues in all regions of the spectrum, this procedure also reduced the overall sensitivity. The combination of large number of remaining rogues plus low S/N at long wavelengths pushed us to only use data to 32 $\mu$m.

We have not used any rogue cleaning algorithms, such as IRSClean\footnote{http://ssc.spitzer.caltech.edu/dataanalysistools/tools/irsclean/}. The sky subtraction procedure we used eliminates the slowly-varying rogues and these are the ones described within IRSClean.

Rogue pixels and low sensitivity are also a problem on LL2 (14 to 20 $\mu$m), resulting in noticeably higher noise in this region. For wavelengths beyond 19.35 $\mu$m, we replaced the values from LL2 by those of the same spectral region in the bonus order, which presents less rogue pixels.  Still, the region between 19 and 20 $\mu$m tends to be noisier than the other wavelengths in most of the spectra. 

For each target we calculated the offset distance between its predicted coordinates and those of the slit, and rejected those observations with offsets larger than 1.5". This corresponds to the 2$\sigma$ pointing error in the dispersion direction for medium accuracy peak-up \footnote{http://isc.astro.cornell.edu/IRS/TechnicalReports/irstr03001.pdf}.Telescope mispointing larger than this results in spectral traces with spurious curvature.

The goal of the S18.7.0 pipeline is to provide spectra with systematic tilts in the spectra smaller than $\sim$5\%. In certain regions, these systematics can be reduced even further, by applying additional corrections to the calibrated data provided by the pipeline. When these corrections can be described by a polynomial, the coefficients are listed in Table \ref{correc}. The corrections that were applied to the pipeline-processed data are as follows:

\begin{itemize}

\item Teardrop correction: Excess emission centered at 13.2 - 15 $\mu$m is observed in the two-dimensional spectral images, to the left of the SL1 spectral trace. In the S18.7.0 calibration this spectral region is not included when deriving the flux calibration. When extracting the spectra using the default point source extraction, the amplitude of the feature is $\sim$10\% larger than the expected spectral trace\footnote{http://ssc.spitzer.caltech.edu/irs/irsinstrumenthandbook/102/\# \_Toc253561116}. We compared the effect of the teardrop on 90 dwarf stars with spectral types from A to M and found the strength of the feature to be proportional to the spectrum brightness. This fact allow us to derive a single polynomial correction which we then apply to all the data (Table \ref{correc}). 

\item Slit position correction: To make use of the better knowledge of the in-orbit performance of the spectrograph, the Spitzer Science Center (SSC) changed the slit position tables starting on 19 June 2006. These tables determine the orientation of the slit on the focal plane. The S18.7.0 pipeline provides an average calibration, applied equally to data before and after this date. We used archival observations of HR~7341 to derive corrections to the spectra ($<2$\%) depending on the date they were taken (Table \ref{correc}). 

\item Residual nod correction: The calibration derived by the SSC and included in the S18.7.0 pipeline is obtained by averaging multiple staring observations of a single source. For a given observation involving two nods, the spectrum from the source falls on two different detector positions. The calibration of the average spectrum in effect corrects for residual flat-field errors. In some of the Atlas sources we discard one of the nods, which may result in spectrum tilt errors of the order of $\sim$2\%. In order to be able to use only one nod, we derive nod-specific polynomial corrections based on HR~7341 (Table \ref{correc}).  

\item Residual model corrections: Even after these corrections, significant order curvature is present in sources that should have relatively flat spectra, like A0 stars. The cause of this curvature is unknown. The largest excursion from the observations with respect to stellar templates occurs at $\sim$8$\mu$m. A possible source of error is the way the MARCS model of HR~7341 describes the absorption wings of the strong SiO fundamental feature. We compared archival observations of the A0 star $\alpha$ Lac to a template model for this star \citep{coh99} to determine residual corrections to the spectra (Table \ref{correc}).

\item The 24 $\mu$m flux deficit: Spectra of faint sources ($<$100 mJy) processed with S18.7.0 present a flux deficit at 24 $\mu$m \citep{sloan2004}, as well as anomalous slopes at shorter wavelengths\footnote{http://ssc.spitzer.caltech.edu/irs/irsinstrumenthandbook/96/\#\_Toc272477717}. For these sources, we determined an additional multiplicative correction, based on 120 sec ramps of $\alpha$ Lac and its corresponding stellar template \citep{coh99}.

\end{itemize}

In addition, fringing is observed in the extracted spectra of HD~269858, RSGC 2 \#2, RSGC 2 \#5, BD+23$^o$1138, and NGC 7419 \#435. We have corrected it by applying a 5 wavelength element (0.7 $\mu$m) boxcar smoothing from 20 to 30 $\mu$m.

After applying these corrections, multiple spectra of the same source are averaged. Order mismatches are corrected by matching the spectra to the highest value. Remaining tilt errors, calculated by comparing individual nod observations for the same target,  are $\sim2\%$. The error in the overall flux level is that of the standard IRS calibration ($\sim$5\% to $\sim$10\%). 

\section{Reliability \label{relab}}

The Atlas files provided electronically contain an error value for each wavelength, which is intended to represent the random 1$\sigma$ error at that wavelength.  This is the error  provided by the SSC 18.7.0 pipeline and propagated along the reduction procedure. However, the treatment of errors remains incomplete in this pipeline \footnote{http://ssc.spitzer.caltech.edu/irs/irsinstrumenthandbook/}. For example, errors in the flat field are not included, and Poisson errors are calculated after dark subtraction. A full study of the validity of the errors remains to be performed. For the purpose of the Atlas, the relative value of the errors seems appropriate to the characteristics of the spectra: overall, the errors are larger in spectral regions that are clearly noisier than others. Absolute noise levels are more suspect, with errors changing substantially between neighboring wavelength bins, specially for faint sources ($<$10mJy). The errors provided here should be considered carefully, before propagating them into further calculations. 

The default (tapered-column) pipeline extraction tends to extend the influence of the rogues over neighboring wavelength bins,  giving the appearance of unresolved emission or, in background-subtracted data, absorption lines. Sky subtraction, as well as averaging of multiple frames, eliminates the rogue pixels that vary on timescales longer than the exposure time. Furthermore, the observations included here have all been performed in the IRS Staring pointing mode, which means that two spectra are available for each pointing. Before averaging we have compared the spectra to each other, to check for anomalously high or low pixels in one nod (pixels with values much larger or smaller than the noise), and replaced them by the value of the spectrum in the other nod. Still, especially at longer wavelengths (where there is more detector damage),  bad pixels do remain. 

As was already mentioned, due to the combination of large number of rogue pixels and low spectral sensitivity, the spectral regions from $\sim$14 -- 20 $\mu$m  and $>30\mu$m may be considerably noisier than the rest. Note for example the F2V star HD~164259, in the Appendix, which presents correlated noise in these ranges. In spite of its less-than-ideal noise characteristics, this star is included in the Atlas because the only other F2V in the sample is the blue straggler HD~88923 \citep{and95}.

The processing insures that the spectra do not have strong spurious emission or absorption lines in large S/N regions, that could be confused with true spectral features. 

At $\sim$6 $\mu$m, an absorption feature is sometimes observed (see the F5V spectrum in the Appendix). For repeated observations of the same star, the exact position and depth of this feature changes with nod and AOR. Because at 6 $\mu$m the PSF is narrower than one pixel, the presence of the feature is likely due to intra-pixel variations, important only at the shortest wavelengths \citep{leb2010}. To obtain the final spectra, we have rejected from the average those spectra in which this feature is largest. However, the ``absorption feature" remains present in some spectra. Care must be exercised in interpreting narrow features in this region.

\section{General Description \label{contents}}

As Figure \ref{graph} shows,  the data span almost 12 orders of magnitude in luminosity and 2 orders of magnitude in temperature. The Atlas is most complete at spectral types later than G and luminosity classes III and V. 

The spectra have not been corrected for extinction. The reddening correction is negligible for nearby (intrinsically faint) stars, as 1 mag of extinction in the Johnson V band corresponds to $\sim$0.05 mag at 8 $\mu$m \citep{schlegel98}. On the other hand, distant (intrinsically bright) stars show broad interstellar absorption features at 10 and 18 $\mu$m, and the continuum slope is likely affected by extinction (section \ref{supergiants_gen_desc}). 

The spectra of most dwarf and giants without circumstellar material are relatively featureless, although for objects with early A spectral types, Hydrogen lines are observed for all luminosity classes. Besides these, the most noticeable photospheric features correspond to water vapor and silicon monoxide in late type dwarfs and giants, as well as methane and ammonia features at the latest spectral types.

Figure \ref{sptvscolor} shows that the [8]-[24] {\it Spitzer} colors (IRAC and MIPS, see \citealt{reach2005, engel2007}) are poor predictors of spectral type for most luminosity classes, as expected. Within a color uncertainty of 0.1 mag, the spectral slopes of most of the sample can be described with Engelke or Blackbody functions. Even the development of the SiO feature (section \ref{giantssection}) results in a color difference $<$ 5\% between M dwarfs and giants. The exceptions are stars for which the excess begins at wavelengths $<24\mu$m. As Figure  \ref{sptvscolor} shows, this is the case with most of the supergiants. 

In Table \ref{big_table} we have also classified the stars according to the {\it ISO}-based, KSPW system described by \citet{kra02} and \citet{slo03}. When the class or subclass to which the star belongs is uncertain, we have marked it as `:'. The KSPW is a three-level system that takes into account not only the features present in the spectra but also the spectral slope. Therefore, the fact that the spectrum presents evidence of extinction becomes part of the classification (unlike the optical classification). As with other spectral classifications, the KSPW system is tied to the instrument and sample that defines it, in this case, {\it ISO}-SWS, 2.4 -- 45.2 $\mu$m, R$\sim$1000 observations of mostly evolved stars. The range and resolution of the spectra presented here are smaller, and our coverage is more extensive, which makes the application of this classification problematic to implement. Regardless of these issues, the KSPW system provides a shorthand to describe the spectra. Our attempts to classify the SASS members within the KSPW aim to highlight the areas where more work is needed before a complete, general purpose IR stellar classification is available.

Within the KSPW system, group 1 describes naked stars, with spectra dominated by ``photospheric emission with no apparent influence from circumstellar dust"  \citep{kra02}. Within group 1, objects are further subdivided as 1.N (no molecular bands), 1.NO (CO and/or SiO absorptions), or 1.NC (molecular absorption bands indicative of carbon-rich photosphere, HCN, and C$_2$H$_2$ at 7-8 $\mu$m). As the {\it Spitzer} IRS range does not cover the CO band at 4.4 $\mu$m, we use only the SiO absorption at 8$\mu$m to distinguish between 1.N and 1.NO \citep{her02}. None of the stars in the SASS presents the absorption features characteristic of 1.NC objects \citep{mat2006}.  Stars with interstellar silicate dust absorption but no noticeable excess have been marked as 1.SA (photospheric slope, silicate absorption). Other extensions and interpretations of the system are mentioned in the following sections. 

In the discussion that follows, we concentrate on a few of the classes present in the Atlas. We will discuss dwarfs, giants, and supergiants, followed by WR stars. 

\subsection{Main sequence \label{ms}}

As was mentioned, at these resolutions and sensitivities the spectra of main sequence stars are relatively featureless until M0V (Figure \ref{allV}), except for the presence of Hydrogen lines.  For example, in the high quality spectrum of HD~213558 (A0V), we observe H (9-6) at 5.91 $\mu$m, H (13-7) at  6.29  $\mu$m, H (12-7) at  6.77  $\mu$m,  the blend of H (6-5) and H (8-6) at 7.50 $\mu$m, H (10-7) at 8.76 $\mu$m, H (13-8) at 9.38 $\mu$m, H (12-8) at 10.50 $\mu$m, H (9-7) at 11.30 $\mu$m, H (7-6) at 12.40 $\mu$m, and H (13-9) at 14.20 $\mu$m. This particular spectrum is the coadd of 48 spectra per wavelength bin. The Humpreys line 5.91 $\mu$m and the 7.50 $\mu$m blend are observed in all other luminosity classes, for early A stars.

The fundamental SiO band (blue edge at 7.55 $\mu$m) first becomes noticeable at M0V and remains strong until it becomes confused with the H$_2$O band at 6.75 $\mu$m. The red wing of the CO fundamental absorption band, at 4.4 $\mu$m, is noticeable already at G1V, although it is difficult to determine the exact spectral type at which the feature first appears, as most of the band falls beyond the blue edge of the {\it Spitzer} spectra. 

For spectral types later than M1V, the same sample of stars included here, with the exception of HD~191849, has been published and analyzed as a group by \citet{cus06} and our description follows theirs. The water vapor bands at 5.8 $\mu$m and 6.75 $\mu$m are noticeable as early as in M0.5V (HD~28343). In later spectral types the water bands result in a spurious ``emission feature" at $\sim$6.25 $\mu$m (e.g.\ the M8 dwarf V1298 Aql). Silicate dust clouds \citep{cus06, ste09} are partially responsible for the broad and shallow absorption feature at  $\sim$9 $\mu$m observed in mid-L spectral types (for example, in the L4.5 object 2MASSJ2224-0158).  The $\nu_4$ fundamental band of CH$_4$ at 7.65 $\mu$m appears in the latest L dwarfs and becomes stronger through the T sequence, and the combination of H$_2$O and CH$_4$ absorption from 4 to 9 $\mu$m suppresses the flux in the spectra of the T dwarfs. The $\nu_2$ fundamental band of NH$_3$ centered at $\sim$10.5 $\mu$m appears in the spectra of the early- to mid-type T dwarfs. Methane and ammonia bands were first detected in brown dwarfs by \citet{roe2004}. 

The KSPW classification does not contemplate the development of methane and ammonia bands at these very late spectral classes, and the lack of the full spectral range precludes a full classification. We have marked their KSPW class as ``:". 

\subsection{Giants \label{giantssection}}

The spectra of the earliest giants in the Atlas show emission lines of [OIV] at 25.87 $\mu$m, HeI, and HI (e.g. HD~205021, B1IIIev in Figure \ref{all_giants}). For HD~24912, HD~36861 and HD~205021 (a $\beta$ Ceph variable and binary system, see \citealt{wheel2009}), the spectral slopes are not photospheric, indicating the presence of circumstellar material and/or line-of-sight extinction. For the rest of the class III, the most noticeable features are the SiO fundamental $v=1-0$ band and the H$_2$O $\nu_2$ band at $\sim$6.75 $\mu$m (Figure \ref{all_giants}).  

The SiO band is observed as early as K0III, and its equivalent width (EQW) increases with spectral type (Figure \ref{eqw}). The EQW plateaus around M1III and remains strong until M6III, the latest giant spectral type in the Atlas. Water is clearly observed at M0III (although it may appear at earlier types) and it remains present at later spectra types. The presence of both features, as well as the increase of SiO equivalent width at later spectral types was noted by \citet{her02} in their study of SWS {\it ISO} spectra.

The silicon monoxide and water vapor features exhibit different behaviors with gravity: the SiO lines become stronger with lower gravity (positive luminosity effect, see Figure \ref{eqw}), while the H$_2$O lines become weaker with lower gravity (negative luminosity effect). For example, as was already mentioned, the water absorption bands at 5.8 $\mu$m and 6.75 $\mu$m result in a false emission feature at $\sim$6.3 $\mu$m in late type dwarfs such as DX Cnc (M6.5V), while only the 6.75  $\mu$m feature is clearly observed in giants like HD~8680 (M6III). 

The EQW values that we measure are comparable to the ones presented by \citet{her02}. In principle, both our measurements and theirs may be systematically high, as the silicate 10 $\mu$m interstellar absorption could contribute to the measured EQW. However, the broad silicate dust absorption at 10 $\mu$m is not evident in any of the giant spectra. For comparison, an silicate dust feature as strong as that of HD~14433 would contribute only 0.05 $\mu$m to the total EQW measurent. Even such a weak silicate feature would be evident in the giant spectra, on top of the SiO gas feature. We therefore conclude that the systematic increase in EQW due to extinction is less $<$0.05 $\mu$m.

The SiO fundamental band has been modeled by \citet{ari97}, using the \citet{jor92} version of the MARCS code. For spectral types later than $\sim$M0, those models fail to reproduce the data presented in Figure \ref{eqw}. In particular, from the Atlas data the maximum EQW for the SiO fundamental band is around 0.25 $\mu$m for M5 giants. For objects with temperatures $\sim$3000K (typical for mid-M) the models predict EQW values between $\sim$0.5 and $\sim$0.7 $\mu$m. For class V stars the models also predict EQWs much larger than observed.  In summary, while the values that we measure may be systematically high, the predicted values are much higher. 

Because the primary calibrator of the IRS is the MARCS model of the K0III star HR~7341 \citep{dec04}, it is fair to ask whether the discrepancies between models and observations mentioned above impact on our data. The observed SiO EQW for HR~7341 is the K0III point included in Figure \ref{eqw}. While the models presented by \citet{ari97} cover spectral types later than this, it is clear that discrepancies with the data become smaller at earlier spectral types (Figure 4 from \citealt{ari97}). The latest \citet{dec04} model available for a giant is that of the K5III star $\gamma$ Dra\footnote{http://ssc.spitzer.caltech.edu/irs/calibrationfiles/decinmodels/}. In this model, the EQW is predicted to be 0.14 $\mu$m, comparable to what we measure for K5 giants. $\gamma$ Dra is not available in the Atlas.

These arguments suggest that errors in the EQW of the SiO feature of HR~7341 are small. However, errors in the shape of the silicate feature may still be partially responsible for the residual model correction that we need to apply to reproduce featureless spectra (Section \ref{reduc}). The study of the full effect of the discrepancies from the \citet{ari97} models on the calibration of the IRS remains is beyond the scope of this paper. 

A measurement of the water EQW in the Atlas giants is not very instructive. Because the blue edge of the 6.75 $\mu$m water band abuts the red wing of the CO fundamental absorption \citep{dec04}, EQW measurements in the former are likely to be affected by opacity changes in the latter. Figure \ref{h2o} shows that, in spite of this, the overall strength of the water vapor feature is remarkably constant over the whole M-giant range. As with SiO, the opacity saturates at around M1III.  

While the presence of water lines in the spectra of M giants is well known, the physical region in which they are formed remains in doubt. Although hydrostatic stellar atmospheres models do not predict H$_2$O features for stars hotter than M6III, water vapor in mid-K to early-M giants was observed in {\it ISO} near-infrared spectra (e.g. \citealt{tsuji2001}). The {\it ISO} features are attributed to a molecular layer situated within a few stellar radii beyond the photosphere but different from the cool, expanding circumstellar envelope. This idea of a molecular layer ("MOLsphere"; see \citealt{mats1999} and references therein) has been used to explain discrepancies between synthetic spectra and near-infrared observations. However, a growing body of spectroscopic data for a range of red giants and supergiants suggests that classical model atmospheres are inadequate \citep{tsuji2003, mcd2010}. It is unclear whether the water vapor features observed in giants and dwarfs have a common origin.

\subsection{Supergiants \label{supergiants}}
\subsubsection{General Description \label{supergiants_gen_desc}}
Most Class I stars included in this Atlas do not have a purely photospheric spectrum. Early-type stars (Figure \ref{new_early_super}) present a rapidly rising (in units of F$_\nu\lambda^2$) continuum, as well as atomic emission lines. Both characteristics are evidence of wind emission or a circumstellar gas disk (e.g. \citealt{hillier1983, lenorzer2002}). 

The early O stars in the Atlas are described in SIMBAD as ``emission line" stars, based on the optical catalog of \citet{wack1970}. In our sample, most stars with spectral types earlier than A present emission lines of helium and hydrogen. 

The Atlas includes one example of an OC class member, the O9.7Ia star HD~152424. Members of this class present typical main-sequence abundances, as opposed to the morphologically-normal majority of OB stars which already have CNO-cycled material mixed into their atmospheres and winds  (e.g. \citealt{walborn2000}). At the wavelengths presented here, the HI+HeI complexes at 7.5 and 12.4 $\mu$m are comparable in EQW to those of earlier-type stars.

The KSPW system defines Group 2 as sources in which ``the SEDs are primarily photospheric at shorter wavelengths, but they also show noticeable or significant dust emission at longer wavelengths"  \citep{kra02}. Although none of the early-type supergiants with non-photospheric spectra present any photospheric or dust features, their overall spectrum decreases with wavelength (in F$_\nu$ vs. $\lambda$) from the blue-most edge. We therefore classify them as ``2.:".

Spectra of early and mid-type supergiants present broad silicate absorption at 10 $\mu$m (Figures \ref{new_early_super} and \ref{midsuper}) and some present 18 $\mu$m absorption (see the spectrum of Cyg OB2 No.12 in Figure \ref{new_early_super}). These are indicative of line-of-sight extinction, which likely plays a role in the overall shape of the spectrum. The models by \citet{ari97} predict that the SiO equivalent width should be very large at these low gravities. Indeed, the K2Iab star HD~45829 presents absorption at 8 $\mu$m suggestive of SiO absorption. However, the spectral coverage of the data presented here is not complete enough to test model predictions. 

\subsubsection{M supergiants \label{msuperg}}

Figure \ref{latesuper} shows the five M-supergiant spectra available in the Atlas. All present strong excesses but can be divided in two groups.  NGC 7419 \#435 (M2Iab), NGC 7419 \#139 (M3.5Iab), and BD+23$^o$1138 (M5Ia) show weak Polycyclic Aromatic Hydrocarbon (PAH) emission lines and broad emission bands at 7 and 12 $\mu$m. RSGC2 \#2 (M3Iab) and RSGC2 \#5 (M4Iab) present broad bands at 11 $\mu$m and 11.5 $\mu$m respectively, and at 18 $\mu$m. RSGC 2 is one of the largest clusters of red supergiants (RSGs) known, with a group of 26 associated stars at the base of the Scutum-Crux arm \citep{dav07}. NGC 7419 has 5 RSGs \citep{bea94}. No previous mid-IR spectroscopy is available in the literature for any of the targets presented here.

The presence of PAHs in RSG is well known. \citet{sylvester94} first reported PAHs on M-supergiant spectra in the h and $\chi$ Persei cluster, although always on top of silicate features. The morphology of sharp PAH features on top of broad, redder plateaus, has been observed in the Orion bar by \citet{bregman1989}, who conclude that a mixture of large (400 or more C atoms) and small (20 to 30 C atoms) PAH clusters is consistent with their 3-14 $\mu$m observations. In the case of the cool supergiants presented here, the appearance of PAHs in the spectrum may be due to a ``Pleiades effect:" Excitation of highly processed PAHs on interstellar material by the stellar radiation field, as has been observed by \citet{sloan2008}. In the classification scheme of \citet{peeters2002} these PAH features would correspond to Class A. The contrast of the features presented here is not high enough for a definite classification. 
 
We classify NGC 7419 \#435, NGC 7419 \#139, and BD+23$^o$1138  as  ``2.U:", where the ``U" stands for Unidentified Infrared Features, now commonly associated with PAHs.

RSGC2 \#2 and RSGC2 \#5 do not present sharp PAH features, and for both, colder dust dominates the emission. While the broad 18 $\mu$m feature could be identified with silicate emission, the emission feature at shorter wavelengths peaks at $\sim$11 $\mu$m for RSGC2 \#2 and $\sim$11.5 $\mu$m for RSGC2 \#5,  too red to be amorphous silicates and bluer that the 12 $\mu$m plateaus from the previous group. The SiC emission feature at 11 $\mu$m is an obvious candidate, although the narrow C$_2$H$_2$ feature at 13.7 $\mu$m, common in other carbon-rich objects, is not observed \citep{buc2006}. On the other hand, for oxygen-rich stars, similar features have been explained as optically thin shells of amorphous alumina dust grains (Al$_2$O$_3$; see \citealp{egan2001}). However, this is based on the analysis of AGBs, not supergiants. \citet{sloan1998} conclude that supergiants generally produce dust shells composed of amorphous silicates and that the presence of alumina in these spectra is rare. The true shape of the complex may also be confused by silicate absorption at 10 $\mu$m. In the KSPW system group 3 consists of ``sources ... dominated by emission from warm dust" \citep{kra02}.  We classify these objects as ``3.:" as dust emission dominates the SED.

\subsection{Wolf-Rayet stars \label{WR}}

Figures \ref{WN}, \ref{WC}, and \ref{WO} show sequences of WNs, WCs, and WOs. For those stars with spectra shown only to 14 $\mu$m, the LL slit is contaminated by extended nebulosity. This is the case for all the WOs in the sample, as well as for WR~23 and WR~145. 

The WR spectra present the well-known pattern of lines of HeI and HeII, as well as forbidden lines of ionized metals \citep{morris2004}. At this wavelength range, the WN and WC classes are primarily distinguished by the strong emission of [Ni IV] (8.4 $\mu$m), [Ne II] (12.8 $\mu$m), and [Ne III] (15.5 $\mu$m) in WC stars.  The characteristic flat-top shape of the [Ne III] line, indicative of saturation in the outer WR wind \citep{morris2000}, is evident even at these low spectral resolutions. The almost perfect F$_\nu \propto \lambda^{-1}$ continuum shown in the early-type WN types, as well as some WC stars, is indicative of optically thick, free-free emission, from a constant velocity wind \citep{hillier1983}, and the lack of silicate absorption indicates that extinction is negligible. The WO stars form an extension of the WC ``early" sequence (WCE: WC4-6; See \citealt{crow2007}). For the sample included here, the two early-type WO stars are distinguished from WCE by the strong [S IV] line at 10.54 $\mu$m.

\citet{vander1996} report that 50\% of WC8 and 90\% of WC9 stars present dust spectral features which in some cases dominate the 5 -- 30 $\mu$m continuum. None of the WC stars included here presents strong dust features. WR53 (WC8d) and WR 103 (WC9d) do show silicate absorption and strong [Ne II] lines.

The binary system Brey 3a (part of the Large Magellanic Cloud -- LMC) shows dust continuum but at temperatures ($<$300 K)  lower than those indicated by \citet{vander1996}. Brey 3a also show strong [Ne II] (12.8 $\mu$m) and [S III] (18.7 $\mu$m), lines not detected in other WC stars in the Atlas (although observed in the WC6 star WR146, see \citealt{willis97}). While the spectral classification listed here is that of \citet{bre99}, we note that \citet{moffat1991} and \citet{hey1992} argue that this is not a WC9 star but an object in transition between an Of or Of? star and a WR star. Some emission from the nearby (5.2") cool giant GV 60 may be contributing to the IRS spectrum \citep{evd2001}.

WR stars in the SASS present a problem within the KSPW system as they generally do not show dust features. As in the case of the supergiants, we classify most of them within Group 2, on the basis of their non-photospheric but decreasing spectral slope (in F$_\nu$ vs. $\lambda$). Furthermore, \citet{kra02} describes the W subclass as spectra in which ``the continuum emission peaks at $\sim$6--12 $\mu$m, usually with apparent silicate absorption at 10 $\mu$m. The `W' stands for Wolf-Rayet, since these spectra are always produced by Wolf-Rayet stars."  However, none of our WR objects resembles the templates described for this class in \citet{kra02}. We therefore classify most of them as ``2.:".

\subsection{Luminous Blue Variables (LBVs) and other transition stars \label{lbv}}

Several stellar classes represent the transition between main sequence O stars and the highly evolved WR stars \citep{morris96, crow1995}. These transition classes include LBVs, early-type O supergiants,  and B[e] stars, some examples of which are present in the Atlas. For these, the nature of the circumstellar material is likely diverse. For example, both dense winds in LBVs and circumstellar clouds in B[e], have been implicated in describing spectra of transition stars \citep{lenorzer2002}. As pointed out by  \citet{morris96} a significant degree of overlap exists in the infrared spectral morphology of these transition stars.

LBVs (S-Dor type, Hubble-Sandage type) are very luminous, unstable hot supergiants, which present irregular eruptions. During outflow, the expanding photosphere is cool and may look like that of A supergiants \citep{hump1994}.  The Atlas includes eight stars that are classified in the literature as LBV or LBV candidates:  V1429~Aql (B[e]Ia), HD~326823 (B1.5Ie), Cyg OB2 No.12 (B5Ie), HD~183143 (B7Ia), HD~160529 (A2Ia), HD~269227 (WN9h; R84, LMC), HD~269858 (Ofpe/WN9; R127; LMC), and 2MASSJ0545-6714 (WN11h; LMC). 

Figure \ref{trans} shows four of these, giving an idea of the spectral diversity of these objects, the result of the different geometric orientations of the surrounding nebulae,  different phases in the evolution of the LBV, or different nature of the circumstellar material, as mentioned before. The above list of LBVs includes three WR stars. HD~269227 (see \citealt{mun2009} and references therein) is a well studied binary LBV candidate unique in the whole Atlas sample because of the presence of the strong amorphous silicate feature. Lower-temperature dust from the surrounding envelope is observed in HD~269858 and 2MASS J0545-6714. In these three objects, as in Brey 3a, strong [Ne II] (12.8 $\mu$m) and [S III] (18.7 $\mu$m) are detected. 

The spectra of the mid-B objects Cyg OB2 No.12 and HD~183143 are dominated by silicate absorption, although HeI lines at 9.7 and 12.4 $\mu$m are also present. On the other hand, the spectrum of the B[e]Ia star V1429~Aql presents a pattern of emission lines (strong HeI and HeII but weak metallic lines) similar to early-type O stars like the O5f+ star HD~14947, although it has been suggested (e.g.\  \citealt{muratorio2008}) that the (optical) emission lines are emitted from a circumstellar disk and that the system is binary. For HD~326823, the most noticeable characteristic of the spectrum presented here is the SiC dust emission feature at 11.5 $\mu$m. \citet{marcolino2007} conclude that the system is a severely hydrogen-depleted and helium-rich, pre-WN8 star. HD160529, the latest spectral type among the LBVs presented here, shows strong line emission at [Ne II] (12.8 $\mu$m) and [S III] (18.7 $\mu$m). \citet{sta03} considers it intermediate between an LBV and a normal supergiant. 

\subsection{Other groups}
\begin{itemize}
\item Blue stragglers: These are thought to originate from normal main-sequence stars that have undergone a recent increase in mass \citep{mathieu2009}, due to collisions or mass transfer between stars in multiple systems (see, e.g. \citealt{perets2009, mathieu2009}). The Atlas contains three objects classified as field blue stragglers: HD 88923 (F2V, see \citealt{and95}), HD 106516 (F6V, \citealt{car01}), HD 35863 (F8V, \citealt{abt84}). Their spectra are similar to those of other F stars in the Atlas: mostly featureless within the noise, although HD~88923 has a slightly non-photospheric slope. \citet{car01} suggest that the field blue-straggler phenomenon primarily involves mass transfer between objects, and that all field blue stragglers are part of binary systems. In our small sample, only HD~106516 is recognized as being part of a binary system \citep{ccdm}. For the three stars presented here, the mass-transfer mechanism does not produce any strong signal at IRS wavelengths: no strong gas lines, dust features, nor extra dust or gas continuum. That lack of any strong IR signal is surprising, and sets limits to the rate of mass transfer between the binary members. Such detailed analysis, however, is beyond the scope of this paper. 

\item Cepheids and other variables: The Atlas includes pulsating variables classified in the literature as Cepheid, $\delta$ Cep type, $\beta$ Cep type, $\gamma$ Dor type, RR Lyr, and $\delta$ Scu type (See comments column in Table \ref{big_table}). Of the 11 stars identified as pulsating variables, 5 present non-photospheric continua: HD~90772 (A9Ia, Cepheid), HD~205021 (B1IIIev,  $\beta$ Cep prototype), HD~27396 (B3V,  $\beta$ Cep type), HD~27290 (F0V, $\gamma$ Dor prototype), and V836 Oph (M4III, Mira type). Pulsating variables with luminosity classes III and brighter also present 10 $\mu$m silicate absorption. Otherwise, the range is bereft of strong spectral features, with the exception of emission HeI+HI lines at 7.48 and 12.379 $\mu$m for HD~205021.

Observations of HD~205021 by \citet{wheel2009} indicate that this is a binary system with a classical Be star secondary. The existence of this companion was suspected from episodic variability of the H$\alpha$ line, and it is also the likely responsible for the emission lines (see for example the spectra of the B[e] star V1429~Aql, in this Atlas). On the other hand, the spectra of HD~27396 (a ``slowly pulsating B star"; see \citealt{chap98}) and HD~27290 ($\gamma$ Dor) suggest the presence of a dusty debris disk (see e.g. \citealt{chen2006}). The M4 giant, Mira-type star V836 Oph shows strong excess continuum as well as dust features at 13 $\mu$m and 20 $\mu$m. The 10 - 15 $\mu$m region is similar to the early-type SE classes from \citet{sloan1998},  which are dominated by thin shells of alumina dust \citep{egan2001}.

\end{itemize}
\section{Summary \label{summary}}

The {\it Spitzer} Atlas of Stellar Spectra is composed of 159 stellar spectra obtained during the cryogenic phase of the {\it Spitzer Space Telescope} (see Table \ref{big_table} and Figures  \ref{panel_WR1} to \ref{panel_V8}). The goal of the SASS is for the spectra to serve as a general stellar reference and to aid in the interpretation of SOFIA, Herschel, and JWST spectra. Most of the stars were observed with both of the IRS low-resolution gratings. All the spectra from SASS have been uniformly reduced and are available online from IRSA, Vizier, and the first author's webpage\footnote{web.ipac.caltech.edu/staff/ardila/Atlas/}. The spectra have resolutions R$ =\lambda/\Delta \lambda \sim$ 60 - 130 and nominal wavelength ranges from 5-  35 $\mu$m, although fluxes beyond 32 $\mu$m fall on a low-sensitivity, significantly damaged regions of the detector, and present very high levels of unstable, rapidly varying pixels.

All the stars included here were observed in the IRS Staring pointing mode, which results in two spectra per observation. This redundancy helps to distinguish real from spurious spectral features. The nods were subtracted from each other to produce sky-subtracted spectra. The spectra were reduced with the S18.7.0 point source pipeline, provided by the Spitzer Science Center, and the pipeline products were further processed to reduce systematic errors. The data were corrected for telescope mispointing, 'teardrop', and residual errors in the calibration. None of the spectra was corrected for interstellar extinction.

\begin{figure} 
\epsscale{1}
\plotone{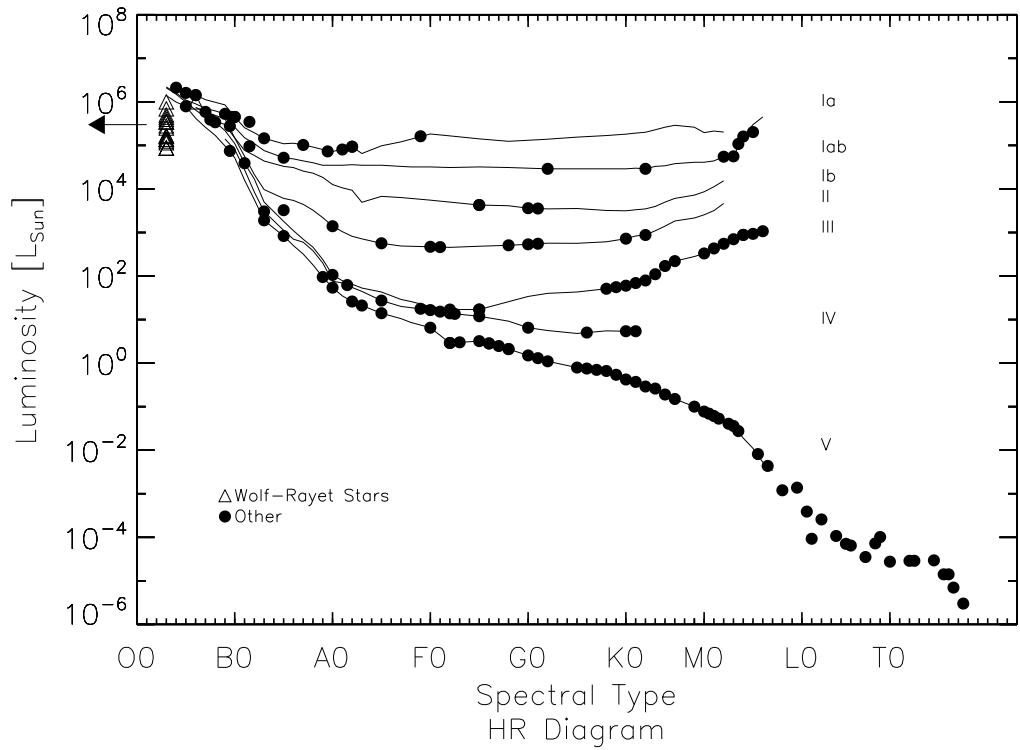}
\caption{A summary view of the contents of the Atlas. Note that for most classes the vertical axis does not represent the actual luminosity of an object, but its predicted luminosity based on the spectral type. For Wolf-Rayet stars and those with types later than M6, the luminosities have been taken or estimated from the literature for the individual objects.  \label{graph}}
\end{figure}

\begin{figure}
\epsscale{1}
\plotone{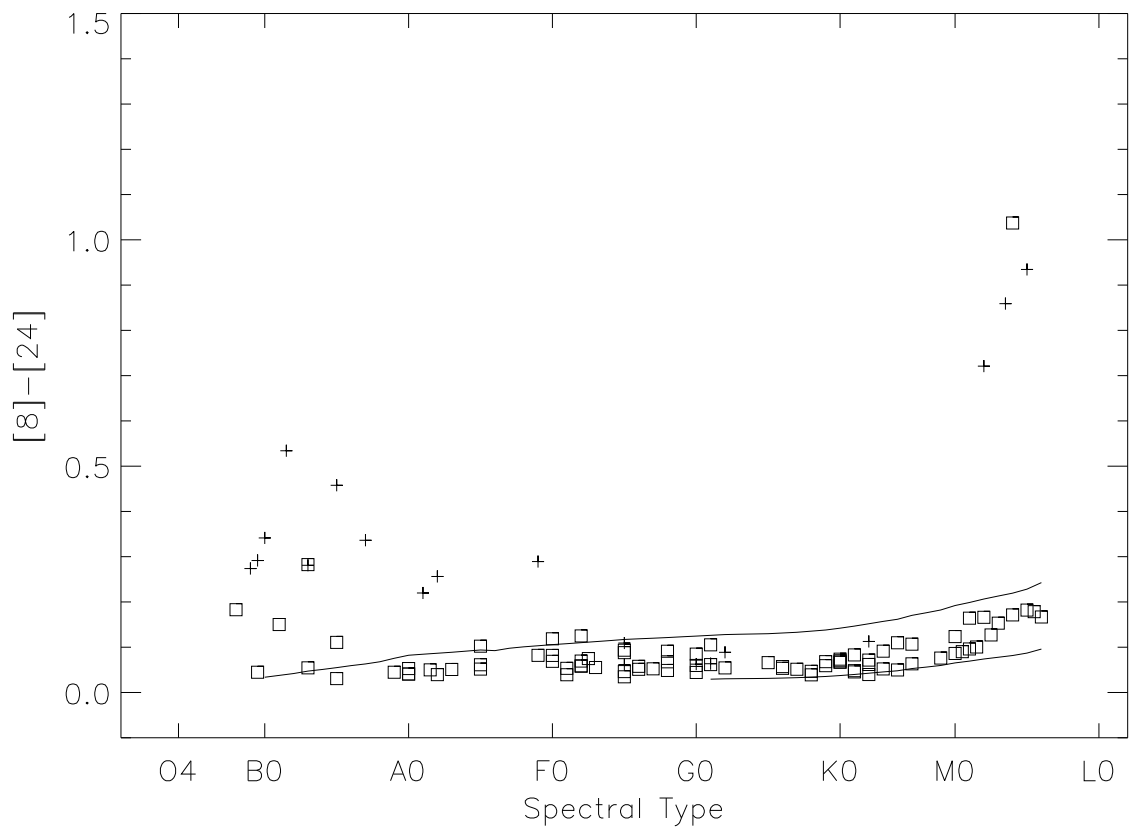} 
\caption{Comparison between different continuum models. The solid lines are models assuming the spectra of pure Engelke functions (bottom solid line) or pure Blackbody functions (top solid line).  $+$: Luminosity Class I; $\square$ : Luminosity Classes II-V. In the plot, the largest [8]-[24] color for a Class I star corresponds to BD+23$^o$1138 (M5Ia). The largest from Classes II-V corresponds to V836Oph (M4III). Not shown: RSGC 2 \#2 (M3Iab, [8]-[24]=2.85 mag) and RSGC 2 \#5 (M4Iab, [8]-[24]=2.97 mag). \label{sptvscolor}}
\end{figure}

\begin{figure}
\epsscale{1}
\plotone{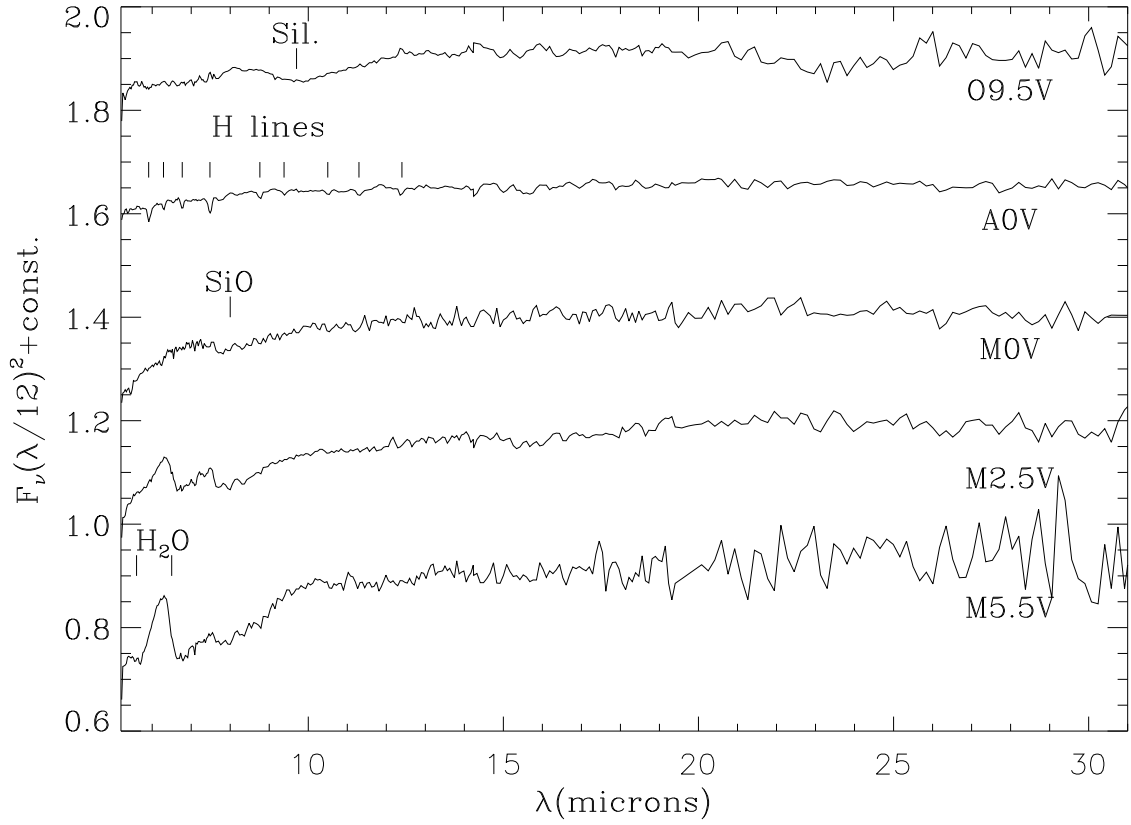} 
\caption{Spectra of dwarfs. From top to bottom, they are HD~149757 (O9.5V), HD~213558 (A0V), HD~85512 (M0V), HD~180617 (M2.5V), GJ~65AB (M5.5V). \label{allV}}
\end{figure}

\begin{figure}
\epsscale{1}
\plotone{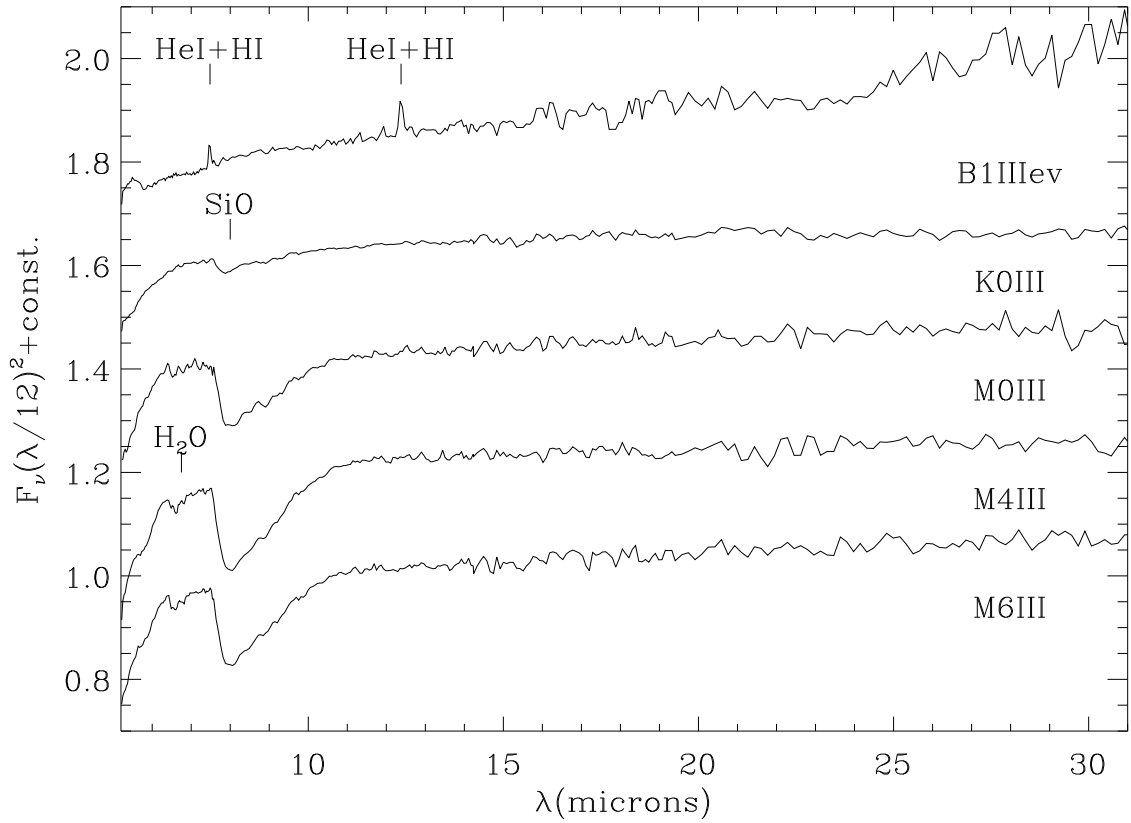} 
\caption{Spectra of class III objects. From top to bottom they are HD~205021 (B1IIIev),  HD~181597 (K0III), HD~107893 (M0III), HD~46396 (M4III), HD~8680 (M6III). \label{all_giants}}
\end{figure}

\begin{figure}
\epsscale{1}
\plotone{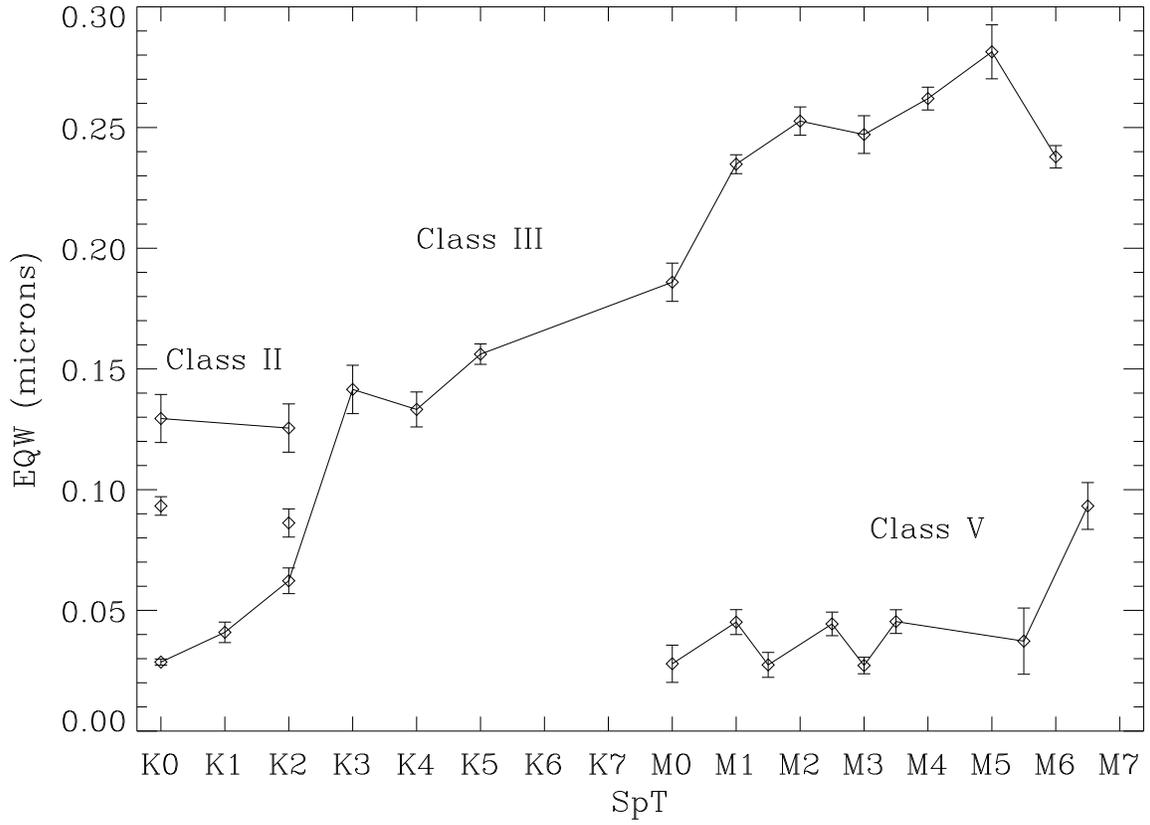} 
\caption{Equivalent width of the SiO fundamental absorption. The two points not joined by a line are HD121146 (K0IV) and HD45829 (K2Iab). \label{eqw}}
\end{figure}

\begin{figure}
\epsscale{1}
\plotone{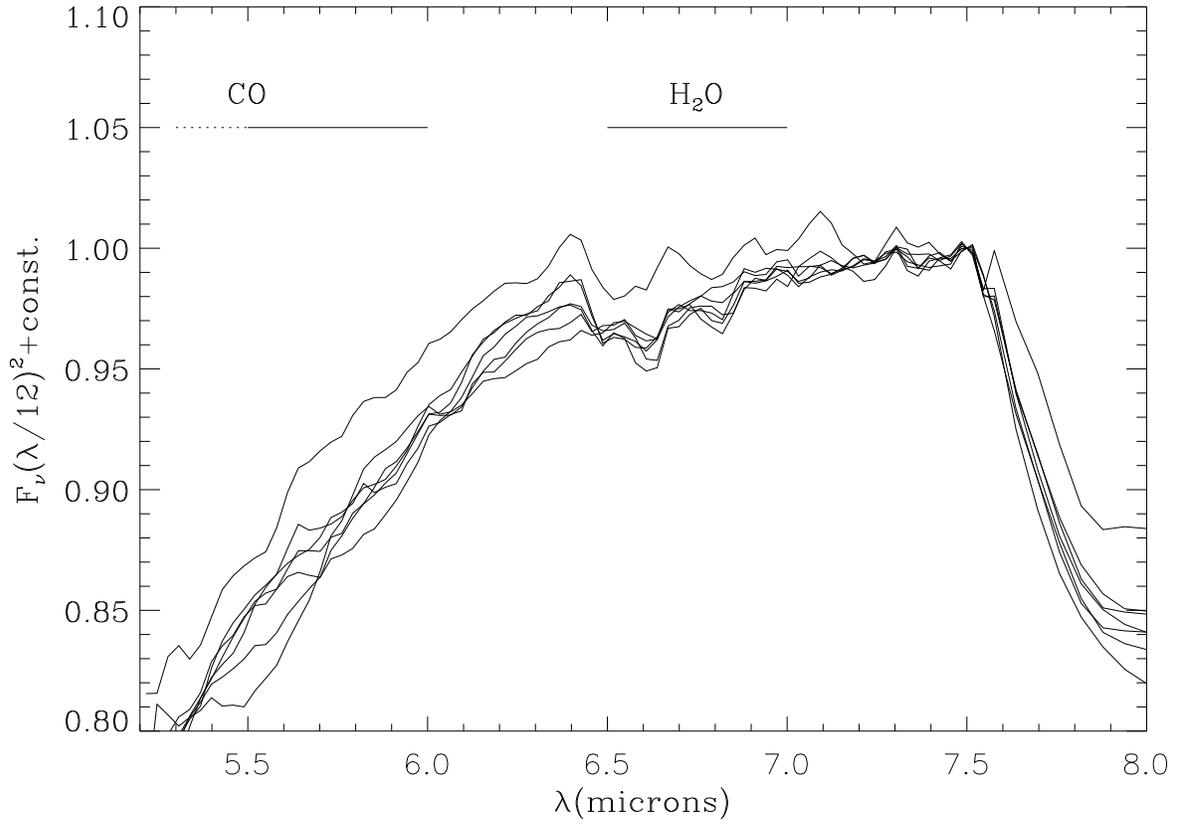} 
\caption{The H$_2$O bending mode in late giants (M0III to M6III). All the spectra are normalized to 7.5 $\mu$m. The strength of the feature remains constant for all types later than M0III (topmost line at wavelengths smaller than 7.5 $\mu$m). \label{h2o}}
\end{figure}

\begin{figure}
\epsscale{1}
\plotone{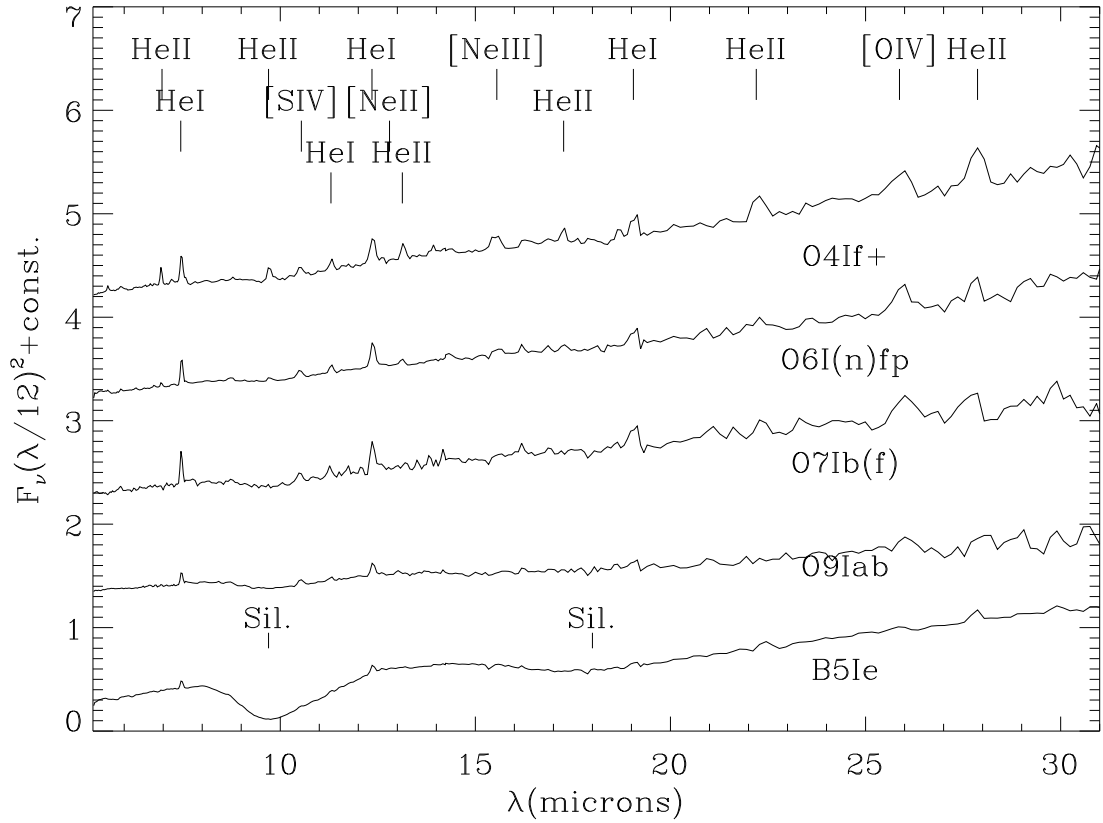} 
\caption{Spectra of early-type supergiants. From bottom to top, they are HD~190429 (O4If+), HD~210839 (O6I(n)fp), HD~192639 (O7Ib(f)), HD~154368 (O9Iab), and Cyg OB2 No. 12 (B5Ie).  \label{new_early_super}}
\end{figure}

\begin{figure}
\epsscale{1}
\plotone{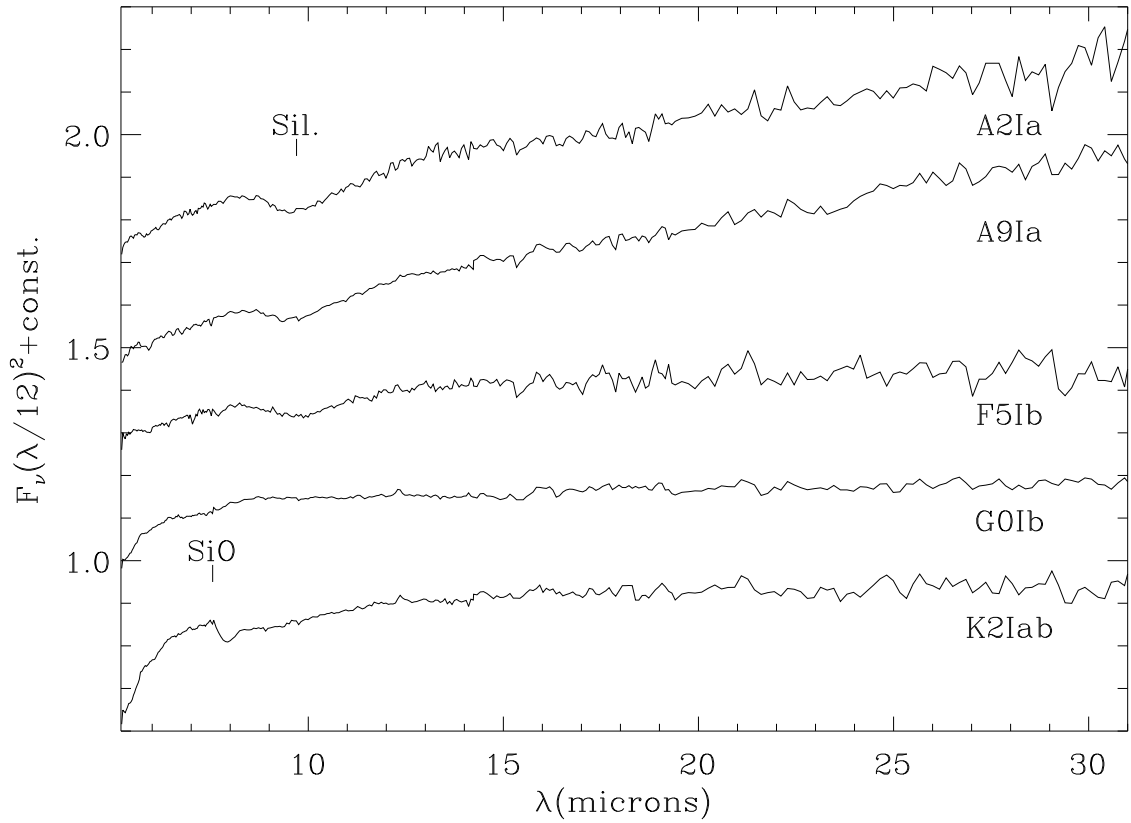} 
\caption{Spectra of mid-Supergiants. From top to bottom they are HD~14433 (A2Ia), HD~90772 (A9Ia), HD~127297 (F5Ib), HD~52973 (G0Ib), HD~45829 (K2Iab). \label{midsuper}}
\end{figure}

\begin{figure}
\epsscale{1}
\plotone{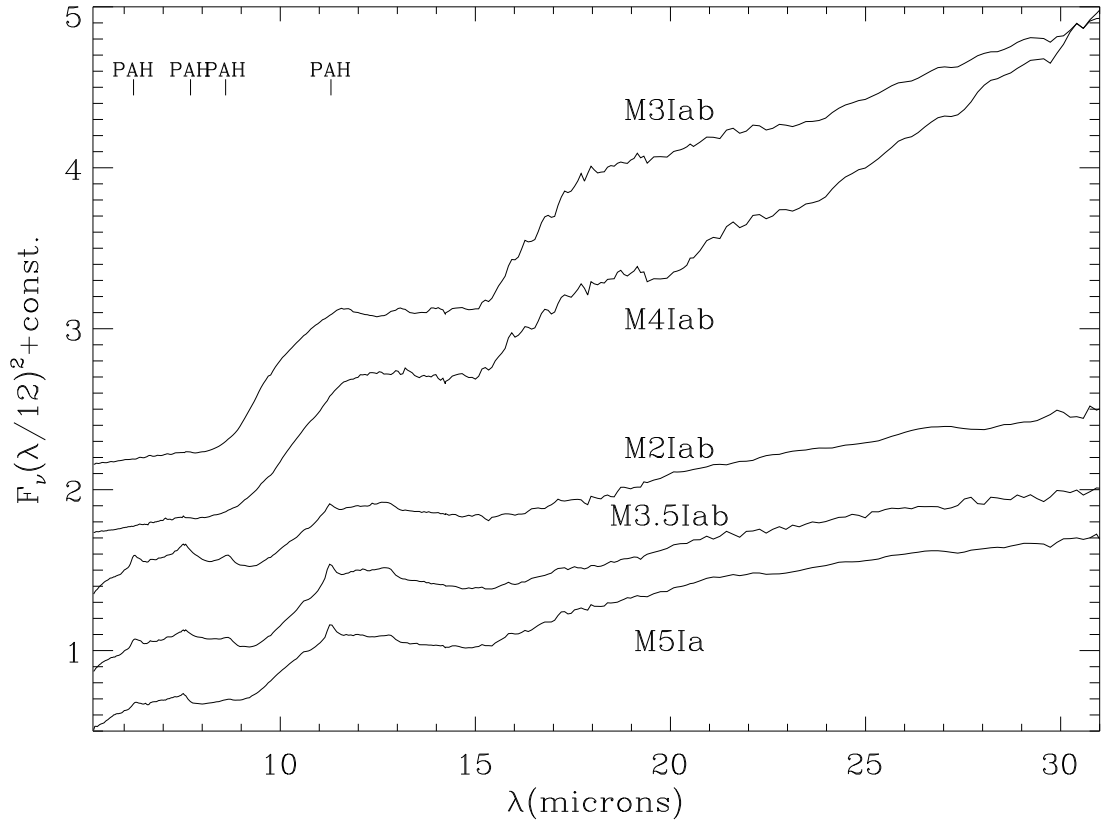} 
\caption{Spectra of late Supergiants. From top to bottom they are RSG 2 \#2 (M3Iab), RSG 2 \#5  (M4Iab), NGC 7419 \#435 (M2Iab), NGC 7419 \#139 (M3.5Iab), BD+23$^o$1138 (M5Ia). \label{latesuper}}
\end{figure}

\begin{figure}
\epsscale{1}
\plotone{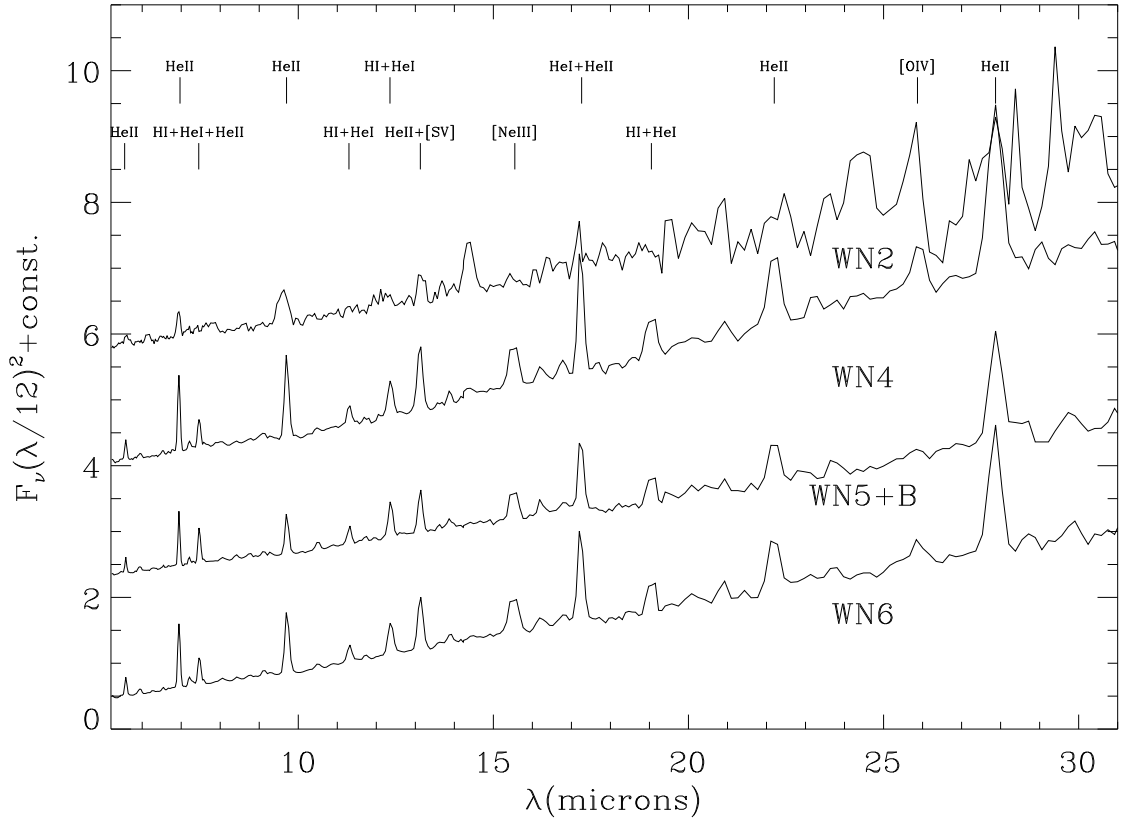} 
\caption{Spectra of WN stars. From top to bottom they are WR~2 (WN2), WR~1 (WN4), WR~138 (WN5+B), WR~134 (WN6). In this, as in figures \ref{WC} and \ref{WO}, line identifications are taken from \citet{morris2000} and \citet{morris2004}. \label{WN}}
\end{figure}

\begin{figure}
\epsscale{1}
\plotone{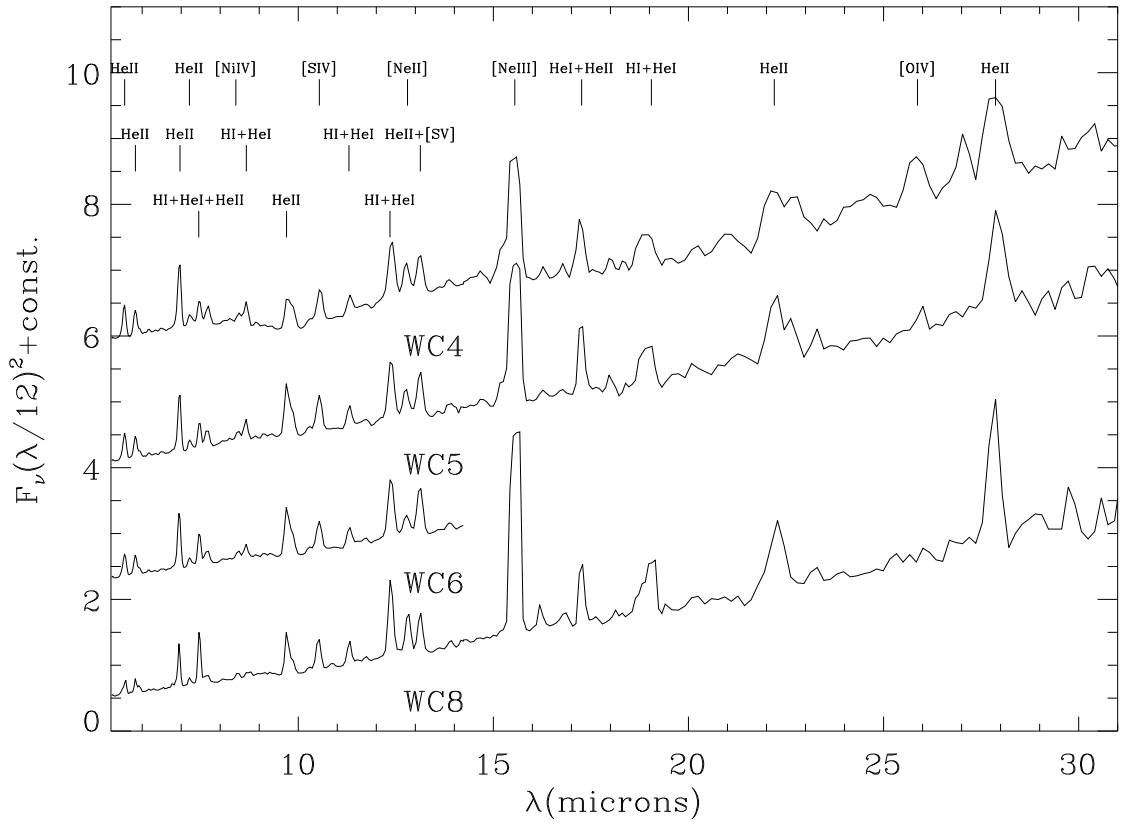} 
\caption{Spectra of WC stars. From top to bottom they are WR~144 (WC4), WR~111 (WC5), WR~23 (WC6), WR~135 (WC8). \label{WC}}
\end{figure}

\begin{figure}
\epsscale{1}
\plotone{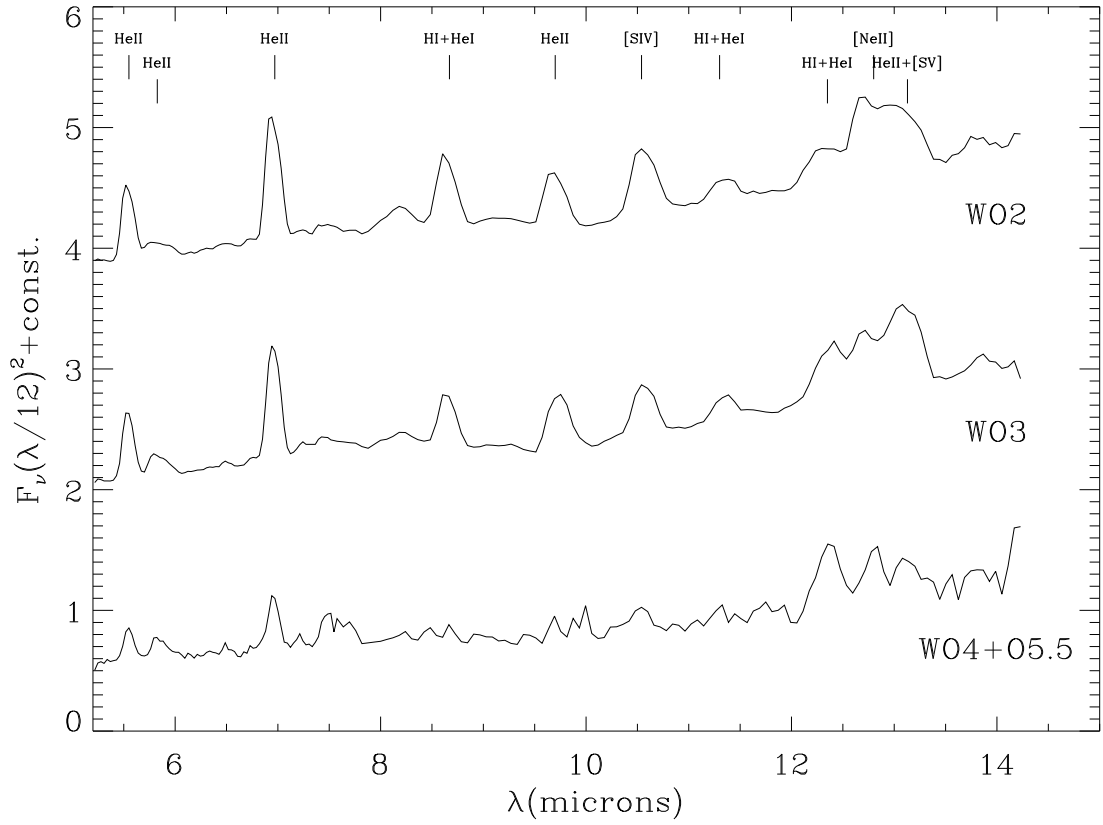} 
\caption{Spectra of WO stars. From top to bottom they are WR~142 (WO2), WR93b (WO3), WR30a (WO4+O5.5). \label{WO}}
\end{figure}

\begin{figure}
\epsscale{1}
\plotone{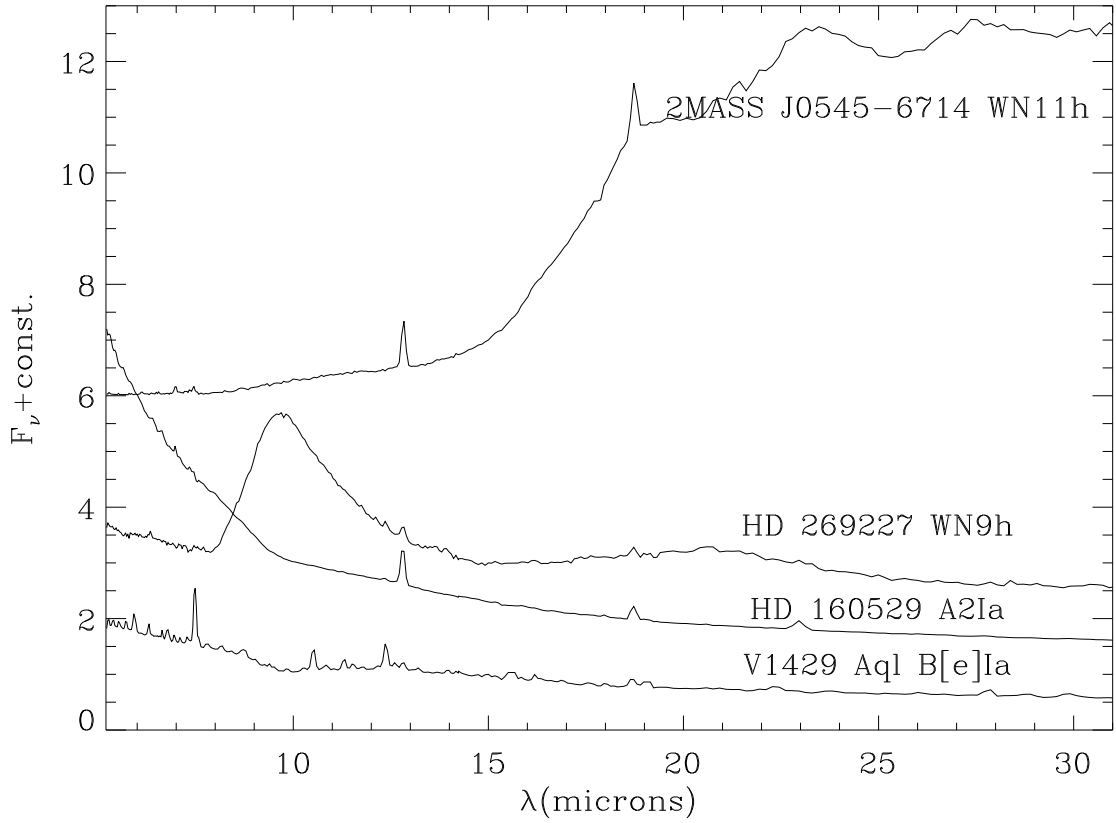} 
\caption{Spectra of LBV stars. The ordinate axis is in flux density units, for all the spectra to fit in the plot. \label{trans}}
\end{figure}

Our primary goal was to provide stellar prototypes for key places of the HR diagram (see Figure \ref{graph}), and we include naked photospheres as much as possible. We did not include known young stars with circumstellar material, stars known to harbor debris disks, or objects classified in SIMBAD as RS CVn, Be stars, or eclipsing binaries. None of those kinds of objects has a particular prototype that could be chosen to represent the class as a whole. When multiple observations of targets with a given spectral type were present in the archive, we selected the highest S/N representative for inclusion in the Atlas. 

In the case of very massive and/or evolved stars, pulsations and winds will result in IR excess. Extinction by interstellar dust along the line of sight will also result in a non-photospheric spectrum.  In the case of supergiants, most spectra included in the Atlas are clearly non-photospheric at the wavelengths considered here.

Beyond prototypes for each class, the Atlas includes a few stars specifically selected for their intrinsic interest, either because they represent optical prototypes of some variable types, or because they have not been well studied in the IR. These were included regardless of their IR excess and even if the Atlas already contained another star with the same spectral type. In general, we avoided classes that have been well described with {\it IRAS}, {\it ISO}, or {\it Spitzer}, such as AGB stars.

The spectra of most dwarf and giants without circumstellar material are relatively featureless, although for objects with early A spectral types, Hydrogen lines are observed for all luminosity classes. Besides these, the most noticeable photospheric features correspond to water vapor and silicon monoxide in late type dwarfs and giants, as well as methane and ammonia features at the latest spectral types.

The fundamental SiO band (8 $\mu$m) first becomes noticeable at M0V and remains strong until it becomes confused with the H$_2$O band at 6.75 $\mu$m. The red wing of the CO fundamental absorption band, at 4.4 $\mu$m, is noticeable already at G1V, although it is not possible to say exactly at which spectral type it first appears, as most of the band falls beyond the blue edge of the {\it Spitzer} spectra. The water bands at 5.8 $\mu$m and 6.75 $\mu$m are noticeable in dwarfs as early as in M0V (HD~85512). In later spectral types the water bands result in a false emission feature at $\sim$6.25 $\mu$m. The $\nu_4$ fundamental band of CH$_4$ at 7.65 $\mu$m and the $\nu_2$ fundamental band of NH$_3$ centered at $\sim$10.5 $\mu$m are observed in the latest spectral types. 

The spectra of the earliest giants in the Atlas show emission lines of [OIV] at 25.87 $\mu$m, HeI, and HI. For the rest of the class III, the most noticeable features are the SiO fundamental $v=1-0$ band and the H$_2$O $\nu_2$ band at 6.75 $\mu$m. The SiO band is observed as early as K0III, and its EQW increases with spectral type, until around M1III. It remains strong until M6III, the latest giant spectral type in the Atlas. The EQW values that we measure are comparable to the ones presented by \citet{her02}, but too large compared with predictions from published models (i.e. \citealt{ari97}). Water is clearly observed at M0III (although it may appear at earlier types) and it remains present at later spectra types.  The silicon monoxide and water vapor features exhibit different behaviors with gravity: the SiO lines become stronger with lower gravity, while the H$_2$O lines become weaker with lower gravity.  Water EQW measurements are likely to be affected by opacity changes in the red wing of the CO fundamental absorption. In spite of this, we conclude that the overall strength of the water feature is remarkably constant over the whole M giant range. As with SiO, the opacity saturates at around M1III.  

For luminosity classes fainter than bright giants, the [8]-[24] {\it Spitzer} colors are poor predictors of spectral type. Within a color uncertainty of 0.1 mag, the spectral slopes of most of the sample can be described with Engelke or blackbody functions. 

The early supergiants in the Atlas present emission lines of helium and hydrogen. The spectra of some mid-supergiants present broad silicate absorption features at 10 and 18 $\mu$m, indicative of line-of-sight extinction. The SASS includes five intriguing M supergiant spectra.  NGC 7419 \#435 (M2Iab), NGC 7419 \#139 (M3.5Iab), and BD+23$^o$1138 (M5Ia) show PAH emission lines and broad emission bands at 7 and 12 $\mu$m. RSGC2 \#2 (M3Iab) and RSGC2 \#5 (M4Iab) present broad bands at 11 $\mu$m and 11.5 $\mu$m respectively, and at 18 $\mu$m. No previous mid-IR spectroscopy is available in the literature for these M supergiants.

The presence of PAHs may be due to chance alignment of the star with a background interstellar cloud (i.e. a Pleiades effect).  The origin of the dust features for RSGC2 \#2 and RSGC2 \#5 remains unknown.  They may be due to alumina dust grains or even SiC.

The SASS includes spectra for three blue stragglers: HD 88923 (F2V), HD 106516 (F6V), HD 35863 (F8V). Their spectra are similar to those of other F stars in the Atlas: mostly featureless within the noise. For the three stars presented here, the mass transfer mechanism responsible for their apparent youth does not produce any strong signal at IRS wavelengths. 

Pulsating variables included in the Atlas are classified in the literature as Cepheid, $\delta$ Cep type, $\beta$ Cep type, $\gamma$ Dor type, RR Lyr, and $\delta$ Scu type. Of the 11 stars identified as pulsating variables, 5 present non-photospheric continua. These excesses present diverse morphologies, from the featureless, cold excesses of  HD~27290, to the structured dust continuum of  the Mira Ceti star V836 Oph.

The WR spectra present the well-known pattern of lines of HeI and HeII. The WN and WC classes are primarily distinguished by the strong emission of [Ni IV], [Ne II], and [Ne III] (15.5 $\mu$m) in WC stars.  The flat-top shape of the [Ne III] line is evident even at these low spectral resolutions. The early-type WN types, as well as some WC stars, show a F$_\nu \propto \lambda^{-1}$ slope, indicative of an optically thick wind. 95 $\mu$m. 

The SASS also includes several LBVs which make up a very diverse group of spectra, perhaps the result of the different geometric orientations of the surrounding nebulae,  different phases in the evolution of the LBV, or different nature of the circumstellar material.  

When possible, we have classified the SASS stars within the KSPW system, which is based on {\it ISO}-SWS observations of mostly evolved stars (\citealt{kra02, slo03}).  In some cases, the SASS spectral range reaches only to 14 $\mu$m, which makes the classification impossible. Furthermore the KSPW system emphasizes dust features to discriminate spectra and needs to be
extended to address the gas continua of WR stars and early
O giants.

This last point shows that the KSPW system is incomplete, the result of the small sample and unique instrument involved in its definition. We recall here that the core of the MK system is based on the Henry Draper Catalog and Extension, which contains over a quarter of a million stars \citep{can1993}, two orders of magnitude larger than the largest IR samples. Our attempts to classify the SASS members within the KSPW aim to highlight the areas where more work is needed before a complete, general purpose IR stellar classification is available.

\clearpage

\clearpage
\clearpage



\clearpage

\begin{deluxetable}{cc}
\tablecolumns{2}
\small
\tablewidth{0pt}
\tablecaption{Wavelength ranges used \label{trim}}
\tablehead{\colhead{Order} & \colhead{Range ($\mu$m)} }
\startdata
SL2	(Short-low, order 2) &5.2 - 7.55 \\
SL1	(Short-low, order 1) & 7.58 - 14.29 \\
LL2\tablenotemark{1} (Long-low, order 2) 	& 14.24 - 20.59 \\
LL1	(Long-low, order 1) & 20.59 - 35.00 \\

\enddata
\tablenotetext{1}{From 19.35 to 20.59 $\mu$m, the bonus order is used instead of LL2. In this region the bonus order presents fewer rogue pixels.}

\end{deluxetable}

\clearpage
\begin{deluxetable}{cccl}
\tablecolumns{4}
\tabletypesize{\tiny}
\tablewidth{0pt}
\tablecaption{Polynomial corrections to the spectral slopes\label{correc}}
\tablehead{\colhead{Correction} & \colhead{Order} & \colhead{$\lambda_0$} & Coefficients \\
 &  & $\mu$m & }
\startdata
Teardrop	& SL1 ($>$ 13$\mu$m )	& 0	& 1.73716, -0.132743, 0.00584572 \\
Nod 0, old	& SL2 & 6	&0.990966, -0.00669176, -0.000115076 \\
Nod 0, new	&SL2 &	6	&1.00474, -0.00658626, -0.00244433 \\
Nod 1, old	&SL2 &	6	&0.992103, 0.0144322, -0.00376203 \\
Nod 1, new	&SL2 &	6	&1.00370, 0.00993892, -0.00270484 \\
Nod 0, old	&SL1&	10	&0.984629, 0.00165273, -0.00102893,0.000355763\\
Nod 0, new	&SL1&	10	&1.00018, -0.00244389, -0.000367869, 0.000233202\\
Nod 1, old	&SL1&	10	&0.998734, 0.00903313, -0.00118942, -0.00137535, 0.000198369\\
Nod 1, new	&SL1&	10	&1.00611, 0.00543378, -0.000617655, -0.00149233, 0.000244382\\
Nod 0, old	&LL2&	18	&0.987009, 0.000834591, -0.000564992\\
Nod 0, new	&LL2&	18	&1.00121, 0.00117174, -0.000130143\\
Nod 1, old	&LL2&	18	&0.997453, -0.000512385, 0.000749858\\
Nod 1, new	&LL2&	18	&1.00857,  -0.00104855, 0.000283562\\
Nod 0, old	&LL1&	26	&1.00047,  -0.00172408\\
Nod 0, new	&LL1&	26	&1.00800, 0.000159851\\
Nod 1, old	&LL1&	26	&0.990219, 0.00229134\\
Nod 1, new	&LL1&	26	&1.0, 0.000368189\\

$\alpha$ Lac, old	&SL2&	6&	0.993109, -0.000274152, 0.00304373\\
$\alpha$ Lac, new	&SL2&	6&	0.99988661,   -0.0031066054, 0.0019065623\\
$\alpha$ Lac, old	&SL1&	10&	1.00100,  -0.00186966, 0.000672951\\
$\alpha$ Lac, new	&SL1&	10&	1.0019434,   -0.0012843547, 0.00030117893\\
$\alpha$ Lac, old	&LL2&	18&	0.989320, -0.000477111\\
$\alpha$ Lac	&LL1&	26&	0.996167, -0.000998179, -0.000334123\\
Faint source deficit	&LL2 &	18  &	1.0, -0.00056316124 \\
Faint source deficit	&LL1 &	26  &	1.0, 0.0026269022, 0.00084026423 \\

\enddata
\tablecomments{Spectra reduced using the S18.7.0 SS pipeline are divided by the polynomials (p=a$_0$+a$_1$($\lambda$-$\lambda_0$)+a$_2$($\lambda-\lambda$$_0$)$^2$+...) whose coefficients (a$_0$, a$_2$, a$_2$, etc) are listed in the fourth column. The corrections marked as 'old' are applied to spectra obtained before 19-June-2006. Nod 0 is arbitrarily defined as the nod position with the spectral trace closest to the X=0 coordinate of the array. \\}
\end{deluxetable}

\clearpage
\appendix
\section{Atlas Contents}

\begin{figure} 
\epsscale{1}
\plotone{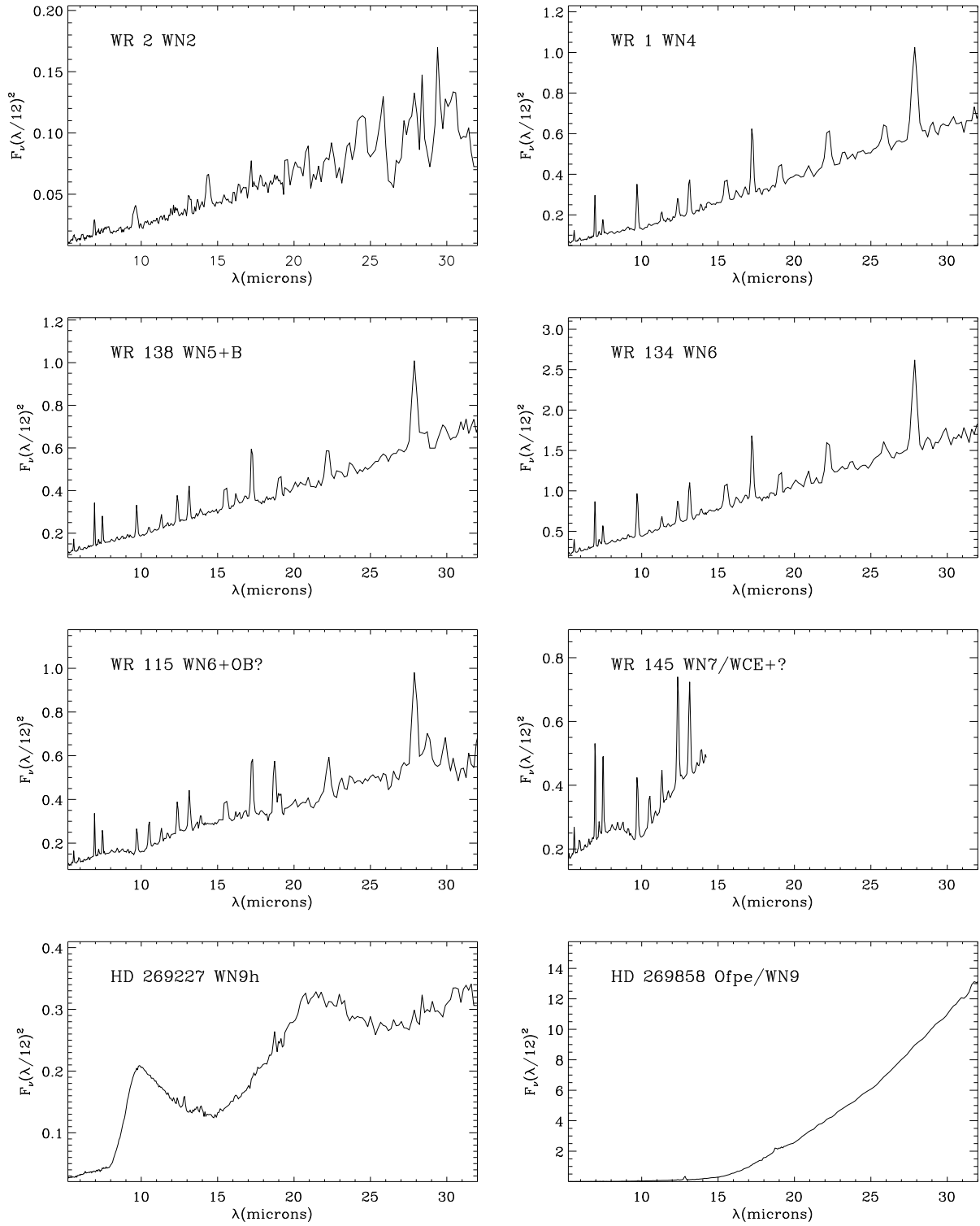}
\caption{Wolf-Rayet stars, WN2 to O/WN.  \label{panel_WR1}}
\end{figure}

\begin{figure}
\epsscale{1}
\plotone{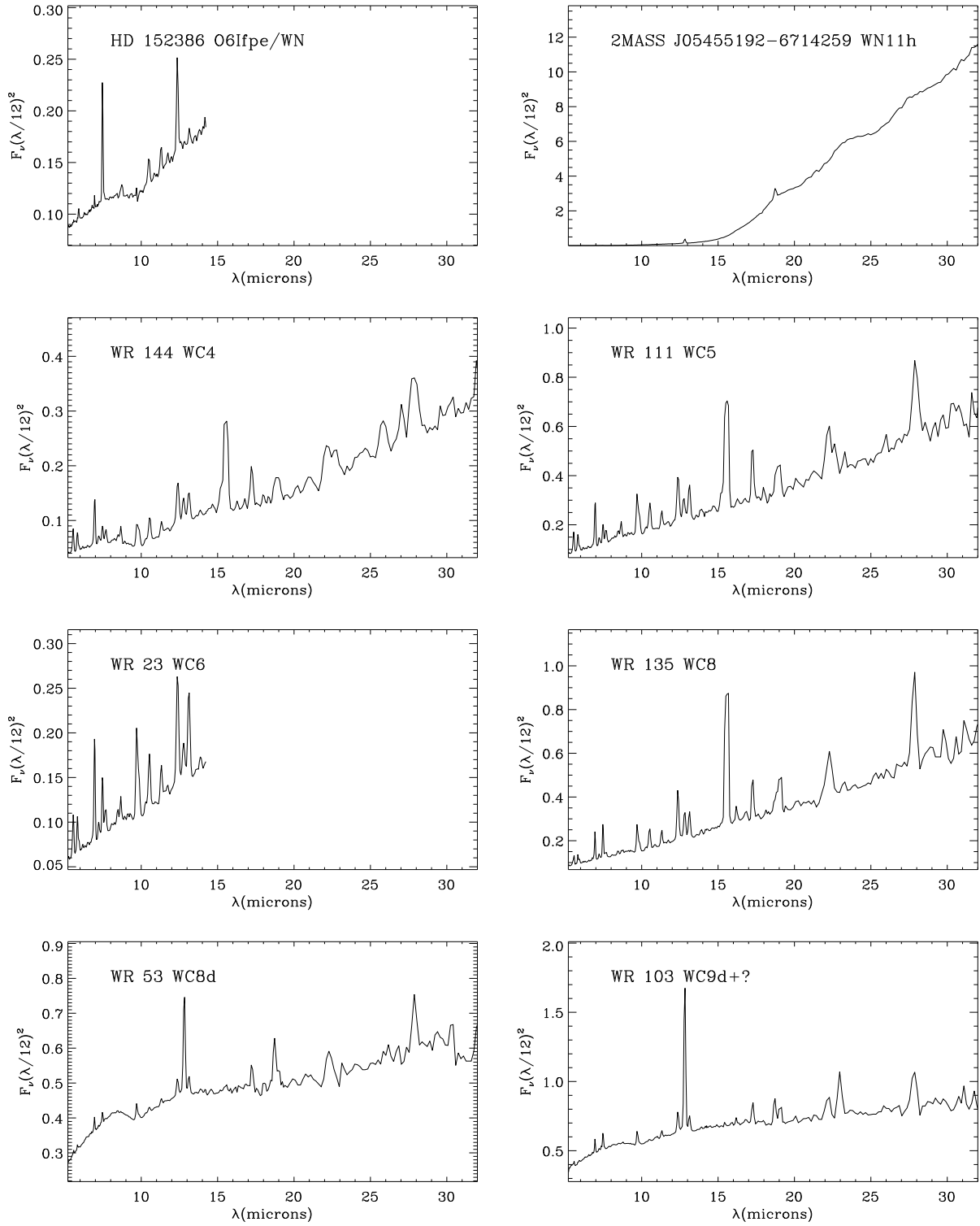}
\caption{Wolf-Rayet stars, O/WN to WC9.  \label{panel_WR2}}
\end{figure}

\begin{figure}
\epsscale{1}
\plotone{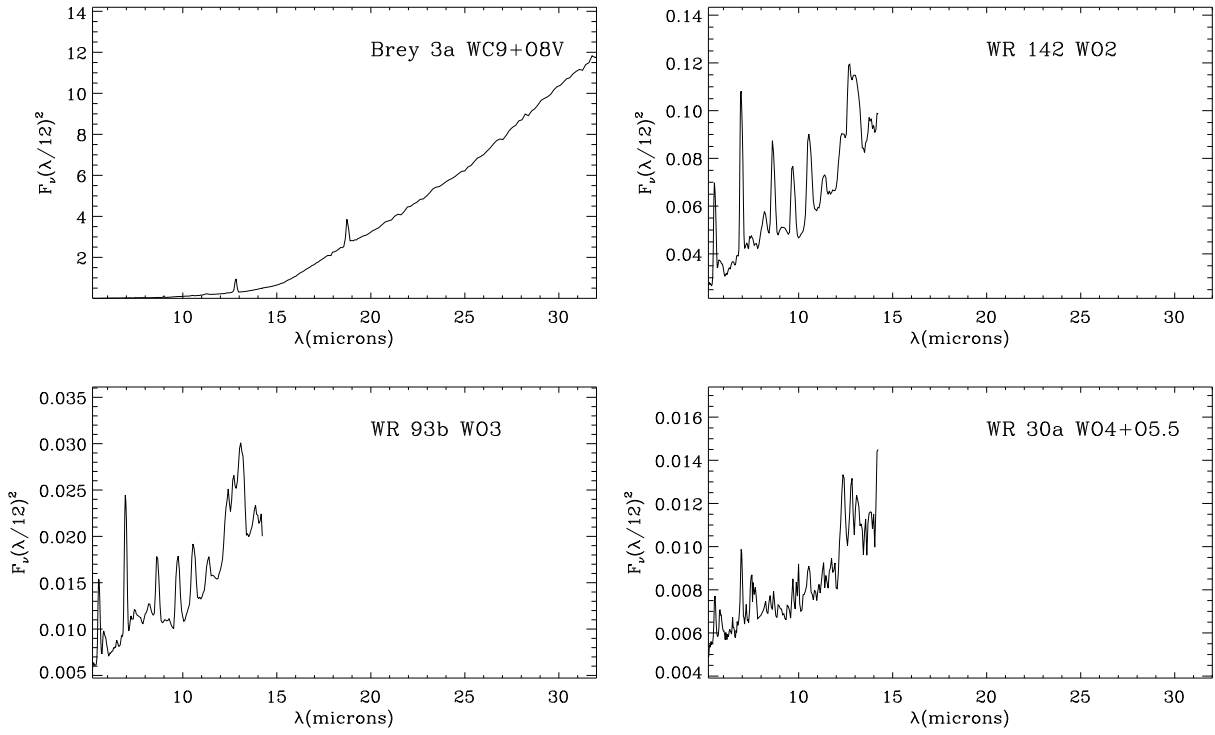}
\caption{Wolf-Rayet stars, WC9+O8V to WO4+O5.5.  \label{panel_WR3}}
\end{figure}

\begin{figure}
\epsscale{1}
\plotone{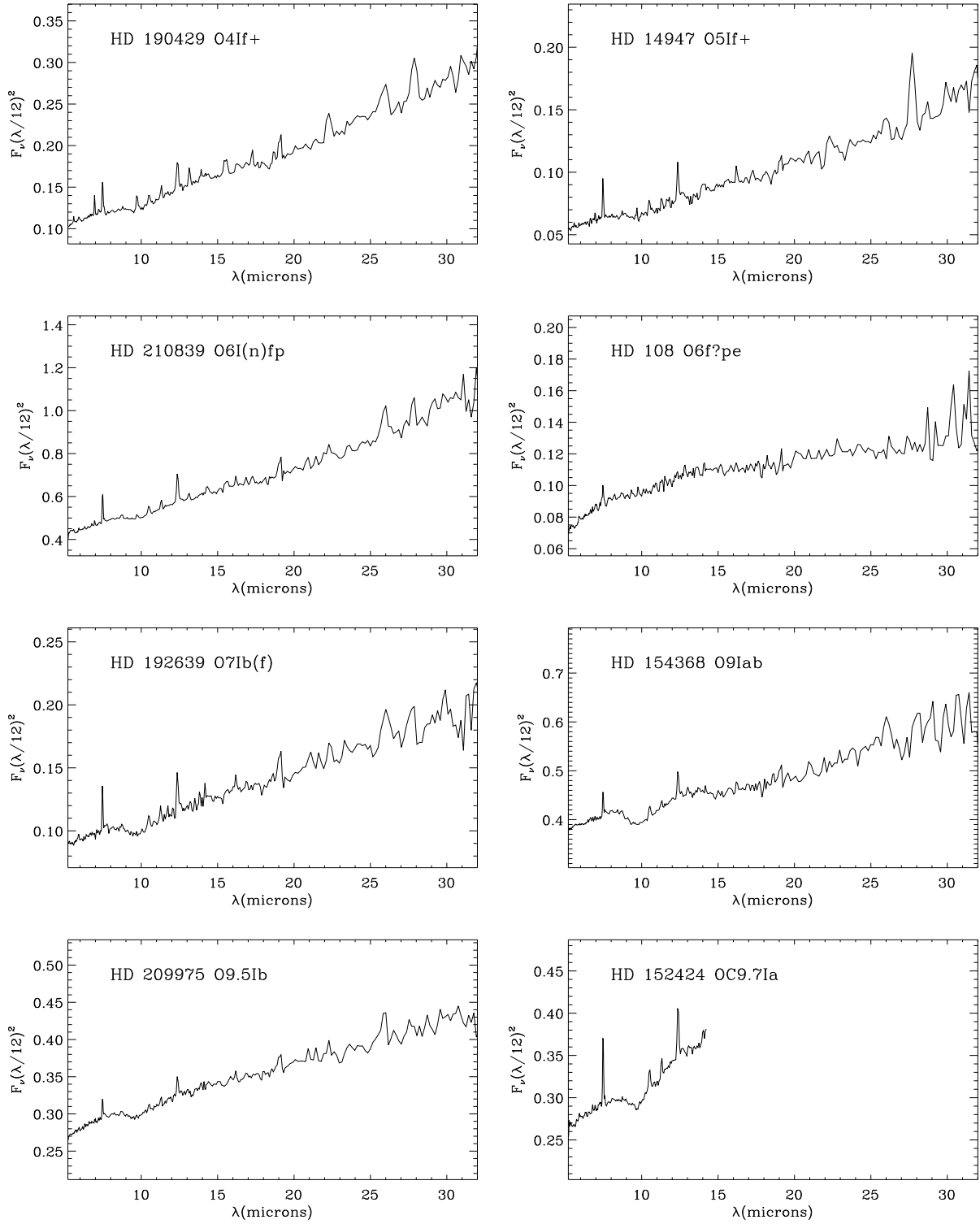}
\caption{Luminosity class I, O4 to OC9.7.  \label{panel_super1}}
\end{figure}

\begin{figure} 
\epsscale{1}
\plotone{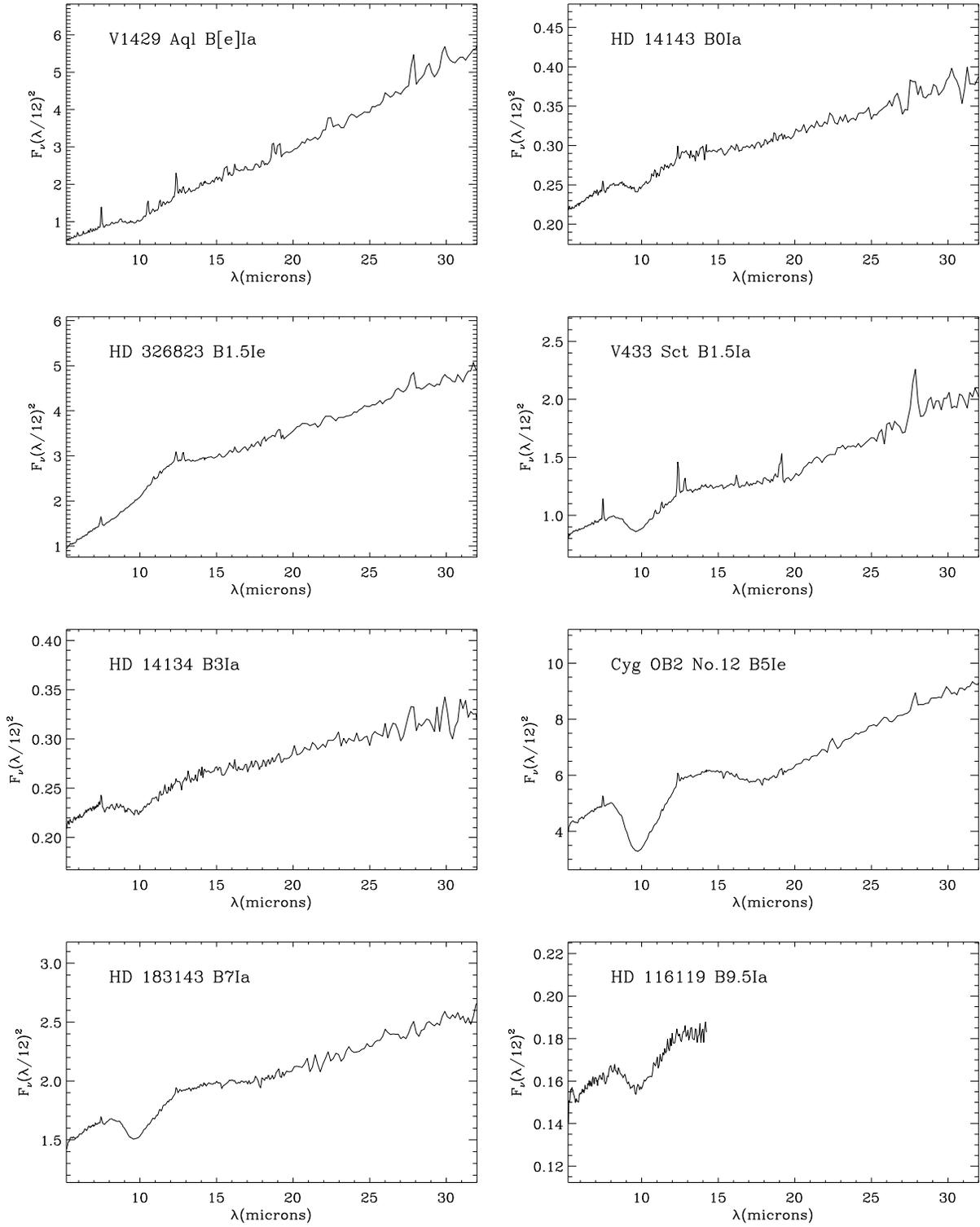}
\caption{Luminosity class I, B to B9.5.  \label{panel_super2}}
\end{figure}

\begin{figure}
\epsscale{1}
\plotone{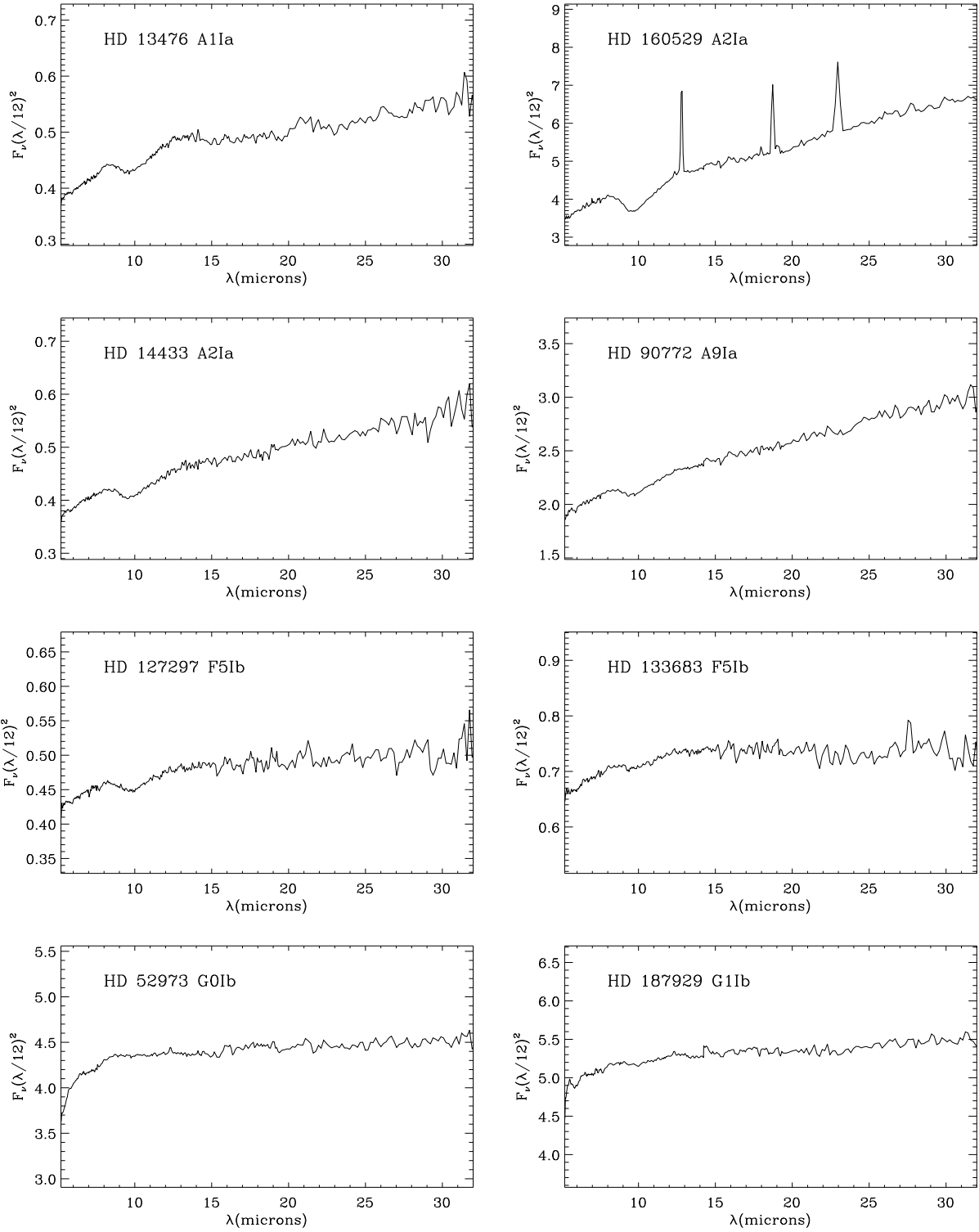}
\caption{Luminosity class I, A1 to G1.  \label{panel_super3}}
\end{figure}

\begin{figure}
\epsscale{1}
\plotone{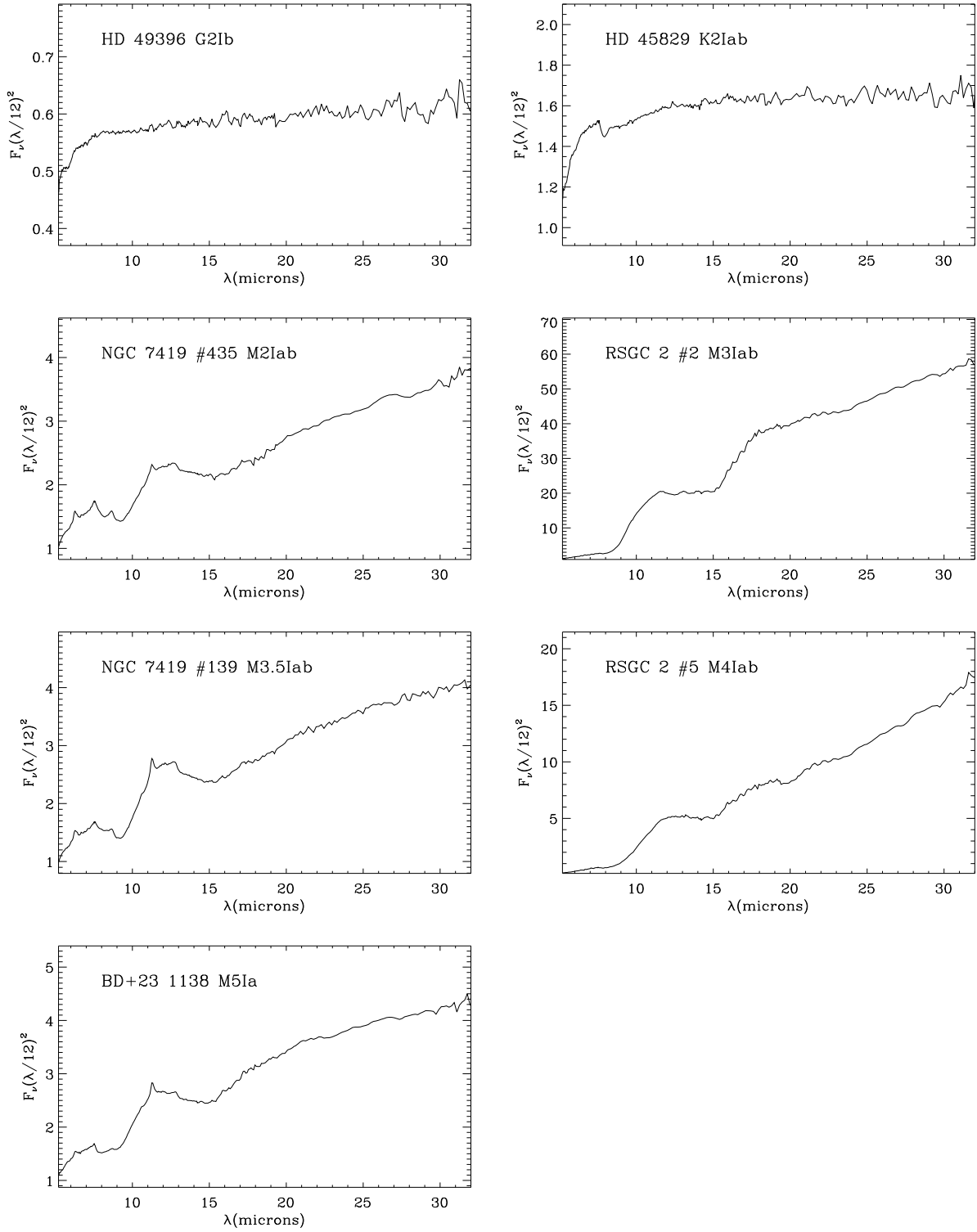}
\caption{Luminosity class I, G2 to M5.  \label{panel_super4}}
\end{figure}

\begin{figure}
\epsscale{1}
\plotone{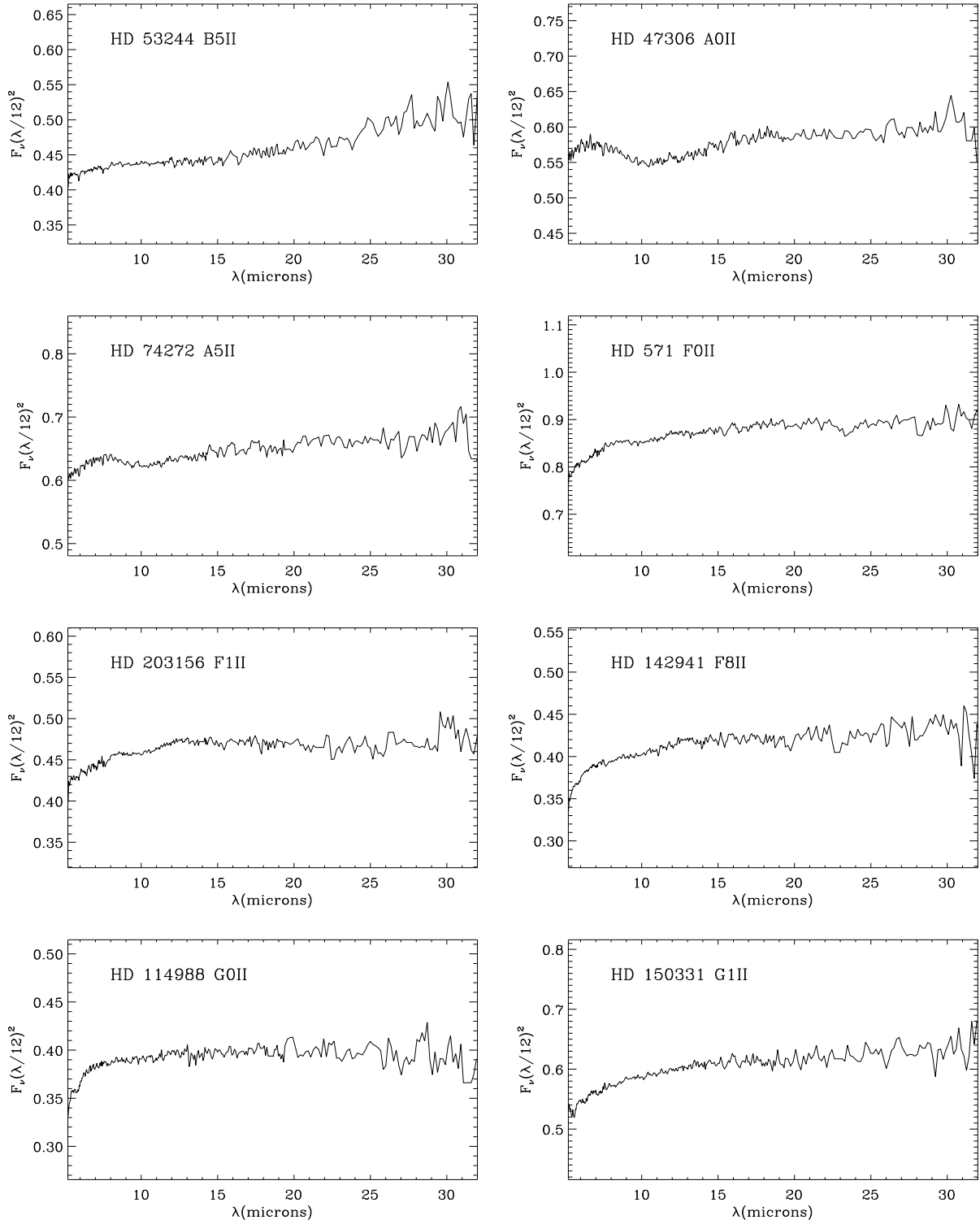}
\caption{Luminosity class II, B5 to G1.  \label{panel_II1}}
\end{figure}

\begin{figure}
\epsscale{1}
\plotone{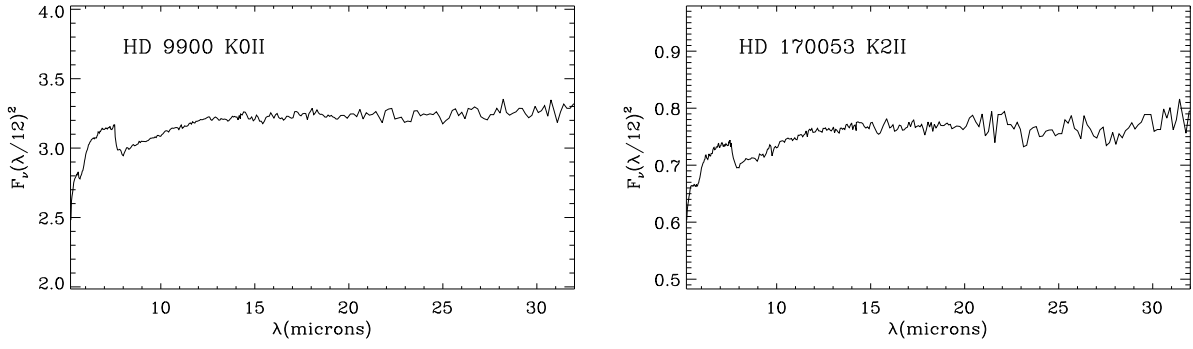}
\caption{Luminosity class II, K0 to K2.  \label{panel_II2}}
\end{figure}

\begin{figure}
\epsscale{1}
\plotone{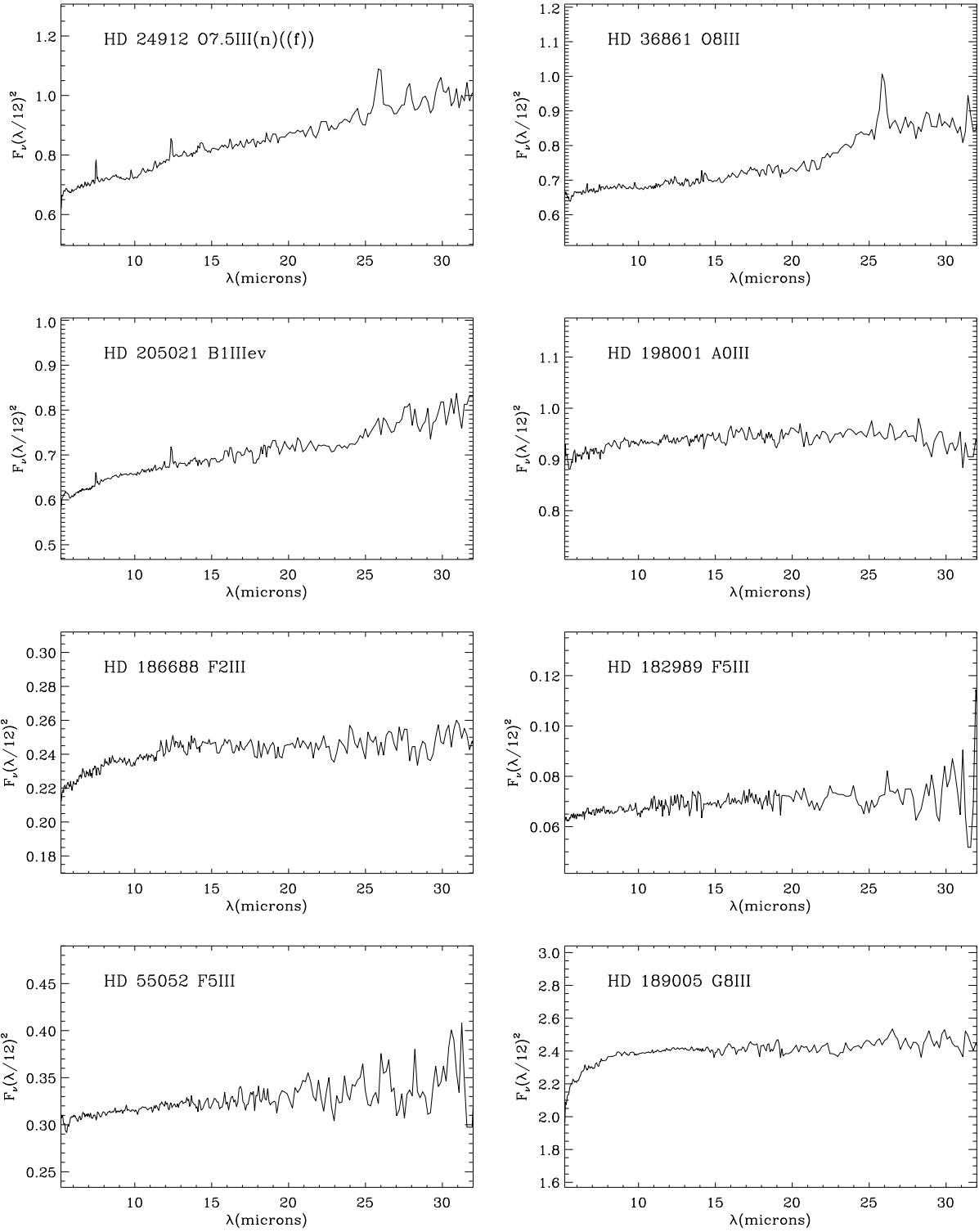}
\caption{Luminosity class III, O7.5 to G8.  \label{panel_III1}}
\end{figure}

\begin{figure}
\epsscale{1}
\plotone{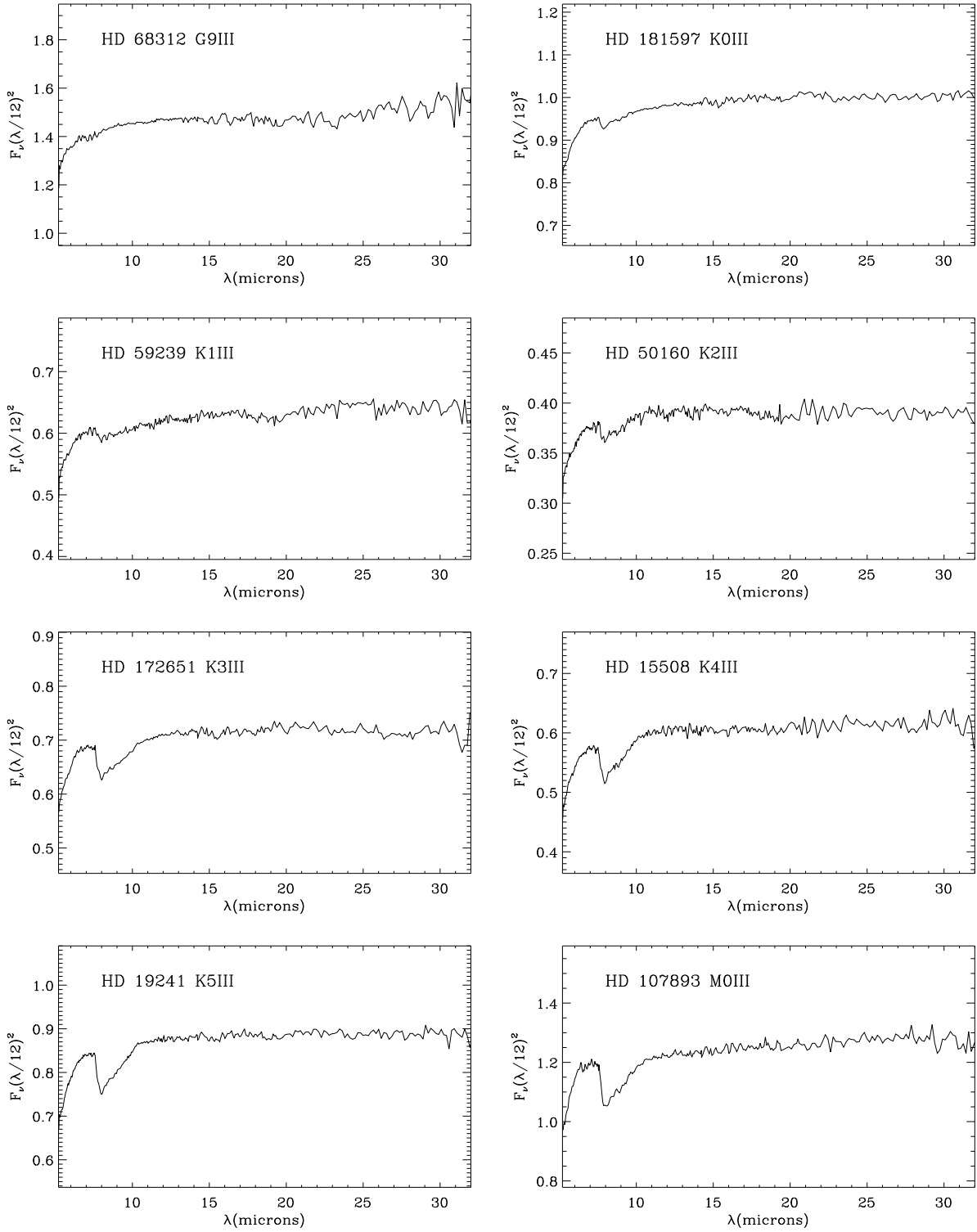}
\caption{Luminosity class III, G9 to M0.  \label{panel_III2}}
\end{figure}

\begin{figure}
\epsscale{1}
\plotone{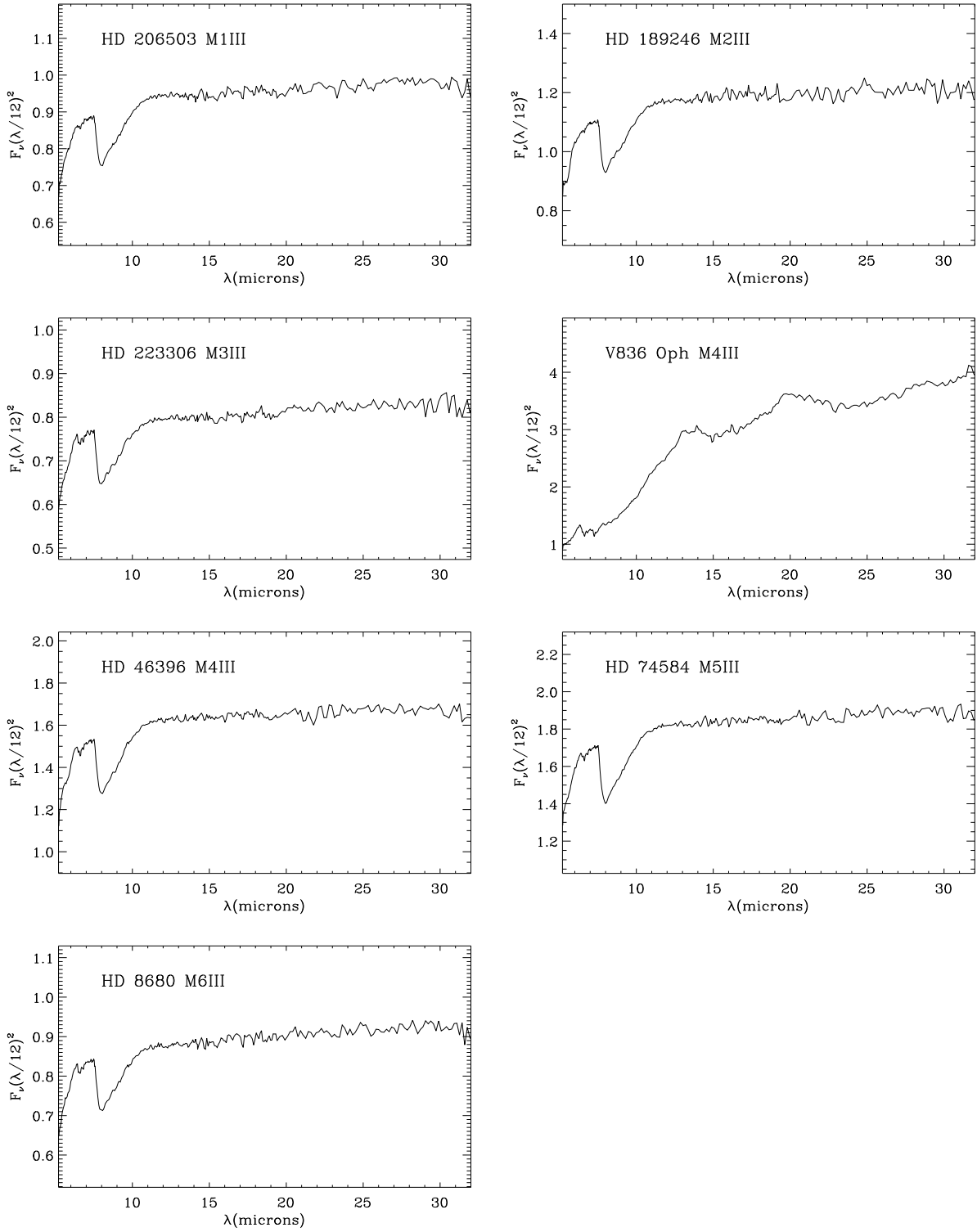}
\caption{Luminosity class III, M0 to M6.  \label{panel_III3}}
\end{figure}

\begin{figure}
\epsscale{1}
\plotone{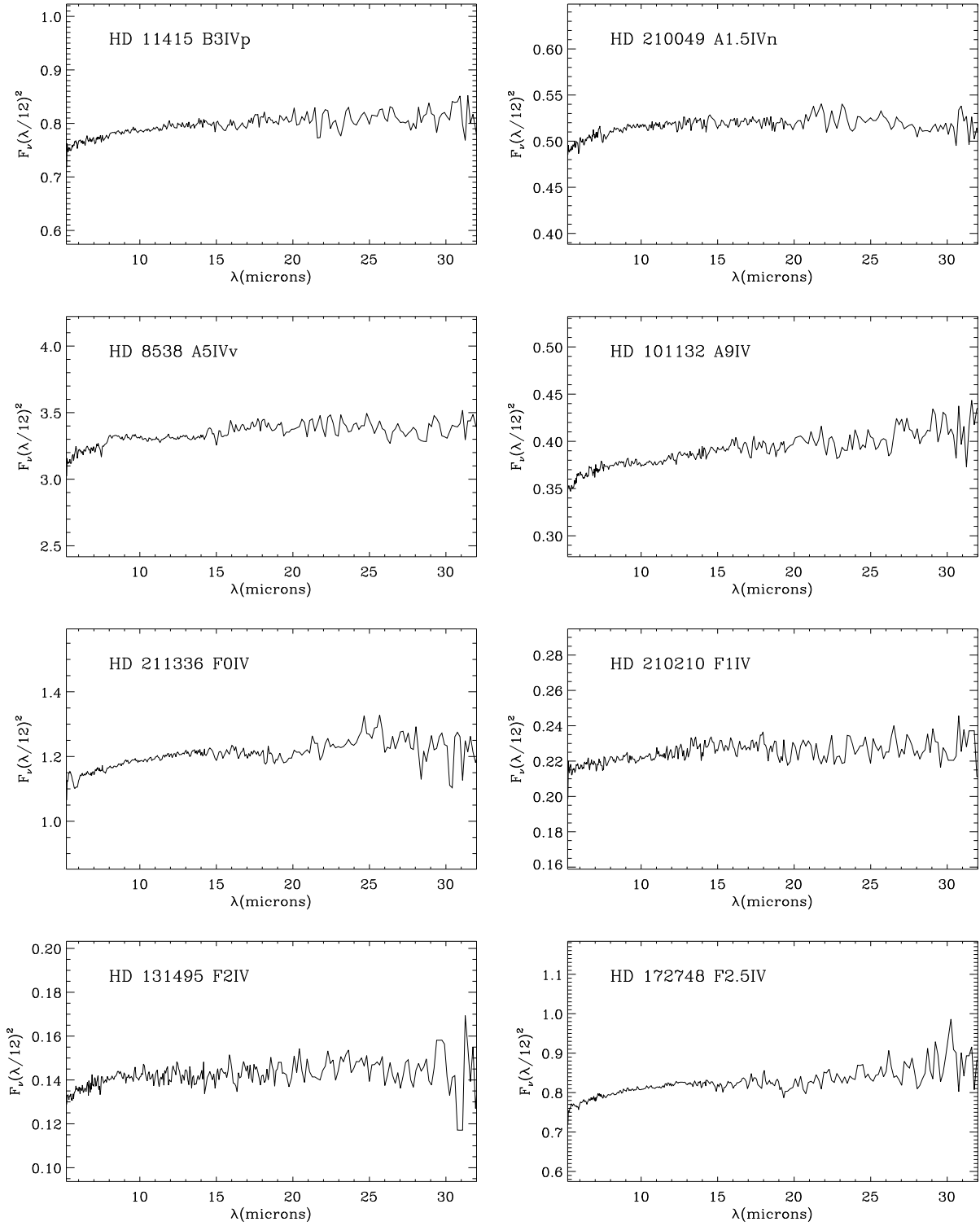}
\caption{Luminosity class IV, B3 to F2.5.  \label{panel_IV1}}
\end{figure}

\begin{figure}
\epsscale{1}
\plotone{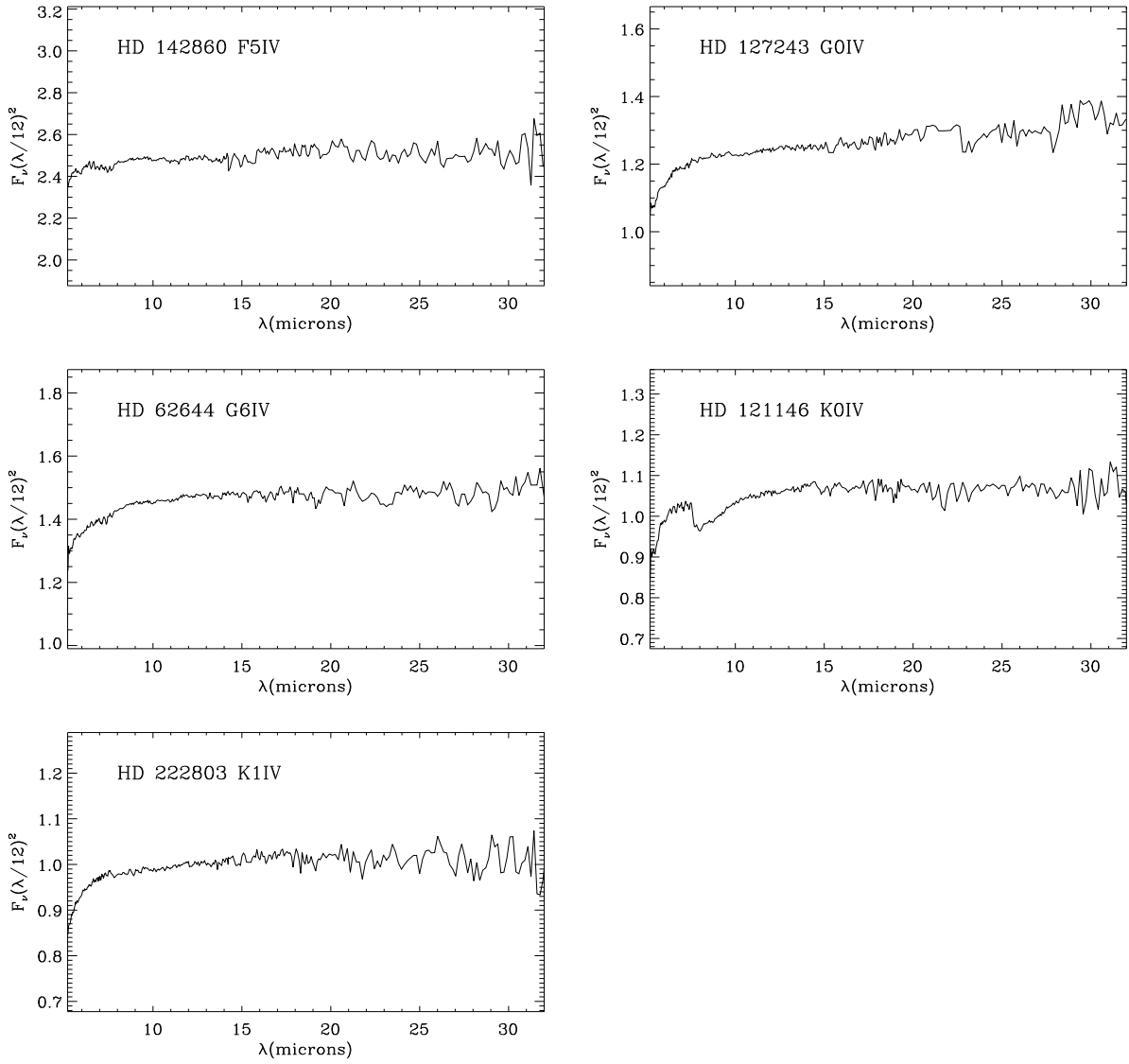}
\caption{Luminosity class IV, F5 to K1.  \label{panel_IV2}}
\end{figure}

\begin{figure}
\epsscale{1}
\plotone{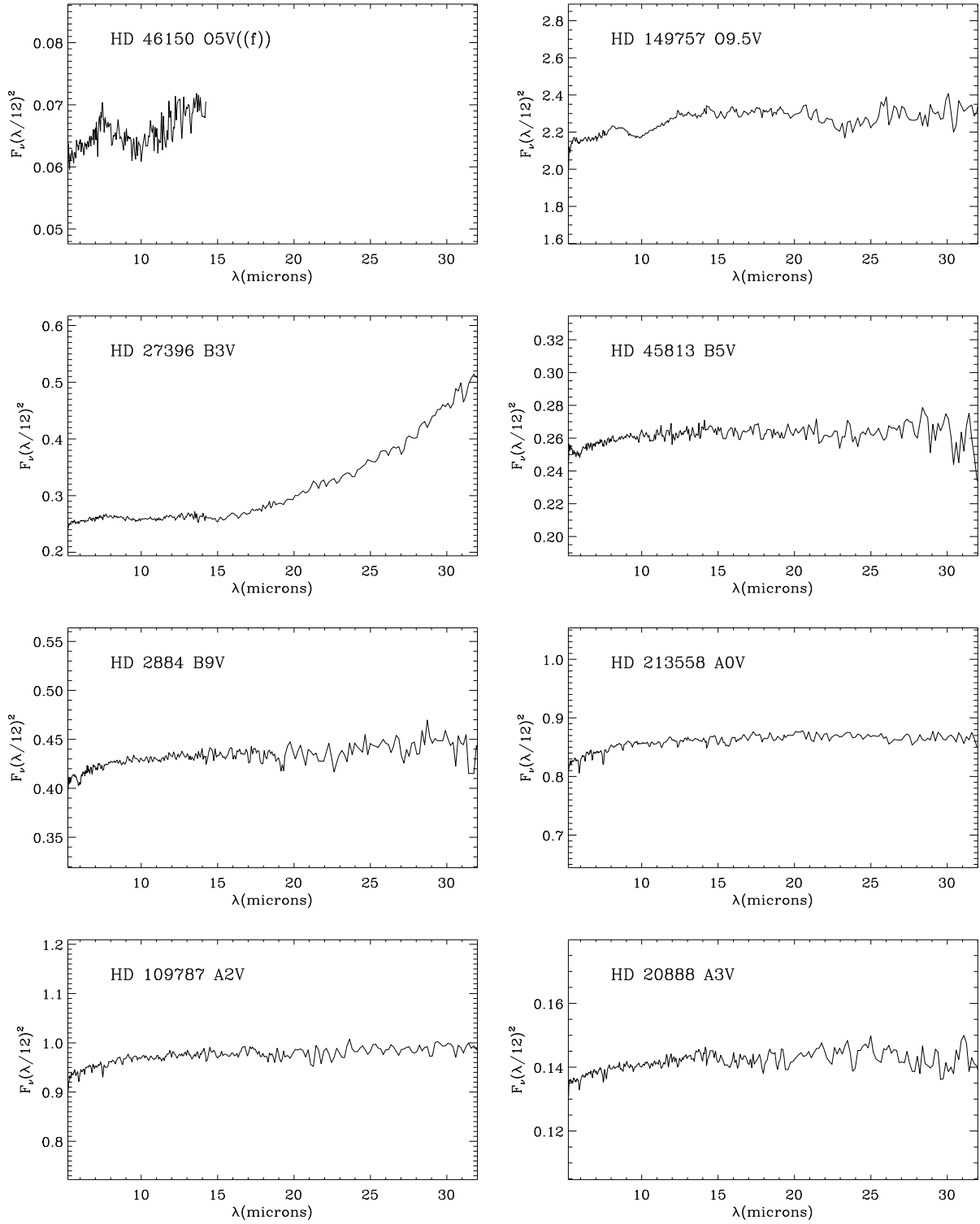}
\caption{Luminosity class V, O5 to A3.  \label{panel_V1}}
\end{figure}

\begin{figure}
\epsscale{1}
\plotone{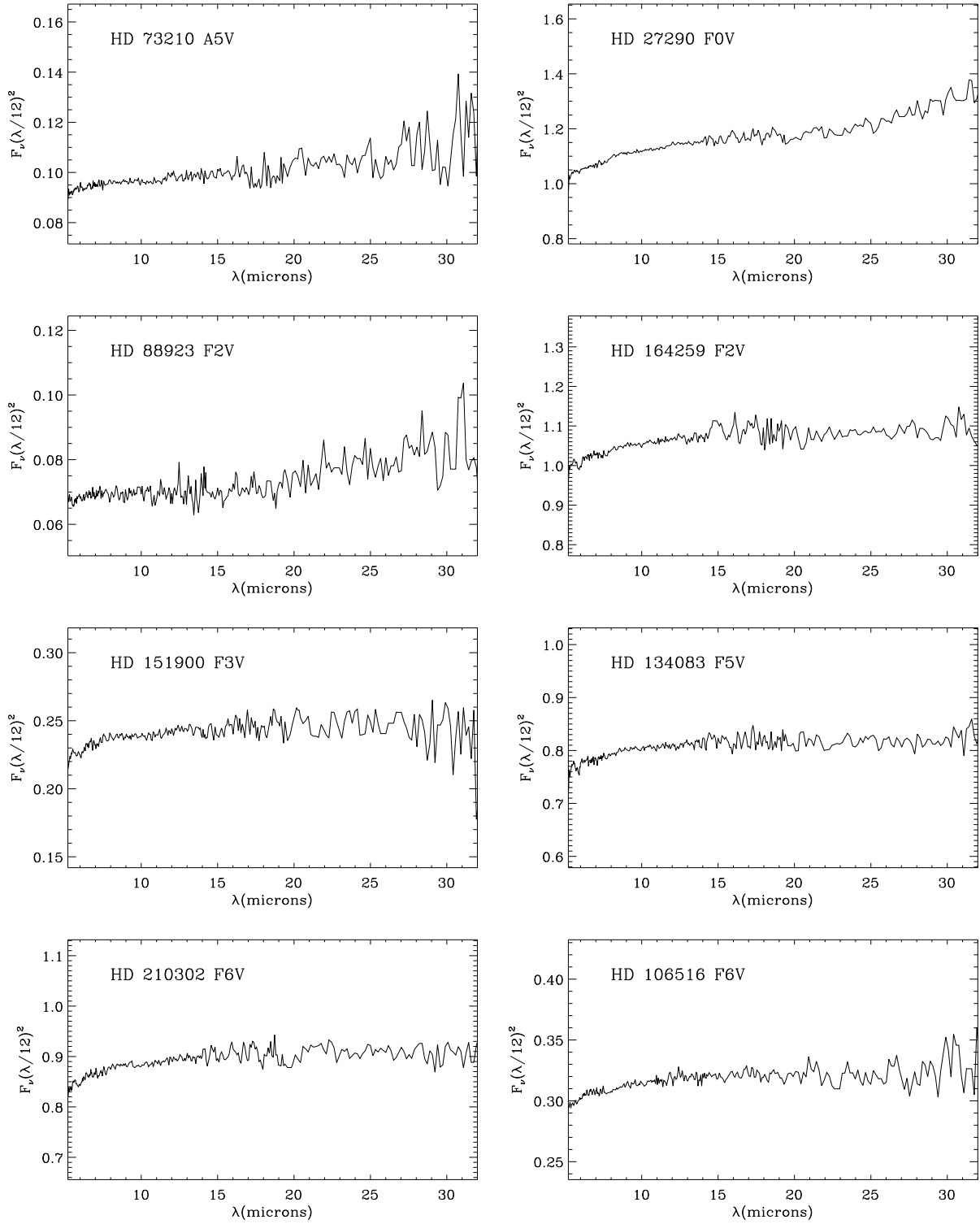}
\caption{Luminosity class V, A5 to F6.  \label{panel_V2}}
\end{figure}

\begin{figure}
\epsscale{1}
\plotone{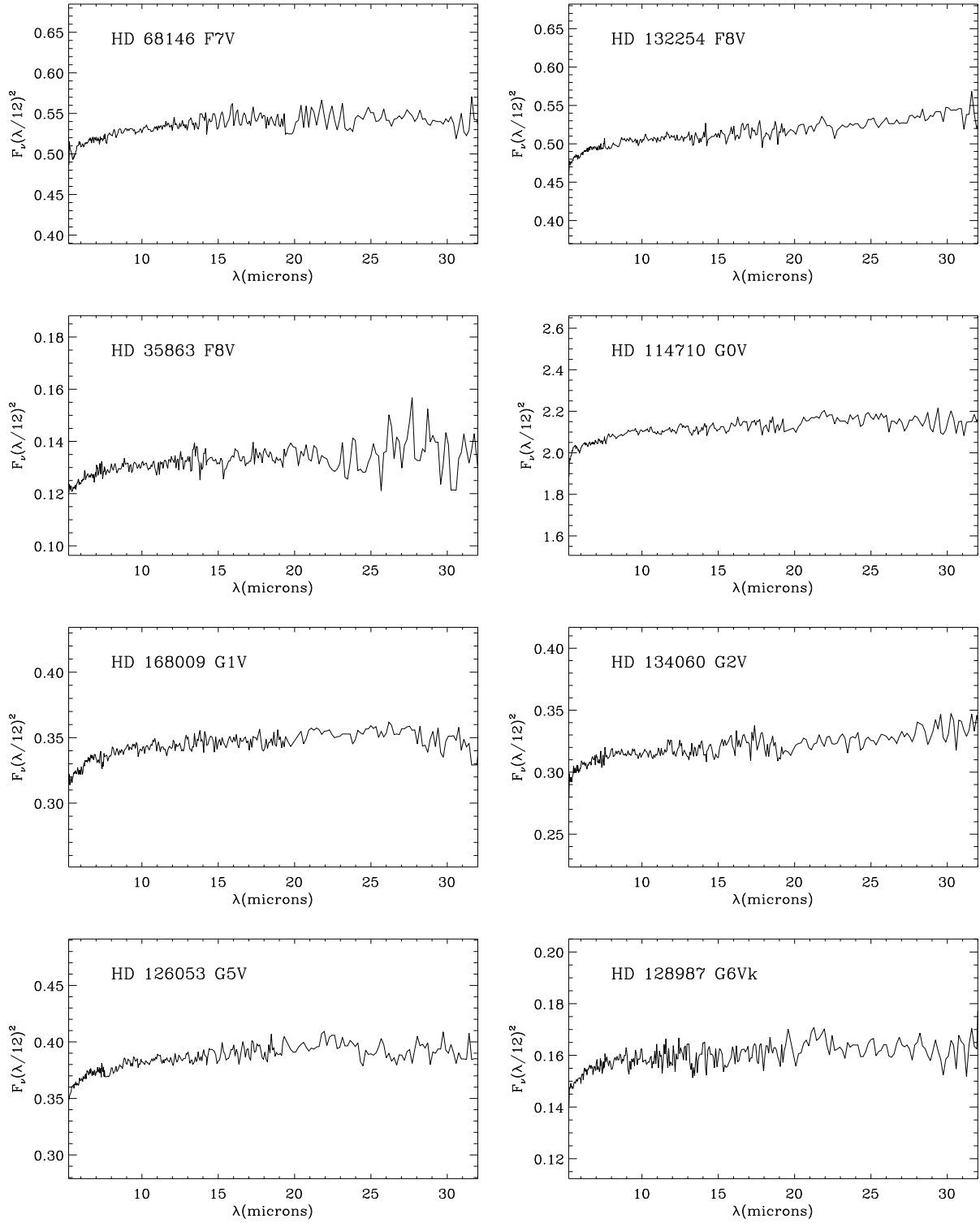}
\caption{Luminosity class V, F7 to G6.  \label{panel_V3}}
\end{figure}

\begin{figure}
\epsscale{1}
\plotone{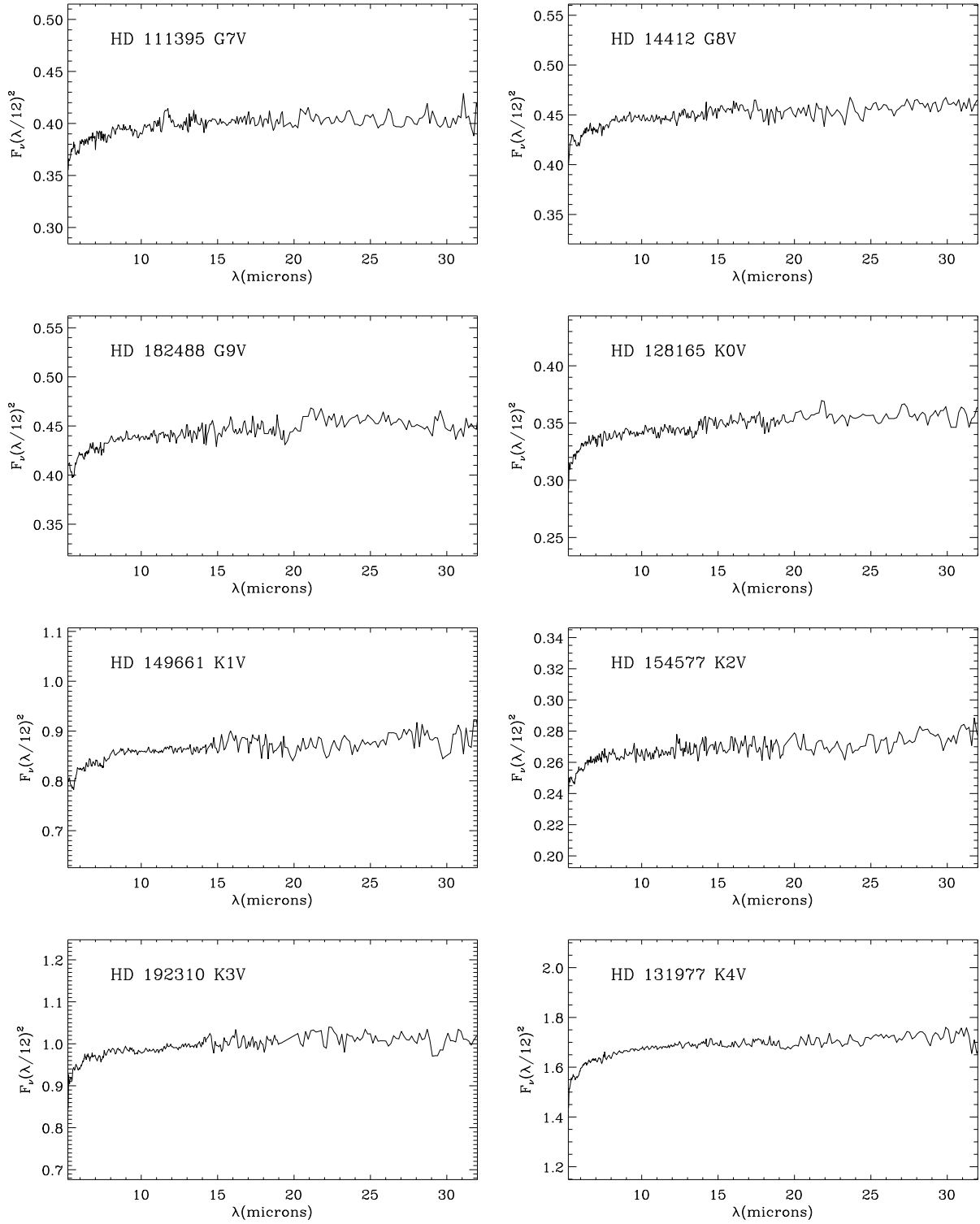}
\caption{Luminosity class V, G7 to K4.  \label{panel_V4}}
\end{figure}

\begin{figure}
\epsscale{1}
\plotone{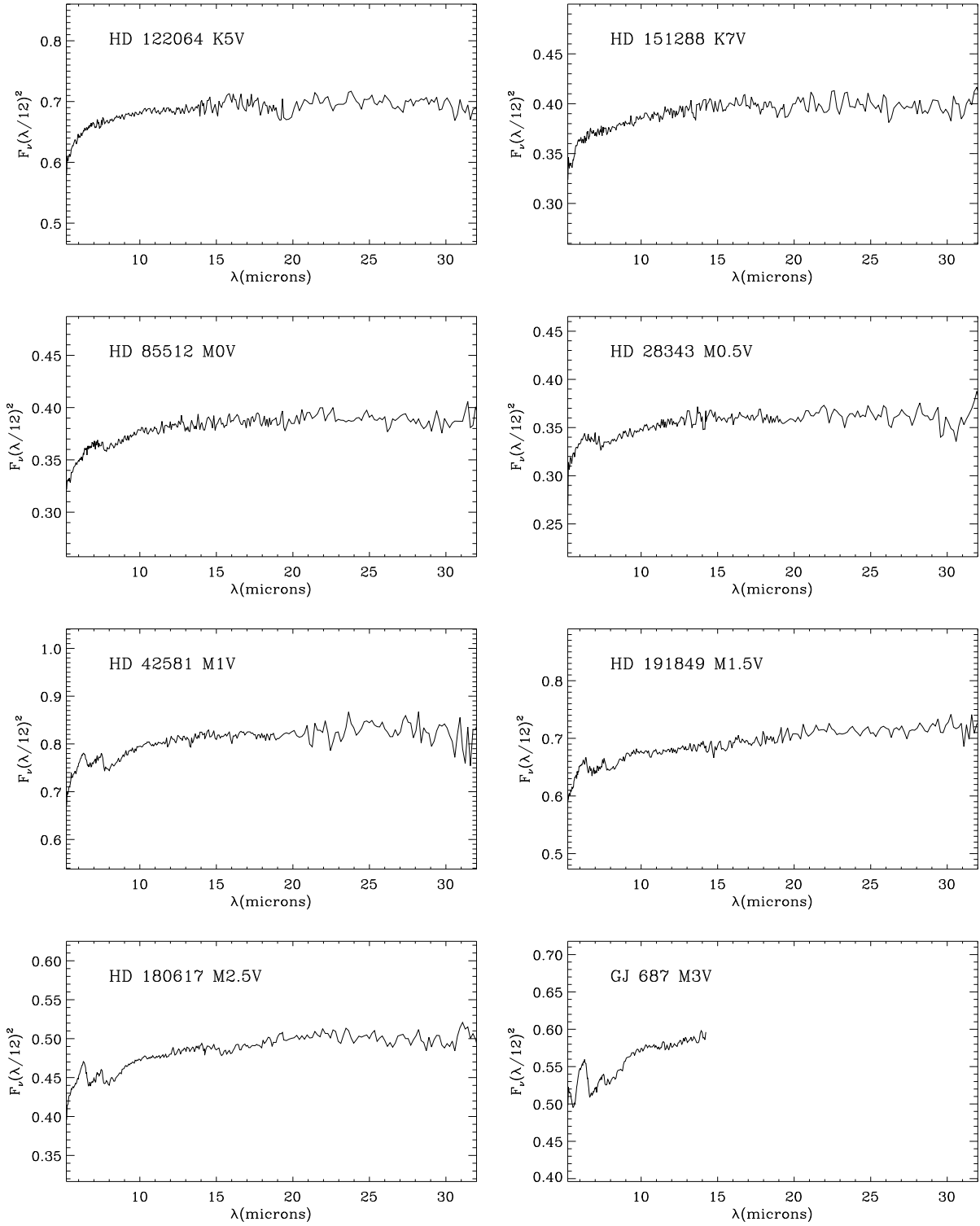}
\caption{Luminosity class V, K5 to M3.  \label{panel_V5}}
\end{figure}

\begin{figure}
\epsscale{1}
\plotone{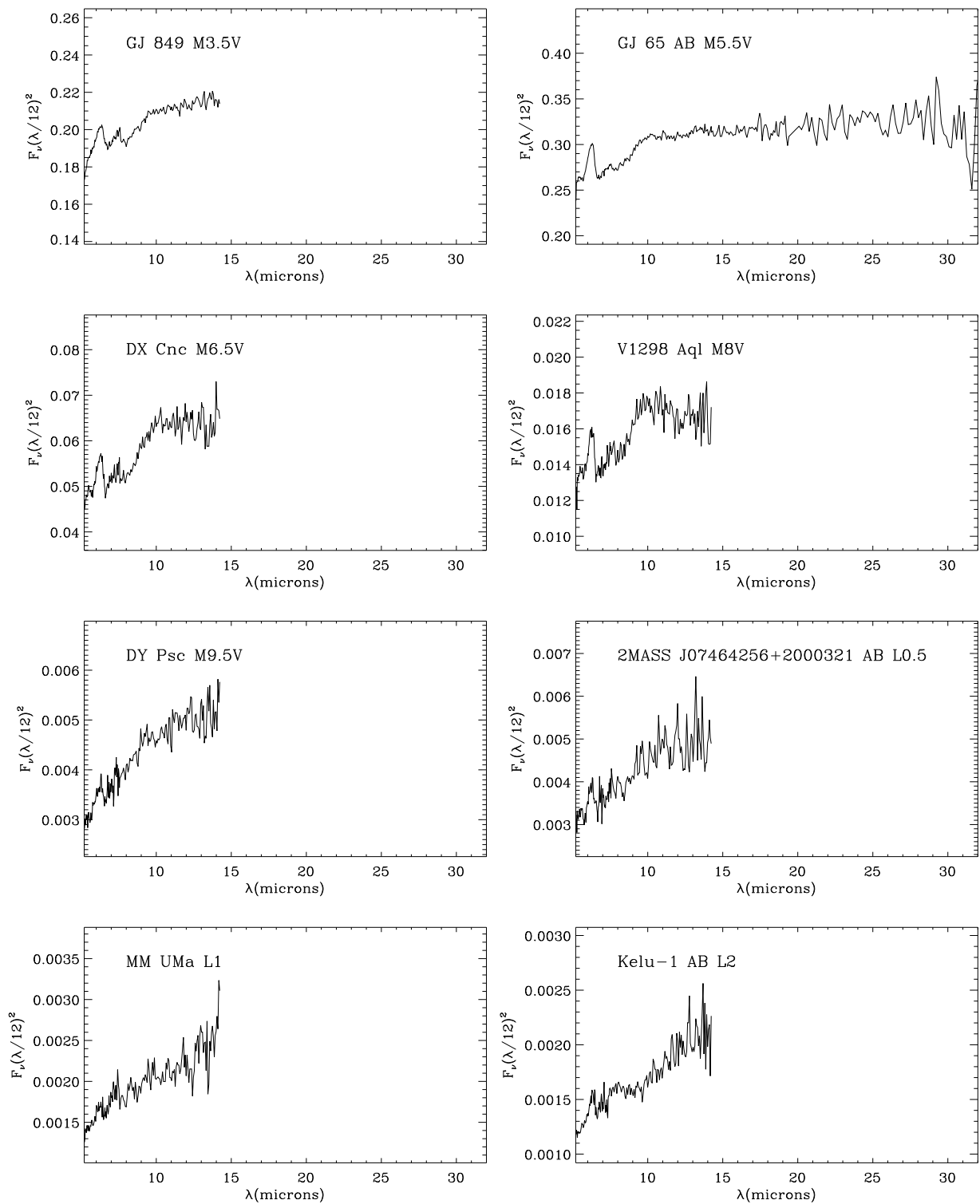}
\caption{Luminosity class V, M3.5 to L2.  \label{panel_V6}}
\end{figure}

\begin{figure}
\epsscale{1}
\plotone{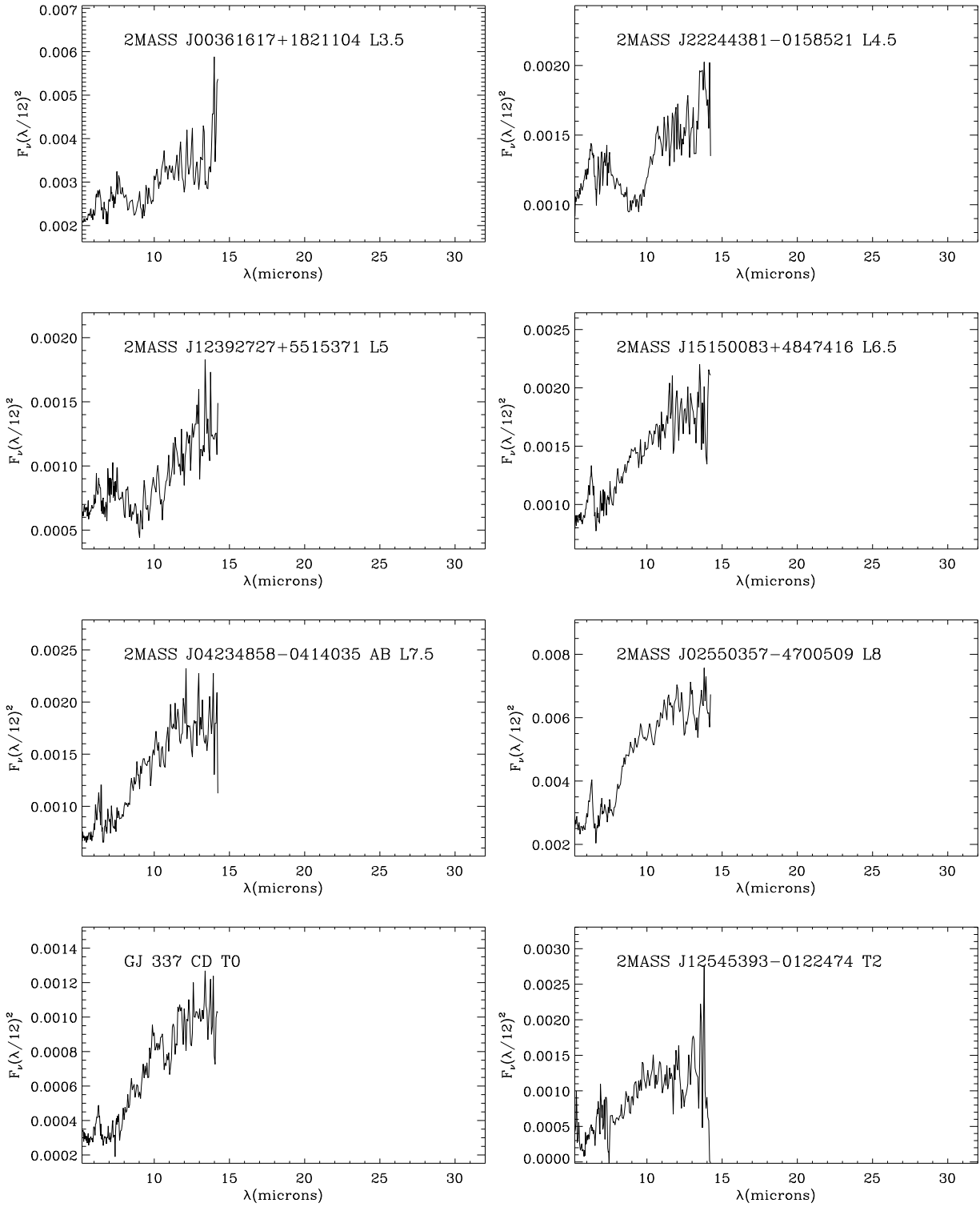}
\caption{Luminosity class V, L3.5 to T2.  \label{panel_V7}}
\end{figure}

\begin{figure}
\epsscale{1}
\plotone{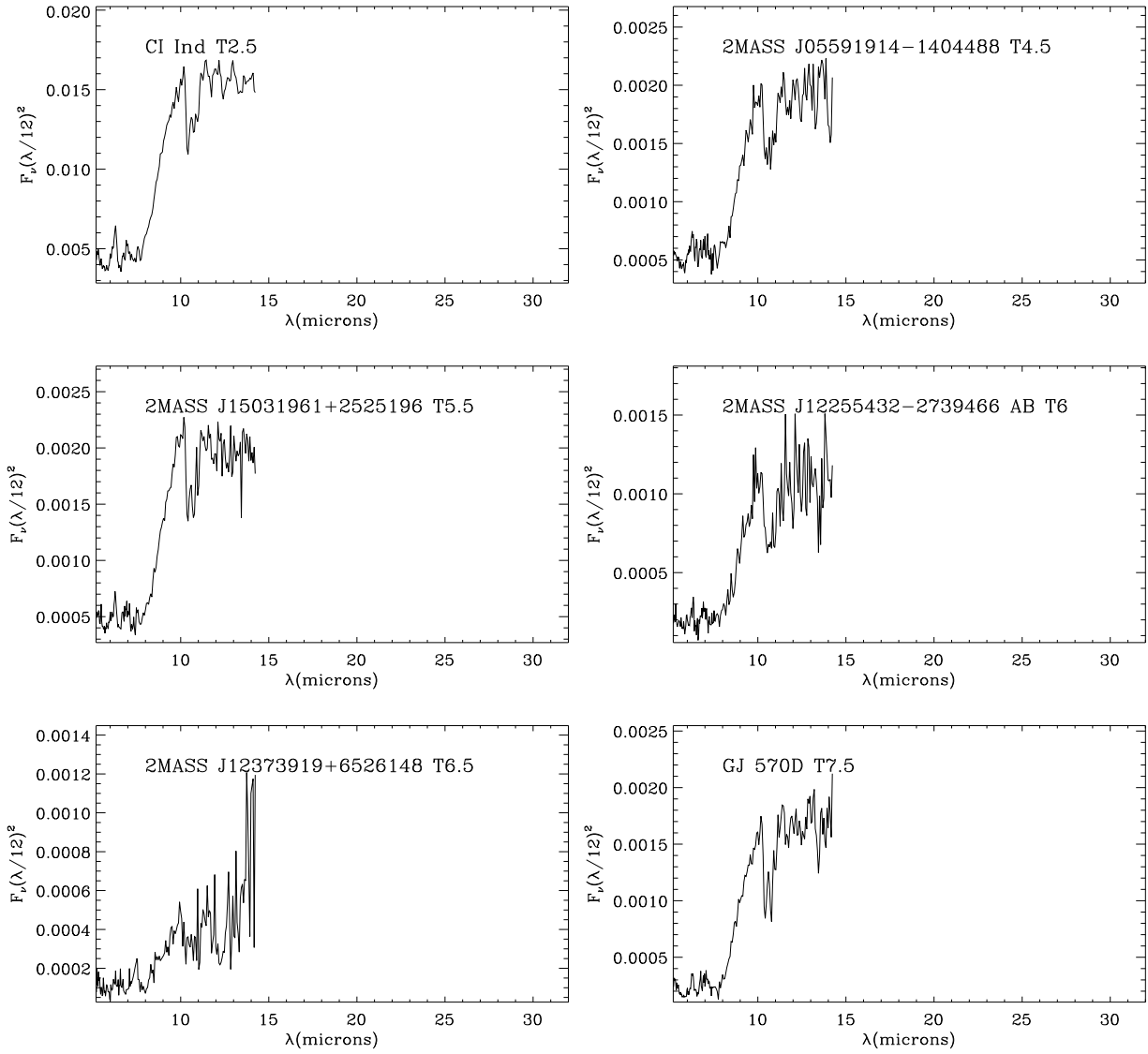}
\caption{Luminosity class V, T2.5 to T7.5.  \label{panel_V8}}
\end{figure}

Table \ref{big_table} lists the stars included in the Atlas and the ancillary data associated with them. Figures \ref{panel_WR1} to \ref{panel_V8} present the Atlas spectra organized by class. The ordinate in all plots is F$_\nu (\lambda/12)^2$, in Jy.  Therefore, a pure Rayleigh-Jeans slope should look like a straight horizontal line.  

The spectra are available electronically from the NASA/IPAC Infrared Science Archive (IRSA), Vizier, and from the first author's webpage (http://web.ipac.caltech.edu/staff/ardila/Atlas/). All the electronic files contain the following information:

\begin{itemize}
\item Target: The primary target name, as selected for the Atlas. Priority is given to the HD name, if it exists.
\item RA: Right Ascension. ICRS 2000.0, from SIMBAD (Epoch 2000.00).
\item Dec: Declination. ICRS 2000.0, from SIMBAD (Epoch 2000.00).
\item AOR\#: The AORs from the {\it Spitzer} archive used in the construction of the final spectrum, separated by spaces. Not all the orders available in a given AOR may have been used. 
\item  DATE\_OBS\#: The observation date associated with each AOR above.
\item  Date (UTC): The creation date for the file.
\item manual scaling: The multiplicative factor applied to each order, to match the overall spectrum.
\item scale sl2: Multiplicative factor applied to SL2.
\item scale sl1: Multiplicative factor applied to SL1.
\item scale ll2: Multiplicative factor applied to LL2.
\item scale ll1: Multiplicative factor applied to LL1.
\end{itemize}

In addition, each electronic file provides four columns of data:
\begin{itemize}
\item Column 1 - Order: The order in which the wavelength is detected (SL2, SL1, LL2, or LL1)
\item Column 2 - Wavelength: Wavelength in microns
\item Column 3 - Flux: Flux density in Jansky
\item Column 4 - Error: One-sigma error in Jansky
\end{itemize}

\begin{deluxetable}{llllrrlllcl}
\rotate
\tablecolumns{11}
\tabletypesize{\tiny}
\tablewidth{0pt}
\tablecaption{Atlas Contents\label{big_table}}
\tablehead{\colhead{NAME}	&	\colhead{NAME}	&	 \colhead{SpT}	& \colhead{KSPW} &	\colhead{B-V}	&	 \colhead{J}	&\colhead{Flux 8 $\mu$m}	&	\colhead{Flux 24 $\mu$m}	&	\colhead{Metallicity} 	& \colhead{AORs} & \colhead{Comments} \phm{    }\\
  & & & & (mag) & (mag)  & (Jy) & (Jy) & ([Fe/H]) & &	}
\startdata
\cutinhead{Wolf-Rayet Stars}
WR 2 & HD 6327 & WN2 & 2.: & 0.19 & 10.07 & 4.84E-02 & 2.17E-02 & \nodata & 1 & Non-photo; Em. lines; SpT ref. 1\\
WR 1 & HD 4004 & WN4 & 2.: & 0.54 & 8.21 & 2.81E-01 & 1.31E-01 & \nodata & 1 & Non-photo; Em. lines; SWS (7); SpT ref. 1 \\
WR 138 & HD 193077 & WN5+B & 2.: & 0.28 & 6.97 & 4.02E-01 & 1.32E-01 & \nodata & 1 & Non-photo; Em. lines; SpT ref. 1 \\
WR 134 & HD 191765 & WN6 & 2.: & 0.19 & 6.72 & 9.18E-01 & 3.47E-01 & \nodata & 1 & Non-photo; Em. lines; SWS (7); SpT ref. 1 \\
WR 115 & HIP 90299 & WN6+OB? & 2.: & 1.18 & 7.99 & 3.82E-01 & 1.24E-01 & \nodata & 1 & Non-photo; Em. lines; SpT ref. 1 \\
WR 145 & V1923 Cyg & WN7/WCE+? & : & 1.70 & 7.37 & 6.47E-01 & \nodata & \nodata & 2 & Non-photo; Em. lines; SL only; SpT ref. 1 \\
HD 269227 & BAT99 22 & WN9h & 3.SE & 0.11 & 9.41 & 1.45E-01 & 7.94E-02 & \nodata & 1 & Non-photo; Em. lines; LMC; LBV cand.; SpT ref. 2 \\
HD 269858 & BAT99 83 & Ofpe/WN9 & 4.E: & 0.31 & 9.63 & 4.82E-02 & 1.31E+00 & \nodata & 1 & Non-photo; LMC; LBV; SpT ref. 3 \\
HD 152386 & 2MASS J16550644-4459213 & O6Ifpe/WN & : & 0.44 & 6.62 & 2.83E-01 & \nodata &\nodata & 1 & Non-photo; Sil. abs.; Em. lines; SL only; SpT ref. 4 \\
2MASS J05455192-6714259 & BAT99 133 & WN11h & 4.E: & -0.13 & 12.03 & 2.05E-02 & 1.47E+00 & \nodata & 1 & Non-photo; Em. lines; LMC; LBV; SpT ref. 3 \\
WR  144 & 2MASS J20320302+4115205 & WC4 & 2.: & 2.02 & 9.41 & 1.65E-01 & 5.61E-02 & \nodata & 2 & Non-photo; Em. lines; SpT ref. 1 \\
WR 111 & HD  165763 & WC5 & 2.: & 0.13 & 7.28 & 3.66E-01 & 1.24E-01 & \nodata & 1 & Non-photo; Em. lines; SpT ref. 1 \\
WR 23 & HD 92809 & WC6 & : & 0.31 & 7.89 & 2.45E-01 & \nodata & \nodata & 2 & Non-photo; Em. lines; SL only; SpT ref. 1 \\
WR 135 & HD 192103 & WC8 & 2.: & 0.16 & 7.23 & 3.36E-01 & 1.21E-01 & \nodata & 2 & Non-photo; Em. lines; SWS (7); SpT ref. 1 \\
WR 53 & HD 117297 & WC8d & 2.: & 0.39 & 8.75 & 9.47E-01 & 1.46E-01 & \nodata & 2 & Non-photo; Em. lines; SpT ref. 1 \\
WR 103 & HD 164270 & WC9d+? & 2.: & 0.17 & 7.75 & 1.27E+00 & 2.13E-01 & \nodata & 2 & Non-photo; Em. lines; SpT ref. 1 \\
Brey 3a & BAT99 4 & WC9+O8V & 4.E: & \nodata & \nodata & 7.82E-02 & 1.39E+00 & \nodata & 1 & Non-photo; Em. lines; LMC; SpT ref. 3 \\
WR 142 & 2MASS J20214434+3722306 & WO2 & : & 1.43 & 9.54 & 1.27E-01 & \nodata & \nodata & 1 & Non-photo; Em. lines; SL only; SpT ref. 1 \\
WR 93b & 2MASS J17320330-3504323 & WO3 & : & \nodata & 11.33 & 2.90E-02 & \nodata & \nodata & 1 & Non-photo; Em. lines; SL only; SpT ref. 1 \\
WR 30a & V574 Car & WO4+O5.5 & : & \nodata & \nodata & 1.73E-02 & \nodata & \nodata & 1 & Non-photo; Em. lines; SL only; SpT ref. 1 \\
\cutinhead{Luminosity Class I - Supergiants}
HD 190429 & HIP 98753 & O4If+ & 2.: & 0.15 & \nodata & 2.95E-01 & 6.03E-02 & \nodata & 3 & Non-photo; Em. lines; Sil. abs.; Em. lines; SpT ref. 4 \\
HD 14947 & HIP 11394 & O5If+ & 2.: & 0.35 & 7.04 & 1.57E-01 & 3.25E-02 & \nodata & 1 & Non-photo; Em. lines; Sil. abs.; Em. lines; SpT ref. 4 \\
HD 210839 & lam Cep & O6I(n)fp & 2.: & 0.17 & 5.05 & 1.18E+00 & 2.21E-01 & \nodata & 1 & Non-photo; Em. lines; Sil. abs.; Em. lines; SpT ref. 4 \\
HD 108 & HIP 505 & O6f?pe & 2.: & 0.11 & 6.89 & 2.16E-01 & 3.28E-02 & \nodata & 1 & Non-photo; Em. lines; SpT ref. 5 \\
HD 192639 & HIP 99768 & O7Ib(f) & 2.: & 0.35 & 6.30 & 2.45E-01 & 4.40E-02 & \nodata & 1 & Non-photo; Sil. abs.; Em. lines; SpT ref. 4 \\
HD 154368 & V1074 Sco & O9Iab & 2.SA & 0.39 & 5.02 & 9.94E-01 & 1.43E-01 & \nodata & 1 & Non-photo; Sil. abs.; Em. lines; SpT ref. 6 \\
HD 209975 & 19 Cep & O9.5Ib & 2.SA & 0.08 & \nodata & 7.18E-01 & 1.05E-01 & \nodata & 1 & Non-photo; Sil. abs.; Em. lines; SpT ref. 6 \\
HD 152424 & HIP 82783 & OC9.7Ia & : & 0.35 & 5.32 & 7.16E-01 & \nodata & \nodata & 1 & Non-photo; Sil. abs.; SL only; SpT ref. 4 \\
V1429 Aql & MWC 314 & B[e]Ia & 2.: & 1.41 & 6.09 & 2.16E+00 & 9.78E-01 & \nodata & 1 & Non-photo; Sil. abs.; Em. lines; LBV candidate  \\
HD 14143 & HIP 10816 & B0Ia & 2.SA & 0.50 & 5.53 & 5.93E-01 & 9.08E-02 & \nodata & 1 & Non-photo; Sil. abs.; Em. lines \\
HD 326823 & V1104 Sco & B1.5Ie & 2.CE & 0.70 & 6.71 & 3.67E+00 & 1.05E+00 & \nodata & 1 & Non-photo; Em. lines; LBV; SpT ref. 7 \\
V433 Sct & HIP 90267 & B1.5Ia & 2.SA & 1.20 & 5.03 & 2.33E+00 & 4.26E-01 & \nodata & 1 & Non-photo; Sil. abs.; Em. lines; SpT ref. 8 \\
HD 14134 & V520 Per & B3Ia & 2.SA & 0.44 & 5.56 & 5.60E-01 & 8.11E-02 & \nodata & 1 & Non-photo; Sil. abs.; Em. lines \\
Cyg OB2 No.12 & HIP 101364 & B5Ie & 2.SA & 0.01 & 4.67 & 1.15E+01 & 1.96E+00 & \nodata & 2 & Non-photo; Sil. abs.; LBV candidate \\
HD 183143 & HT Sge & B7Ia & 2.SA & 1.00 & 4.18 & 3.97E+00 & 6.05E-01 & \nodata & 1 & Non-photo; Sil. abs.; LBV candidate \\
HD 116119 & V965 Cen & B9.5Ia & : & 0.60 & 6.12 & 3.93E-01 & \nodata & \nodata & 1 & Non-photo; SL only \\
HD 13476 & HR 641 & A1Ia & 2.SA & 0.50 & 5.38 & 1.03E+00 & 1.41E-01 & \nodata & 1 & Non-photo?; Sil. abs. \\
HD 160529 & V905 Sco & A2Ia & 2.: & 1.04 & 3.55 & 9.59E+00 & 1.60E+00 & \nodata & 1 & Non-photo; Sil. abs.; Em. lines; LBV; SpT ref. 9 \\
HD 14433 & HIP 11020 & A2Ia & 2.SA & 0.58 & 5.08 & 9.96E-01 & 1.41E-01 & \nodata & 1 & Non-photo; Sil. abs. \\
HD 90772 & V399 Car & A9Ia & 2.SA & 0.51 & 3.52 & 5.05E+00 & 7.37E-01 & \nodata & 1 & Non-photo; Sil. abs.; Cepheid \\
HD 127297 & V Cen & F5Ib & 1.SA & 0.77 & 5.18 & 1.10E+00 & 1.36E-01 & \nodata & 1 & Non-photo; Sil. abs.; Cepheid \\
HD 133683 & HR 5621 & F5Ib & 1.N: & 0.64 & 4.53 & 1.69E+00 & 2.00E-01 & \nodata & 1  & \\
HD 52973 & $\zeta$ Gem & G0Ib & 1.N & 0.88 & 2.38 & 1.03E+01 & 1.22E+00 & \nodata & 1 & $\delta$ Cep type \\
HD 187929 & $\eta$ Aql & G1Ib & 1.N & 0.74 & 2.35 & 1.24E+01 & 1.47E+00 & \nodata & 1 & $\delta$ Cep type \\
HD 49396 & HR 2513 & G2Ib & 1.N & 1.07 & 4.93 & 1.36E+00 & 1.65E-01 & \nodata & 1  & \\
HD 45829 & HIP 30970 & K2Iab & 1.NO & 1.61 & 4.31 & 3.62E+00 & 4.49E-01 & \nodata & 1 & SiO  \\
NGC 7419 \#435 & 2MASS J22541609+6049289 & M2Iab & 2.U: & 1.48 & 5.64 & 3.78E+00 & 8.21E-01 & \nodata & 1 & Non-photo; PAHs; SpT ref. 10 \\
RSGC 2 \#2 & Stephenson 2 \#2 & M3Iab & 3.: & \nodata & \nodata & 7.60E+00 & 1.17E+01 & \nodata & 1 & Non-photo \\
NGC 7419 \#139 & 2MASS J22540122+6047417 & M3.5Iab & 2.U: & 3.92 & 6.06 & 3.73E+00 & 9.20E-01 & \nodata & 1 & Non-photo; PAHs; SpT ref. 10 \\
RSGC 2 \#5 & Stephenson 2 \#5 & M4Iab & 3.: & \nodata & \nodata & 1.63E+00 & 2.80E+00 & \nodata & 1 & Non-photo \\
BD+23 1138 & IRAS 05564+2345 & M5Ia & 2.U: & 2.52 & 5.07 & 3.82E+00 & 1.01E+00 & \nodata & 1 & Non-photo; PAHs \\
\cutinhead{Luminosity Class II - Bright Giants}
HD 53244 & $\gamma$ CMa & B5II & 2.F: & -0.11 & 4.59 & 1.05E+00 & 1.30E-01 & \nodata & 1 & Non-photo? \\
HD 47306 & N Car & A0II & 1.SA & -0.03 & 4.30 & 1.38E+00 & 1.62E-01 & \nodata & 1 & Sil. abs. \\
HD 74272 & n Vel & A5II & 1.SA & 0.12 & 4.34 & 1.53E+00 & 1.81E-01 & \nodata & 1 & Sil. abs. \\
HD 571 & 22 And & F0II & 1.N & 0.36 & 4.35 & 2.04E+00 & 2.43E-01 & \nodata & 1  & \\
HD 203156 & V1334 Cyg & F1II & 1.SA & 0.46 & 5.17 & 1.09E+00 & 1.28E-01 & \nodata & 1 & Sil. abs. \\
HD 142941 & S TrA & F8II & 1.N & 0.77 & 4.95 & 9.54E-01 & 1.16E-01 & \nodata & 1  & \\
HD 114988 & HIP 64543 & G0II & 1.N & 0.76 & 5.16 & 9.36E-01 & 1.09E-01 & \nodata & 1  & \\
HD 150331 & HR 6192 & G1II & 1.N & 0.63 & 4.96 & 1.38E+00 & 1.70E-01 & \nodata & 1  & \\
HD 9900 & HR 461 & K0II & 1.NO & 1.40 & 3.30 & 7.44E+00 & 8.85E-01 & \nodata & 1 & SiO \\
HD 170053 & BD+06 3773 & K2II & 1.NO & 1.42 & 5.01 & 1.75E+00 & 2.09E-01 & \nodata & 1 & SiO \\
\cutinhead{Luminosity Class III - Giants}
HD 24912 & Menkhib & O7.5III(n)((f)) & 2.: & -0.02 & 3.99 & 1.74E+00 & 2.48E-01 & \nodata & 1 & Non-photo; Sil. abs.; Em. lines \\
HD 36861 & lam01 Ori & O8III & 2.Fe & -0.16 & \nodata & 1.64E+00 & 2.17E-01 & \nodata & 1 & Non-photo; Em. lines; SpT ref. 4 \\
HD 205021 & Alfirk & B1IIIev & 2.Fe & -0.20 & \nodata & 1.55E+00 & 1.99E-01 & \nodata & 1 & Non-photo; $\beta$ Cep proto; Em. lines \\
HD 198001 & $\epsilon$ Aqr & A0III & 1.N & 0.01 & 3.85 & 2.24E+00 & 2.60E-01 & \nodata & 1  & \\
HD 186688 & SU Cyg & F2III & 1.N & 0.60 & 5.79 & 5.63E-01 & 6.71E-02 & \nodata & 1 & $\delta$ Cep type \\
HD 182989 & RR Lyr & F5III & 1.N & 0.16 & 6.95 & 1.60E-01 & 1.94E-02 & \nodata & 1 & RR Lyrae prototype \\
HD 55052 & 48 Gem & F5III & 1.N & 0.36 & 5.20 & 7.55E-01 & 9.20E-02 & 0.01 & 1  & \\
HD 189005 & 60 SgrA & G8III & 1.N & 0.88 & 3.36 & 5.68E+00 & 6.63E-01 & \nodata & 1  & \\
HD 68312 & HR 3212 & G9III & 2.F: & 0.90 & 3.79 & 3.43E+00 & 4.05E-01 & \nodata & 1 & Non-photo? \\
HD 181597 & HR 7341 & K0III & 1.NO & 1.12 & 4.47 & 2.29E+00 & 2.74E-01 & \nodata & 91 & SiO; SWS (7) \\
HD 59239 & HIP 35809 & K1III & 1.NO & 1.08 & 4.98 & 1.45E+00 & 1.75E-01 & \nodata & 5 & SiO  \\
HD 50160 & HIP 32396 & K2III & 1.NO & 1.24 & 5.54 & 9.04E-01 & 1.07E-01 & \nodata & 4 & SiO \\
HD 172651 & HIP 91655 & K3III & 1.NO & 1.47 & 4.93 & 1.62E+00 & 1.97E-01 & \nodata & 1 & SiO \\
HD 15508 & HIP 11364 & K4III & 1.NO & 1.45 & 5.26 & 1.35E+00 & 1.67E-01 & \nodata & 2 & SiO  \\
HD 19241 & HIP 14188 & K5III & 1.NO & 1.44 & 5.07 & 1.97E+00 & 2.43E-01 & \nodata & 4 & SiO  \\
HD 107893 & HIP 60491 & M0III & 1.NO & 1.67 & 4.84 & 2.77E+00 & 3.47E-01 & \nodata & 1 & SiO; H2O \\
HD 206503 & HIP 107416 & M1III & 1.NO & 1.67 & 4.92 & 2.03E+00 & 2.64E-01 & \nodata & 1 & SiO; H2O \\
HD 189246 & HIP 98501 & M2III & 1.NO & 1.65 & 4.40 & 2.51E+00 & 3.27E-01 & \nodata & 1 & SiO; H2O \\
HD 223306 & DT Tuc & M3III & 1.NO & 1.54 & 5.25 & 1.74E+00 & 2.24E-01 & \nodata & 1 & SiO; H2O \\
V836 Oph & 2MASS J17431137-1000143 & M4III & 2.SEa & 0.10 & 5.61 & 3.20E+00 & 9.30E-01 & \nodata & 1 & Non-photo; Mira Cet type; H2O; dust features \\
HD 46396 & AX Dor & M4III & 1.NO & 1.51 & 4.37 & 3.46E+00 & 4.53E-01 & \nodata & 1 & SiO; H2O \\
HD 74584 & HIP 42617 & M5III & 1.NO & 1.56 & 4.79 & 3.85E+00 & 5.09E-01 & \nodata & 1 & SiO; H2O \\
HD 8680 & BZ Phe & M6III & 1.NO & 1.54 & \nodata & 1.91E+00 & 2.49E-01 & \nodata & 1 & SiO; H2O; SpT ref. 11 \\
\cutinhead{Luminosity Class IV - Subgiants}
HD 11415 & $\epsilon$ Cas & B3IVp & 1.N & -0.12 & 3.86 & 1.88E+00 & 2.21E-01 & \nodata & 1  & \\
HD 210049 & HU PsA & A1.5IVn & 1.N & 0.06 & 4.46 & 1.23E+00 & 1.44E-01 & \nodata & 2  & \\
HD 8538 & $\delta$ Cas & A5IVv & 1.N & 0.13 & 2.34 & 7.94E+00 & 9.31E-01 & \nodata & 1  & \\
HD 101132 & pi Cha & A9IV & 1.N & 0.31 & 4.97 & 9.04E-01 & 1.09E-01 & \llap -0.35 & 1  & \\
HD 211336 & $\epsilon$ Cep & F0IV & 1.N & 0.28 & 3.84 & 2.82E+00 & 3.40E-01 & \nodata & 1  & \\
HD 210210 & HR 8441 & F1IV & 1.N & 0.28 & 5.46 & 5.34E-01 & 6.19E-02 & \nodata & 1  & \\
HD 131495 & HIP 72807 & F2IV & 1.N & 0.39 & \nodata & 3.38E-01 & 3.99E-02 & \llap -0.13 & 1  & \\
HD 172748 & $\delta$ Sct & F2.5IV & 1.N & 0.33 & 4.22 & 1.92E+00 & 2.30E-01 & 0.41 & 1 & $\delta$ Scu prototype \\
HD 142860 & 41 Ser & F5IV & 1.N & 0.48 & 3.15 & 5.96E+00 & 6.88E-01 & \llap -0.19 & 1  & \\
HD 127243 & g Boo & G0IV & 1.N & 0.85 & 3.85 & 2.92E+00 & 3.53E-01 & \nodata & 1  & \\
HD 62644 & HR 2998 & G6IV & 1.N & 0.80 & 3.68 & 3.44E+00 & 4.04E-01 & \nodata & 1  & \\
HD 121146 & HR 5227 & K0IV & 1.NO & 1.19 & 4.35 & 2.44E+00 & 2.91E-01 & \llap -0.16 & 1 & SiO \\
HD 222803 & HR 8993 & K1IV & 1.N & 0.98 & 4.27 & 2.37E+00 & 2.77E-01 & \nodata & 1  & \\
\cutinhead{Luminosity Class V - Dwarfs}
HD 46150 & HIP 31130 & O5V((f)) & : & 0.12 & 6.45 & 1.59E-01 & \nodata & \nodata & 1 & Non-photo; Sil. abs.; SL only; SpT ref. 4  \\
HD 149757 & $\zeta$ Oph & O9.5V & 1.SA & 0.02 & 2.53 & 5.33E+00 & 6.21E-01 & \nodata & 1 & Sil. abs.; SWS (1.N:) \\
HD 27396 & V469 Per & B3V & 2.F & -0.04 & 5.11 & 6.35E-01 & 9.21E-02 & \nodata & 1 & Non-photo; $\beta$ Cep type \\
HD 45813 & $\lambda$ CMa & B5V & 1.N & -0.15 & 5.06 & 6.26E-01 & 7.20E-02 & \nodata & 1  & \\
HD 2884 & $\beta^1$ Tuc & B9V & 1.N & -0.05 & \nodata & 1.03E+00 & 1.20E-01 & \nodata & 1  & \\
HD 213558 & $\alpha$ Lac & A0V & 1.N & 0.03 & 3.83 & 2.05E+00 & 2.38E-01 & \nodata & 24 & Humpreys lines; SWS (1.N:) \\
HD 109787 & $\tau$ Cen & A2V & 1.N & 0.06 & 3.80 & 2.32E+00 & 2.69E-01 & \nodata & 7  & \\
HD 20888 & HR 1014 & A3V & 1.N & 0.14 & 5.78 & 3.38E-01 & 3.96E-02 & \nodata & 5  & \\
HD 73210 & HIP 42327 & A5V & 1.N & 0.16 & \nodata & 2.32E-01 & 2.85E-02 & \nodata & 1  & \\
HD 27290 & $\gamma$ Dor & F0V & 2.F & 0.35 & 3.68 & 2.63E+00 & 3.28E-01 & \nodata & 3 & Non-photo; $\gamma$ Dor proto. \\
HD 88923 & HIP 50315 & F2V & 2.F: & 0.32 & 6.81 & 1.69E-01 & 2.12E-02 & \nodata & 1 & Non-photo?; blue straggler; SpT ref. 12 \\
HD 164259 & $\zeta$ Ser & F2V & 1.N & 0.36 & 3.98 & 2.51E+00 & 2.96E-01 & \llap -0.14 & 1  & \\
HD 151900 & HR 6248 & F3V & 1.N & 0.36 & 5.46 & 5.73E-01 & 6.74E-02 & \llap -0.45 & 1  & \\
HD 134083 & 45 Boo & F5V & 1.N & 0.43 & 4.25 & 1.92E+00 & 2.24E-01 & 0.02 & 1  & \\
HD 210302 & $\tau$ PsA & F6V & 1.N & 0.48 & 3.93 & 2.12E+00 & 2.50E-01 & 0.05 & 1  & \\
HD 106516 & HR 4657 & F6V & 1.N & 0.46 & 5.13 & 7.50E-01 & 8.79E-02 & \llap -0.67 & 1 & Blue straggler \\
HD 68146 & 18 Pup & F7V & 1.N & 0.49 & 4.14 & 1.27E+00 & 1.49E-01 & \llap -0.12 & 1  & \\
HD 132254 & HR 5581 & F8V & 2.F: & 0.53 & 4.69 & 1.21E+00 & 1.44E-01 & 0.02 & 1  & \\
HD 35863 & HIP 26008 & F8V & 1.N & 0.36 & 6.02 & 3.13E-01 & 3.66E-02 & \llap -0.39 & 1 & Blue straggler; SpT ref. 12 \\
HD 114710 & $\beta$ Com & G0V & 1.N & 0.57 & 3.23 & 5.02E+00 & 5.93E-01 & \llap -0.06 & 1  & \\
HD 168009 & HR 6847 & G1V & 1.N & 0.60 & 5.12 & 8.18E-01 & 9.68E-02 & \llap -0.11 & 1  & \\
HD 134060 & HR 5632 & G2V & 1.N & 0.58 & 5.21 & 7.58E-01 & 8.91E-02 & \llap -0.10 & 1  & \\
HD 126053 & HR 5384 & G5V & 1.N & 0.60 & 5.05 & 9.09E-01 & 1.08E-01 & \llap -0.39 & 1  & \\
HD 128987 & KU Lib & G6Vk & 1.N & 0.68 & 5.95 & 3.81E-01 & 4.49E-02 & 0.02 & 1  & \\
HD 111395 & LW Com & G7V & 1.N & 0.70 & 5.12 & 9.47E-01 & 1.11E-01 & \llap -0.05 & 1  & \\
HD 14412 & HR 683 & G8V & 1.N & 0.73 & 5.06 & 1.07E+00 & 1.24E-01 &\llap  -0.46 & 1  & \\
HD 182488 & HR 7368 & G9V & 1.N & 0.79 & 5.37 & 1.05E+00 & 1.25E-01 & 0.14 & 1  & \\
HD 128165 & GJ 556 & K0V & 1.N & 0.99 & 5.44 & 8.20E-01 & 9.75E-02 & 0.04 & 1  & \\
HD 149661 & V2133 Oph & K1V & 1.N & 0.81 & 4.45 & 2.05E+00 & 2.39E-01 & \llap -0.01 & 1  & \\
HD 154577 & GJ 656 & K2V & 1.N & 0.89 & 5.69 & 6.39E-01 & 7.41E-02 & \llap -0.58 & 1  & \\
HD 192310 & HR 7722 & K3V & 1.N & 0.88 & 4.11 & 2.37E+00 & 2.78E-01 & 0.14 & 1 & SWS (1.N:) \\
HD 131977 & KX Lib & K4V & 1.N & 1.11 & 3.66 & 3.99E+00 & 4.67E-01 & \nodata & 1  & \\
HD 122064 & HR 5256 & K5V & 1.N & 1.05 & 5.02 & 1.62E+00 & 1.92E-01 & \nodata & 1  & \\
HD 151288 & GJ 638 & K7V & 1.N & 1.37 & 5.48 & 9.09E-01 & 1.09E-01 & \nodata & 1  & \\
HD 85512 & GJ 370 & M0V & 1.NO & 0.90 & 5.45 & 8.84E-01 & 1.07E-01 & \nodata & 1 & SiO \\
HD 28343 & GJ 169 & M0.5V & 1.NO & 1.38 & 5.67 & 8.19E-01 & 9.95E-02 & \nodata & 1 & SiO \\
HD 42581 & GJ 229A & M1V & 1.NO & 1.51 & 5.10 & 1.85E+00 & 2.26E-01 & \nodata & 1 & SiO; H2O; SpT ref. 13 \\
HD 191849 & GJ 784 & M1.5V & 1.NO & 1.46 & 5.12 & 1.59E+00 & 1.95E-01 & \nodata & 1 & SiO; H2O \\
HD 180617 & V1428 Aql & M2.5V & 1.NO & 1.50 & 5.58 & 1.09E+00 & 1.37E-01 & \nodata & 1 & SiO; H2O; SpT ref. 13 \\
GJ 687 & HIP 86162 & M3V & : & 1.50 & 5.34 & 1.30E+00 & \nodata & \nodata & 1 & SL only; SpT ref. 13 \\
GJ 849 & HIP 109388 & M3.5V & : & \nodata & \nodata & 4.77E-01 & \nodata & \nodata & 1 & SL only; SiO; H2O; SpT ref. 13 \\
GJ 65 AB & BL Cet + UV Cet & M5.5V & 1.NO & 1.85/1.87 & \nodata & 6.70E-01 & 8.83E-02 & \nodata & 1 & SiO; H2O; SpT ref. 13 \\
DX Cnc & GJ 1111 & M6.5V & : & 2.06 & 8.23 & 1.29E-01 & \nodata & \nodata & 1 & SL only; SiO; H2O; SpT ref. 13 \\
V1298 Aql & GJ 752B & M8V & : & 2.12 & 9.91 & 3.55E-02 & \nodata & \nodata & 1 & SL only; SiO; H2O; SpT ref. 13 \\
DY Psc & 2MASS J00242463-0158201 & M9.5V & : & 1.78 & 11.99 & 9.50E-03 & \nodata & \nodata & 1 & SL only; H2O; SpT ref. 13 \\
2MASS J07464256+2000321 AB & USNO-B1.0 1100-00150847 AB & L0.5 & : & 4.00 & 11.76 & 9.23E-03 & \nodata & \nodata & 1 & SL only; H2O; SpT ref. 13\\
MM UMa & 2MASS J11083081+6830169 & L1 & : & \nodata & 13.12 & 4.41E-03 & \nodata & \nodata & 1 & SL only; H2O; SpT ref. 13\\
Kelu-1 AB & V421 HyaAB & L2 & : & \nodata & 13.41 & 3.73E-03 & \nodata & \nodata & 1 & SL only; H2O; SpT ref. 13\\
2MASS J00361617+1821104 & USNO-B1.01 083-00010105 & L3.5 & : & \nodata & 12.47 & 6.20E-03 & \nodata & \nodata & 1 & H2O; SpT ref. 13\\
2MASS J22244381-0158521 & [B2006] J222443.8-015852 & L4.5 & : & \nodata & 14.07 & 2.86E-03 & \nodata & \nodata & 1 & SL only; SpT ref. 13 \\
2MASS J12392727+5515371 &  & L5 & : & \nodata & 14.71 & 1.79E-03 & \nodata & \nodata & 1 & SL only; H2O; SpT ref. 13\\
2MASS J15150083+4847416 & [B2006] J151500.8+484741 & L6.5 & : & \nodata & 14.11 & 2.73E-03 & \nodata & \nodata & 1 & SL only; H2O; SpT ref. 13 \\
2MASS J04234858-0414035 AB & SDSS J042348.56-041403.5 AB & L7.5 & : & \nodata & 14.47 & 2.36E-03 & \nodata & \nodata & 1 & SL only; H2O; CH4; SpT ref. 13 \\
2MASS J02550357-4700509 & DENIS-P J025503.3-470049 & L8 & : & \nodata & 13.25 & 8.56E-03 & \nodata & \nodata & 1 & SL only; H2O; CH4; SpT ref. 13 \\
GJ 337 CD & 2MASS J09121469+1459396 CD & T0 & : & \nodata & 15.51 & 9.94E-04 & \nodata & \llap -0.30 & 1 & SL only;  H2O; CH4; SpT ref. 13 \\
2MASS J12545393-0122474 & SDSS J125453.90-012247.5 & T2 & : & \nodata & 14.88 & 1.58E-03 & \nodata & \nodata & 1 & SL only; H2O; CH4; NH3; SpT ref. 13 \\
CI Ind & $\epsilon$ Ind B & T2.5 & : & \nodata & 11.91 & 1.49E-02 & \nodata & \nodata & 1 & SL only; SpT ref. 13 \\
2MASS J05591914-1404488 & [B2006] J055919.1-140448 & T4.5 & : & \nodata & 13.80 & 1.71E-03 & \nodata & \nodata & 2 & SL only; H2O; CH4; NH3; SpT ref. 13 \\
2MASS J15031961+2525196 & 2MASS J1503196+252519 & T5.5 & : & \nodata & 13.94 & 1.61E-03 & \nodata & \nodata & 1 & SL only; SpT ref. 13 \\
2MASS J12255432-2739466 AB & 2MASSW J1225543-273947 AB & T6 & : & \nodata & 15.26 & 7.14E-04 & \nodata & \nodata & 1 & SL only; H2O; CH4; NH3; SpT ref. 13 \\
2MASS J12373919+6526148 & 2MASS J1237392+652615 & T6.5 & : & \nodata & 16.05 & 3.80E-04 & \nodata & \nodata & 1 & SL only; SpT ref. 13 \\
GJ 570D & 2MASS J14571496-2121477 & T7.5 & : & \nodata & 15.32 & 9.77E-04 & \nodata & \nodata & 1 & SL only; SpT ref. 13 \\

\enddata
\tablecomments{B-V colors were taken from SIMBAD; J magnitudes are taken from 2MASS \citep{skr06}; Fluxes at 8 and 24 $\mu$m are measured directly from the spectra (see text); Metallicity is taken from \citet{hol09}; KSPW column: See section \ref{contents}. The `:' means uncertain classification. In the Comments column: ``Non-photo.'': Spectrum is non-photospheric; ``Sil. abs.'': Silicate Absorption; ``Em. lines'': Emission lines; "SWS": This star is part of the sample that defined the KSPW system, original classification in parenthesis. Spectral types are taken from a hierarchy of sources in the literature (see text), except for those sources where a reference number is given. The references are: 1-\citet{vanderhucht01}; 2-\citet{torres1988}; 3-\citet{bre99}; 4-\citet{mai04}; 5-\citet{wal1972}; 6-\citet{han05}; 7-\citet{van01}; 8-\citet{dej98}; 9-\citet{sta03};  10-\citet{bea94}; 11-\citet{jon72}; 12-\citet{abt84}; 13-IR spectral types; \citet{cus06}}

\end{deluxetable}
\clearpage
\clearpage
\clearpage

\clearpage

\clearpage

\clearpage

\clearpage

\clearpage
\clearpage

\clearpage
\clearpage
\clearpage
\clearpage
\clearpage

\clearpage
\clearpage
\clearpage
\clearpage
\clearpage

\clearpage

\clearpage
\clearpage

\clearpage
\acknowledgments

This work is based on observations and archival data from the {\it Spitzer Space Telescope}, which is operated by the Jet Propulsion Laboratory (JPL), California Institute of Technology (Caltech) under a contract with National Aeronautics and Space Administration (NASA). Support for this work was provided by NASA through an award issued by JPL/Caltech. This research has also made use of: the NASA / Infrared Processing and Analysis Center (IPAC) Science Archive, operated by the JPL, Caltech, under contract with NASA; the SIMBAD database and the Vizier service, operated at CDS, Strasbourg, France; the data products from the Two Micron All Sky Survey (2MASS), a joint project of the University of Massachusetts and IPAC/Caltech, funded by NASA and the National Science Foundation. 

{\it Facilities:} \facility{{\it Spitzer} (IRS)}.



\end{document}